\documentclass[aps, prd, amsmath, floatfix, superscriptaddress, preprintnumbers, preprint]{revtex4-1}

\pdfoutput=1

\usepackage{amsmath}
\usepackage{subfigure}
\usepackage{enumerate}
\usepackage{amssymb}
\usepackage{amsfonts}
\usepackage{mathrsfs}
\usepackage{latexsym}
\usepackage{bm}
\usepackage{amscd}
\usepackage{physics} 
\usepackage{graphicx}
\usepackage{epsfig}
\usepackage{hhline,multirow}
\usepackage{dcolumn}
\usepackage{url}
\usepackage[colorlinks=true,linkcolor=blue,citecolor=blue,urlcolor=blue]{hyperref}
\usepackage{soul}
\usepackage[dvipsnames]{xcolor}
\usepackage{color}
\usepackage{slashed}
\usepackage{float}
\usepackage{comment}
\usepackage{mathtools}

\begin{document}

\preprint{JLAB-THY-25-6}

\title{\mbox{Strangeness in the proton from $W$+\,charm production and SIDIS~data}}

\author{Trey~Anderson}
\affiliation{\mbox{Department of Physics, William \& Mary, Williamsburg, 
        Virginia 23185, USA}}
\affiliation{Jefferson Lab, Newport News, Virginia 23606, USA \\
        \vspace*{0.2cm}
        {\bf JAM Collaboration \\ {\footnotesize \ (PDF Analysis Group)}
        \vspace*{0.2cm} }}
\author{W.~Melnitchouk}
\author{N.~Sato}
\affiliation{Jefferson Lab, Newport News, Virginia 23606, USA \\
        \vspace*{0.2cm}
        {\bf JAM Collaboration \\ {\footnotesize \ (PDF Analysis Group)}
        \vspace*{0.2cm} }}

\begin{abstract}
We perform a global QCD analysis of unpolarized parton distribution functions (PDFs) in the proton, including new $W +$\,charm production data from $pp$ collisions at the LHC and semi-inclusive pion and kaon production data in lepton-nucleon deep-inelastic scattering, both of which have been suggested for constraining the strange quark PDF.
Compared with a baseline global fit that does not include these datasets, the new analysis reduces the uncertainty on the strange quark distribution over the range $0.01 < x < 0.3$, and provides a consistent description of processes sensitive to strangeness in the proton.
Including the new datasets, the ratio of strange to nonstrange sea quark distributions is $R_s = (s+\bar s)/(\bar u+\bar d) = \{ 
0.72^{+0.52}_{-0.34},\, 
0.46^{+0.30}_{-0.20},\, 
0.32^{+0.23}_{-0.15} \}$ 
for $x = \{ 0.01, 0.04, 0.1 \}$ at $Q^2 = 4$~GeV$^2$.
The data place more stringent constraints on the strange asymmetry $s-\bar s$, which is found to be consistent with zero in this range.
\end{abstract}

\date{\today}
\maketitle

\section{Introduction}

The simple picture of matter that has been built up over the past few decades of probing the femtometer scale structure of the proton attributes the bulk of its properties, such as baryon number, charge, or magnetic moment, to the irreducible core of its valence $u$ and $d$ quarks.
In momentum space the distributions of these quarks have been mapped out in considerable detail from deep-inelastic scattering (DIS) and other high-energy scattering observables, over a large range of light-cone momentum fractions $x$. 
On top of this structure, we now understand, lies a teeming sea of virtual quark--antiquark pairs. These pairs do not alter the global quantum numbers of the proton, but can affect some of its properties, such as the spin and magnetic moment, and yield nontrivial structures, such as an excess of $\bar d$ antiquarks over $\bar u$ \cite{Thomas:1983fh, NuSea:2001idv, SeaQuest:2021zxb, Cocuzza:2021cbi} in the proton, and an analogous excess of $\bar u$ in the neutron.

While expected to be suppressed because of its larger mass, the creation of strange-antistrange quark pairs is also expected to play a role in the femtoscopic structure of the proton.
In ordinary matter, the creation of virtual $u \bar u$ and $d \bar d$ pairs is screened by the presence of the valence $u$ and $d$ quarks.
Strange quarks, on the other hand, can directly reveal the properties of $q \bar q$ pairs in the proton, although the practical realization of this has been severely hampered by the difficulty of obtaining reliable empirical information on the $s$ and $\bar s$ distributions in the proton.
Although it is anticipated on general grounds that an asymmetry between $s$ and $\bar s$ quarks would be a unique window into the nonperturbative dynamics of quarks in the proton~\cite{Signal:1987gz, Ji:1995rd, Melnitchouk:1996fj} (for example, in relation to the spontaneous breaking of chiral SU(3) symmetry), in practice even the magnitude of the sum $s+\bar s$ is poorly known.

Historically, the shape of the strange quark PDF was first studied in neutrino-induced DIS from nuclear targets, particularly in the semi-inclusive production of charmed mesons in charged current reactions from the CCFR~\cite{CCFR:1994ikl} and NuTeV~\cite{NuTeV:2007uwm} Collaborations at the Tevatron, as well as from the CHORUS~\cite{Kayis-Topaksu:2011ols} and NOMAD~\cite{NOMAD:2013hbk} experiments at CERN.
An important complication in this process is the modeling of nuclear corrections in neutrino-nucleus DIS, which are currently poorly understood \cite{Kalantarians:2017mkj}, and in the treatment of charm quark energy loss and charmed $D$ meson-nucleon rescattering inside the nucleus~\cite{Accardi:2009qv}.
An alternative avenue that avoids nuclear corrections is semi-inclusive DIS production of kaons, which depends on the strange quark PDFs and strange quark to kaon fragmentation functions (FFs) as an additional nonperturbative input.
Attempts have been made to extract the $s+\bar s$ distribution from kaon SIDIS data by the HERMES Collaboration~\cite{HERMES:2008pug, HERMES:2013ztj} (see also Refs.~\cite{Stolarski:2014jka, Leader:2014oxa, Leader:2015hna}).
More recently, inclusive $W$ and $Z$ boson production has been found to have sensitivity to the strange quark PDF, although mixed results have been obtained from measurements at ATLAS and CMS at the LHC~\cite{CMS:2011bet, CMS:2012ivw, CMS:2013pzl, ATLAS:2016nqi, CMS:2016qqr, ATLAS:2012sjl}.
Inclusive $W$+\,charm production in principle could have even greater sensitivity to the $s$ and $\bar s$ PDFs~\cite{CMS:2013wql, ATLAS:2014jkm, CMS:2018dxg}, in analogy with the neutrino DIS measurements, but still without the complications of nuclear targets.

Aside from the intrinsic value of understanding the structure of the proton sea, the precise determination of strange quark PDFs is key for extracting Standard Model parameters, such as the the Cabibbo-Kobayashi-Maskawa matrix element $V_{cs}$ and the weak mixing angle, $\sin^2\theta_W$, as well as precision measurements on the mass of the $W$-boson that depend on precise knowledge of the strange quark PDF~\cite{ATLAS:2017rzl, Alekhin:2017olj}.
In this paper, we revisit the question of the strange and antistrange quark distributions in the proton with a new analysis that includes all of the above datasets involving proton beams and targets.
For the first time we combine the inclusive and semi-inclusive DIS (SIDIS) structure functions and multiplicities, together with cross section ratios and charge asymmetries in $W/Z$ and $W$+\,charm production in $pp$ collisions at the LHC, simultaneously fitting PDFs and FFs to nearly 6\,000 data points.

We contrast the pulls of different datasets on the $s$ and $\bar{s}$ distributions, finding that both SIDIS and $W$+\,charm data reduce uncertainties relative to the baseline that does not include these datasets, although each of the scenarios is consistent within uncertainties.
Furthermore, we find that SIDIS data alone allow for slightly smaller strangeness at $x \lesssim 0.1$, while the $W$+\,charm data alone generally allow for enhanced strangeness for all $x \lesssim 0.3$.  On the other hand, combining both the SIDIS and $W$+\,charm data reduces the overall uncertainty compared to the results with the individual datasets and gives a suppressed strange to nonstrange quark PDF ratio $R_s = (s+\bar{s})/(\bar{d}+\bar{u})$ for $0.02 \lesssim x \lesssim 0.2$ relative to some previous analyses of the ATLAS $W$+\,charm data~\cite{ATLAS:2014jkm}.

We note that our simultaneous reconstruction of PDFs and FFs from all the available datasets leverages the quark flavor sensitivity of the SIDIS data, uniformly propagating uncertainties across all the involved distributions. 
Some previous analyses, specifically of polarized PDFs~\cite{Borsa:2017vwy, DeFlorian:2019xxt}, have included SIDIS data with fixed FFs via a reweighing procedure, which typically requires inflating the experimental errors to include the uncertainties from FFs. 
Because the FFs are fixed, this hinders simultaneously improving knowledge of the PDFs and FFs from data.
Our analysis eliminates this limitation, allowing us to constrain all the nonperturbative functions needed to describe the observational data.

We begin in Sec.~\ref{s.theory} with a review of the theoretical framework used in this analysis, focusing on the new $W$+\,charm and SIDIS observables, and discussing the nonperturbative modeling for the PDFs and FFs.
Section~\ref{s.datasets} provides a concise summary of the datasets used in the fit, including DIS, Drell-Yan lepton-pair production in $pp$ and $pd$ scattering, $W^\pm$, $Z$, $W+$\,charm, and jet production in $pp$ or $p\bar{p}$ reactions. 
The datasets also include pion, kaon, and charged hadron production in SIDIS, as well as single-inclusive $e^+ e^-$ annihilation (SIA).
In Sec.~\ref{s.results}, we discuss in detail our methodology for implementing the regression problem, including model calibration and the kinematical cuts applied to the data. 
We then present the results of the data-versus-theory comparisons and the extracted PDFs and FFs.
Finally, conclusions are drawn in Sec.~\ref{s.conclusions}, where we discuss future theoretical and experimental steps that could provide further insights into the strangeness content of the proton.

\section{Theoretical framework}
\label{s.theory}

The theoretical basis for this analysis is collinear QCD factorization, to fixed order in the QCD coupling $\alpha_s$, for various high-energy scattering processes which involve PDFs or FFs. 
These include DIS, Drell-Yan lepton-pair production, weak boson and jet production, and $W+$\,charm production, which directly constrain proton PDFs; $\pi$, $K$ and unidentified charged hadron production in SIA, which give information on FFs; and SIDIS, which depends on both PDFs and FFs.
Since the focus of this study is specifically on the role of strangeness in the proton, we will discuss in greater detail the theoretical framework for processes most sensitive to the $s$ and $\bar s$ PDFs in the proton, namely, inclusive $W+$\,charm production in $pp$ collisions and lepton-deuteron SIDIS.
We also describe the parametrizations employed for our PDFs and FFs.

\subsection{Physical processes and factorization}

The associated production of a $W$ boson and a charm quark in $pp$ collisions,
\begin{equation}
    p + p \rightarrow W + c + X,
    \label{eq:wc_react}
\end{equation}
where $X$ represents all other particles in the final state, is expected to be sensitive to the strange content of the proton.
Here the charge state $W^- + c$ originates from the scattering off a strange quark, while the $W^+ + \bar c$ tags an antistrange quark.
The $c$ and $\bar c$ quarks are identified within a jet by a muon produced from its semileptonic decay.
Within the collinear factorization framework the differential cross section can be written as a convolution of the perturbatively calculated partonic cross section $\widehat{\sigma}^{W+c}_{a,b}$ and the nonperturbative PDFs $f_a$ and $f_b$ of partons $a$ and $b$ in the colliding protons, 
\begin{eqnarray}
    \frac{\dd \sigma^{W+c}}{\dd |\eta|} 
    &=& \sum_{a,b} \int\limits_{x_a}\!\!\int\limits_{x_b} \dd{\hat{x}}_a \dd{\hat{x}}_b f_a(\hat{x}_a,\mu_{F}) \, f_{b}(\hat{x}_b,\mu_F) \, \widehat{\sigma}^{W+c}_{a,b}\bigg(\frac{\hat{x}_a}{x_a},\frac{\hat{x}_b}{x_b},\mu_F,\mu_R\bigg),
    \label{eq:wccs}
\end{eqnarray}
where $\eta$ is the pseudorapidity of the lepton from the $W$-boson decay.
The PDFs are functions of the partonic momentum fraction variables $\hat{x}_{a,b}$, with the sum in Eq.~(\ref{eq:wccs}) running over parton flavors $a$ and $b$ for the contributing partonic channels. 
At leading order in the strong coupling $\alpha_s(\mu_R)$, the variables $x_{a,b}$ are related to the pseudorapidity by 
    $x_{a,b} = (M_W/\sqrt{s}) \, e^{\pm \eta}$.
The partonic cross sections $\widehat{\sigma}^{W+c}_{a,b}$ is calculated at next-to-leading order (NLO) in $\alpha_{s}(\mu_R)$ using the Monte Carlo program MCFM~\cite{MCFM}, with the factorization scale $\mu_F$ and renormalization scale $\mu_R$ set to \mbox{$\mu_F = \mu_R = M_W$}.

For the SIDIS of a lepton $\ell$ from a nucleon $N$, producing charged pions $\pi^\pm$, kaons $K^\pm$ or unidentified hadrons,
\begin{equation}
    \ell + N \rightarrow \ell + h^{\pm} + X,
    \label{eq:sidis_react}
\end{equation}
the differential cross section can be written as the double convolution of the partonic cross section $\widehat{\sigma}^h_{a,b}$ with the PDF $f_a$ and the parton $b$ to hadron $h$ FF $D^h_b$,
\begin{eqnarray}
    \frac{\dd \sigma^h}{\dd x_B \, \dd z_h\, \dd Q^2} 
    &=& \sum_{a,b} \int\limits_{z_h}\!\!\int\limits_{x_{\!B}} \dd \hat{x} \, \dd \hat{z} \, f_a(\hat{x},\mu_F) \, D^h_b(\hat{z},\mu_F) \,
    \widehat{\sigma}^h_{a,b}
    \bigg( \frac{\hat{x}}{x_{\!B}},\frac{\hat{z}}{z_h};\mu_F,\mu_R,Q \bigg).
    \label{eq:sidiscs}
\end{eqnarray}
Here $Q^2 \equiv -q^2$ is the squared four-momentum transfer to the nucleon, $x_B = Q^2/2 \, p \cdot q$ is the Bjorken scaling variable, with $p$ and $q$ the target nucleon and virtual photon four-momenta, respectively, and $z_h = p \cdot p_h / p \cdot q$ is the fraction of the virtual photon's energy carried by the fragmenting hadron $h$ with four-momentum $p_h$. 
The invariant mass squared of the unmeasured hadronic final state is given by $W^2_{\rm SIDIS} = (p+q-p_h)^2$.
The partonic cross section $\widehat{\sigma}^h_{a,b}$ is evaluated perturbatively to NLO accuracy, and the factorization and renormalization scales are set as $\mu_F = \mu_R = Q$.

For all processes in our analysis we use the $\overline{\text{MS}}$ scheme for the renormalization group equations, with the strong coupling $\alpha_s$ evolved numerically using the QCD $\beta$-functions with the boundary condition $\alpha_s(M_Z) = 0.118$ at the $Z$-boson mass, $M_Z = 91.18$~GeV.
The PDFs and FFs are evolved to next-to-leading logarithmic accuracy using the DGLAP evolution equations~\cite{Dokshitzer:1977sg, Gribov:1972ri, Altarelli:1977zs} in the zero-mass variable flavor scheme, setting the input scale to the charm quark mass, $\mu = m_c$, for both PDFs and FFs. 
The heavy quark mass thresholds are taken from PDG to be $m_c = 1.28$~GeV and $m_b = 4.18$~GeV~\cite{ParticleDataGroup:2024cfk}.

\subsection{Nonperturbative modeling}
\label{sub.modeling}

We parameterize all PDFs at the input scale $\mu^2 = m_c^2$ using the standard, phenomenologically successful template function
\begin{equation}
    f(x,\mu_{};\bm{a}) = \frac{N}{\mathcal{M}}\,x^{\alpha}(1-x)^{\beta}(1+\gamma \sqrt{x} + \delta x),
    \label{eq:pdf_temp}
\end{equation}
where the set of parameters to be fitted,
    $\bm{a} = \{N,\alpha,\beta,\gamma,\delta\}$, 
includes the normalization coefficient $N$ and shape parameters $\alpha$, $\beta$, $\gamma$ and $\delta$. 
To ensure that the normalization coefficient $N$ is maximally decorrelated from the shape parameters, we normalize the function using 
    $\mathcal{M} = B[\alpha + 2, \beta + 1] + \gamma B[\alpha + \frac{5}{2}, \beta + 1] + \delta B[\alpha + 3, \beta + 1]$,
where $B$ is the beta function (Euler integral of the first kind).

As in most global QCD analyses, we assume isospin symmetry for the PDFs and FFs, so that a $u$-quark PDF in the proton is equivalent to a $d$-quark PDF in the neutron, for instance.
For the PDFs, we parameterize the valence $u$ and $d$ quark distributions directly,
\begin{equation}
u_v \equiv u - \bar u, \qquad\qquad
d_v \equiv d - \bar d,
\end{equation}
along with the gluon distribution, $g$, via Eq.~(\ref{eq:pdf_temp}).
For the sea quark and antiquark distributions we use the {\it ansatz},
\begin{equation}
    \begin{aligned}[c]
        \bar u &= S_1 + \delta \bar u,\\
        s &= S_2 + \delta s,
    \end{aligned}
    \qquad\qquad
    \begin{aligned}[c]
        \bar d &= S_1 + \delta \bar d,\\
        \bar s &= S_2 + \delta \bar s,
    \end{aligned}
\label{eq:pdf_shapes}
\end{equation}
where $S_1$ and $S_2$ contain the bulk of the flavor-independent sea distribution for the light sea and strange sectors, respectively, and $\delta f$ ($f=\bar u, \bar d, s, \bar s$) are additional flavor-dependent distortions away from $S_{1,2}$ that are required to describe the observational data.
Our modeling allows us to explore a range of possibilities for the sea distributions, including the symmetric sea scenario ($S_1 = S_2$ and $\delta f = 0$), light sea quark asymmetry, light and strange sea quark asymmetry, as well as the $s-\bar{s}$ asymmetry.
For the parametrizations of $S_1$, $S_2$ and $\delta f$ we use the same template function as in Eq.~(\ref{eq:pdf_temp}).

As usual, the normalization parameters for the $u_v$, $d_v$, and $\delta s$ distributions are set using valence number sum rules, 
\begin{eqnarray}
\int_0^1 \dd{x} u_v = 2, \qquad
\int_0^1 \dd{x} d_v = 1, \qquad
\int_0^1 \dd{x} (s-\bar s) = 0,
\end{eqnarray}
while the normalization parameter for the gluon PDF is set by the momentum sum rule,
\begin{eqnarray}
\int_0^1 \dd{x} x \, \Big( \sum_q (q + \bar q) + g \Big) = 1.
\end{eqnarray}

The parametrization template in Eq.~(${\ref{eq:pdf_temp}}$) is also used for FFs, but with $x$ replaced by~$z$, the momentum fraction of the parton carried by the produced hadron. 
The FF model used in this analysis allows for maximum parametrization flexibility, while ensuring that the number of independent functions does not exceed that which can be reliably constrained by the available SIA and SIDIS observables (see Ref.~\cite{JAMFF} for further details).
For each hadron species, we have 4 independent functions from SIA and 2 independent functions from SIDIS.
We allow the $g$, $c$, and $b$ FFs to be independent, leaving 3 functions to parameterize the light quarks.
Assuming charge symmetry, for the $q \to \pi^+$ FFs we take
\begin{equation}
    \begin{aligned}[c]
        D^{\pi^{+}}_u &= D^{\pi^{+}}_{\bar{d}},\\
        D^{\pi^{+}}_d &= D^{\pi^{+}}_{\bar{u}},\\
        D^{\pi^{+}}_q &= D^{\pi^{+}}_{\bar{q}},\quad q = s,c,b.
    \end{aligned}
\label{eq:pion_shapes}
\end{equation}
For the $q \to K^+$ FFs we allow the favored $D_u^{K^+}$ and $D_{\bar s}^{K^+}$ FFs to be independent, but set the FFs for the unfavored flavors to be equal,
\begin{equation}
    \begin{aligned}[c]
        D^{K^{+}}_d &= D^{K^{+}}_{\bar{u}} = D^{K^{+}}_{\bar{d}} = D^{K^{+}}_s,\\
        D^{K^{+}}_q &= D^{K^{+}}_{\bar{q}},\quad q = c,b,
    \end{aligned}
\label{eq:kaon_shapes}
\end{equation}
with the heavier charm and bottom quark and antiquark FFs also set equal to each other.
For the unidentified charged hadron FFs, we follow the previous JAM FF analysis~\cite{Moffat:2021dji} and use a residual term, $D^{\text{res}^+}_q$, to parametrize the difference between the total charged hadron $D^{h^+}_q$ and the $D^{\pi^+}_q$ and $D^{K^+}_q$ FFs, 
\begin{equation}
    D^{h^+}_{q} = D^{\pi^+}_{q} + D^{K^+}_{q} + D^{\text{res}^{+}}_{q}.
\label{eq:had_def}
\end{equation}
Since the residual hadrons are dominated by protons, we allow $D^{{\rm res}^{+}}_u$ and $D^{{\rm res}^{+}}_d$ to be independent, but equate the light sea quark FFs,
\begin{equation}
    \begin{aligned}[c]
         D^{\text{res}^+}_{\bar{u}} &= D^{\text{res}^+}_{\bar{d}} = D^{\text{res}^+}_{\bar{s}} =
         D^{\text{res}^+}_{s}, \\
         D^{\text{res}^{+}}_q &= D^{\text{res}^{+}}_{\bar{q}},\quad q = c,b.
    \end{aligned}
\label{eq:had_shapes}
\end{equation}
Finally, we use charge symmetry to relate the FFs for positively and negatively charged hadrons, $D^{h^-}_q = D^{h^+}_{\bar{q}}$, for all flavors $q$.

\section{Datasets}
\label{s.datasets}

In this section we summarize the datasets that are used in the current global analysis.
These include the standard baseline sets of DIS, Drell-Yan lepton-pair production, inclusive weak boson and jet production, as well as recent $W$+\,charm production data from the LHC.
To facilitate better flavor separation of the PDFs, we utilize in addition SIDIS data from muon-deuterium scattering at COMPASS. 
Since the SIDIS process involves also FFs, we simultaneously fit these datasets together with $e^+ e^-$ SIA data, constraining the FF and PDF parameters self-consistently. 
The list of reactions and observables, and their associated connection to PDFs and FFs, is summarized in Table~\ref{tb:reactions}.

\begin{table}[t]
\begin{center}
\caption{Summary of processes, observables, and their relation to PDFs and FFs. The symbols $\dagger$, $\star$, and $\diamond$ indicate the $z_h$ cut used for high-energy SIA data, low-energy SIA data, and BaBar kaon data, respectively.\\}
    \begin{tabular}{l|l|l|l} 
\hline\hline
    ~~Process 
&
    ~Observables 
&
    ~Cuts 
&
    ~PDFs/FFs
\\ 
\hline
    ~~$\ell+(p,d)\to \ell'+X$ 
&
    ~$F_2,\sigma_{\rm red}$
    \cite{Whitlow:1991uw,BCDMS:1989ggw,NewMuon:1996fwh,NewMuon:1996uwk,H1:2015ubc}
&
    ~$Q^2 > m_c^2$ 
&
    ~$f_{i/p}$
\\
    ~~
&
    ~
&
    ~$W^2 > 10$~GeV$^2$
&
\\
    ~$\ell+d\to \ell'+(\pi^{\pm},K^{\pm},h^{\pm})+X$~
&
    ~$\dd{M}^h/\dd{z_h}$~\cite{COMPASS:2016xvm, COMPASS:2016crr}
&
    ~$Q^2 > m_c^2$
&
    ~$f_{i/p}$
\\
    ~~
&
    ~
&
    ~
&
    ~$D^{(\pi^+,\, K^+,\, h^+)}_i$

\\
    ~$p+(p,d)\to \ell\bar{\ell}+X$
&
    ~$\dd\sigma^{p/d}/\dd x_F \, \dd Q^2$~\cite{NuSea:2001idv, SeaQuest:2021zxb}~
&
    ~
&
   ~$f_{i/p}$
\\
    ~$p+(p,\bar{p}) \to W+X$
&
    ~$A_W$, $A_\ell$, $\sigma^{W^+}/\sigma^{W^-}$~
&
    ~
&
    ~$f_{i/p}$
\\
    ~
&
    ~~~~\cite{CMS:2011bet, CMS:2012ivw, CMS:2013pzl, CMS:2016qqr, LHCb:2014liz, LHCb:2015mad, D0:2013lql, CDF:2009cjw, STAR:2020vuq}
&
    ~
&
    ~
\\
    ~$p+\bar{p} \to Z/\gamma^*+ X$
&
    ~$\dd \sigma/\dd y$~\cite{D0:2007djv, CDF:2010vek}
&
    ~
&
    ~$f_{i/p}$
\\
    ~$p+(p,\bar{p}) \to {\rm jet} + X$
&
    ~$\dd \sigma / \dd \eta \, \dd p_T$~\cite{PhysRevLett.101.062001, PhysRevD.75.092006, PhysRevLett.97.252001}
&
    ~$p_T> 8$~GeV
&
    ~$f_{i/p}$
\\
    ~$p+p \to W + c + X$
&
    ~$\dd \sigma / \dd \eta$~\cite{ATLAS:2014jkm, CMS:2013wql, CMS:2018dxg}
&
    ~
&
    ~$f_{i/p}$
\\
    ~$\ell + \bar{\ell}\to (\pi^\pm, K^\pm, h^\pm) + X$
&
    ~$\dd\sigma/\dd z_h$~\cite{ARGUS:1989zdf, BaBar:2013yrg, Belle:2013lfg, BRANDELIK1981357, TASSO:1983cre, TASSO:1988jma, Lu:1986mc, TPCTwoGamma:1988yjh, TOPAZ:1994voc, ALEPH:1994cbg, DELPHI:1998cgx, OPAL:1994zan, SLD:2003ogn, 
    TASSO:1982bkc, TASSO:1981gag, ALEPH:1995njx, 
    DELPHI:1998cgx, OPAL:1998arz}
&
    ~$0.02 < z_h < 0.9$ $\dagger$~
&
    ~$D^{(\pi^+,\, K^+,\, h^+)}_i$
\\
    ~
&
    ~
&
    ~$0.15 < z_h < 0.9$ $\star$~
&
    ~
\\
    ~
&
    ~
&
    ~$0.20 < z_h < 0.9$ $\diamond$~
&
    ~
    \\
    \hline\hline
    \end{tabular}
    \label{tb:reactions}
\end{center}
\end{table}

For inclusive DIS experiments, we use fixed-target data on proton and deuteron targets from SLAC~\cite{Whitlow:1991uw}, BCDMS~\cite{BCDMS:1989ggw}, and NMC~\cite{NewMuon:1996fwh, NewMuon:1996uwk}, along with reduced neutral current and charged current proton cross sections for the combined H1 and ZEUS analysis of HERA collider data~\cite{H1:2015ubc}.
For the SIDIS measurements, we include data on pion, kaon, and unidentified charged hadron production on deuteron targets from the COMPASS Collaboration~\cite{COMPASS:2016xvm, COMPASS:2016crr}.
We also note that in previous JAM analyses~\cite{JAM19, Moffat:2021dji} of the SIDIS data, one of the sources of systematic uncertainty was interpreted as a normalization uncertainty.
This interpretation required a large fitted normalization to adequately describe the data. 
In the present analysis this uncertainty is instead interpreted as a point-to-point correlated systematic uncertainty, allowing the data to be well described without the need of a large fitted normalization for each dataset.

Drell-Yan lepton-pair production data from the E866 (NuSea)~\cite{NuSea:2001idv} and E906 (SeaQuest)~\cite{SeaQuest:2021zxb} $pp$ and $pD$ scattering experiments at Fermilab are used to constrain mostly the $\bar d/\bar u$ PDF ratio in the proton at small and intermediate values of $x$.
Data for weak vector boson mediated processes including $W$-lepton asymmetries from the CMS~\cite{CMS:2011bet, CMS:2012ivw, CMS:2013pzl, CMS:2016qqr} and LHCb~\cite{LHCb:2014liz, LHCb:2015mad} Collaborations at the LHC, and the STAR~\cite{STAR:2020vuq} Collaboration at RHIC, along with $W^{\pm}$ charge asymmetries and $Z/\gamma^*$ rapidity distributions from CDF and D0 at the Tevatron~\cite{D0:2013lql, CDF:2009cjw, D0:2007djv, CDF:2010vek}, are used to further constrain flavor separation including at high $x$. 
Jet production data from CDF and D0~\cite{PhysRevLett.101.062001, PhysRevD.75.092006} and STAR~\cite{PhysRevLett.97.252001} are also included, which are important for constraining the gluon PDF at large $x$. 
For $W$+\,charm quark production, which have been argued could be more sensitive to the strange quark PDFs, we use cross sections from the ATLAS~\cite{ATLAS:2014jkm} and CMS~\cite{CMS:2013wql, CMS:2018dxg} Collaborations at the LHC.  SIA data involving pion, kaon, and unidentified charged hadron from the ARGUS~\cite{ARGUS:1989zdf}, BaBar~\cite{BaBar:2013yrg}, Belle~\cite{Belle:2013lfg}, TASSO~\cite{BRANDELIK1981357, TASSO:1983cre,TASSO:1988jma,TASSO:1981gag,TASSO:1982bkc}, TPC~\cite{Lu:1986mc,TPCTwoGamma:1988yjh}, TOPAZ~\cite{TOPAZ:1994voc}, ALEPH~\cite{ALEPH:1994cbg,ALEPH:1995njx}, DELPHI~\cite{DELPHI:1998cgx}, OPAL~\cite{OPAL:1994zan,OPAL:1998arz}, and SLD~\cite{SLD:2003ogn} Collaborations are included to constrain the FFs.

\begin{figure}[t]
\centering
\includegraphics[scale=0.4]{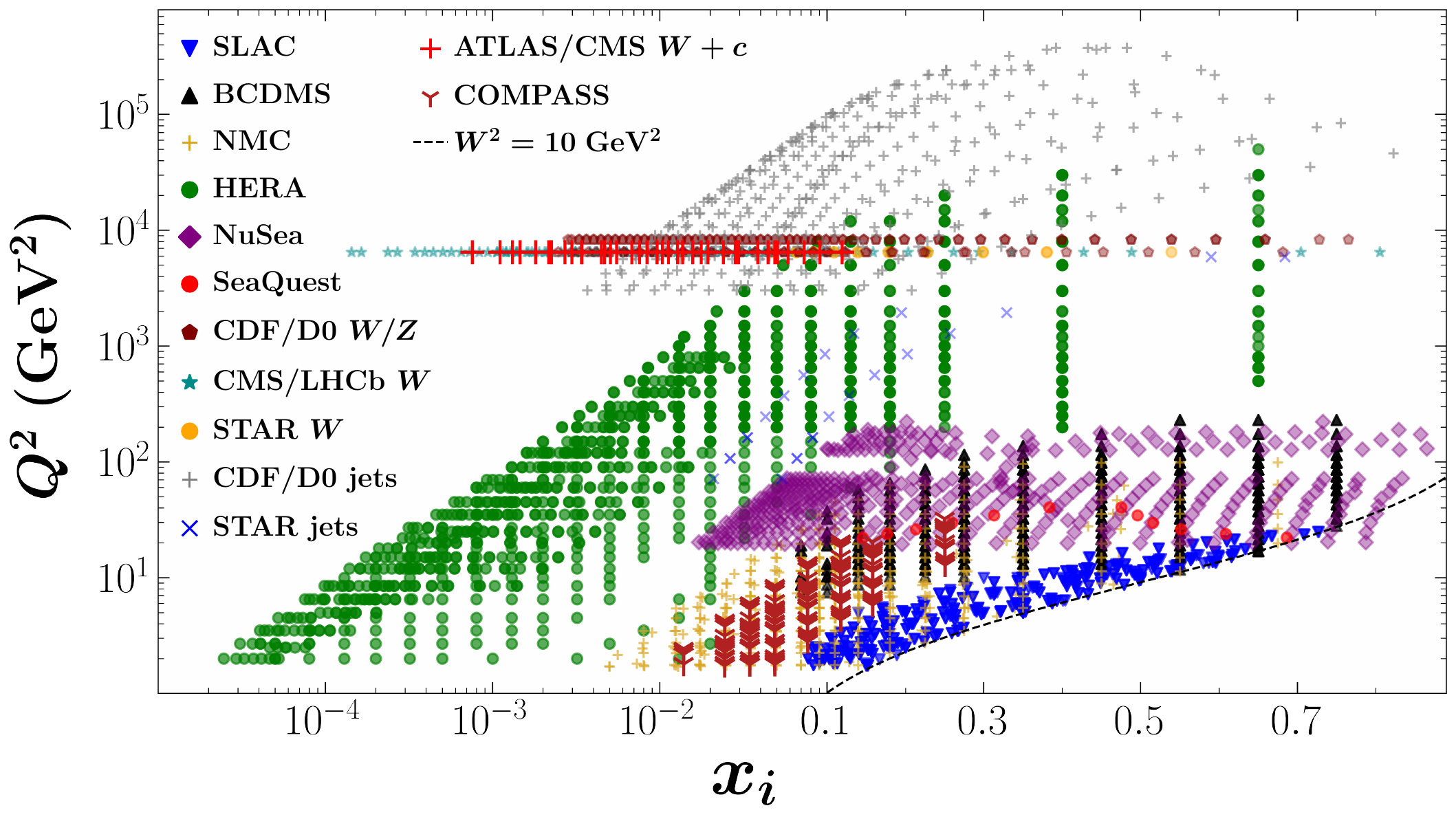}\vspace*{0.5cm}
\includegraphics[scale=0.4]{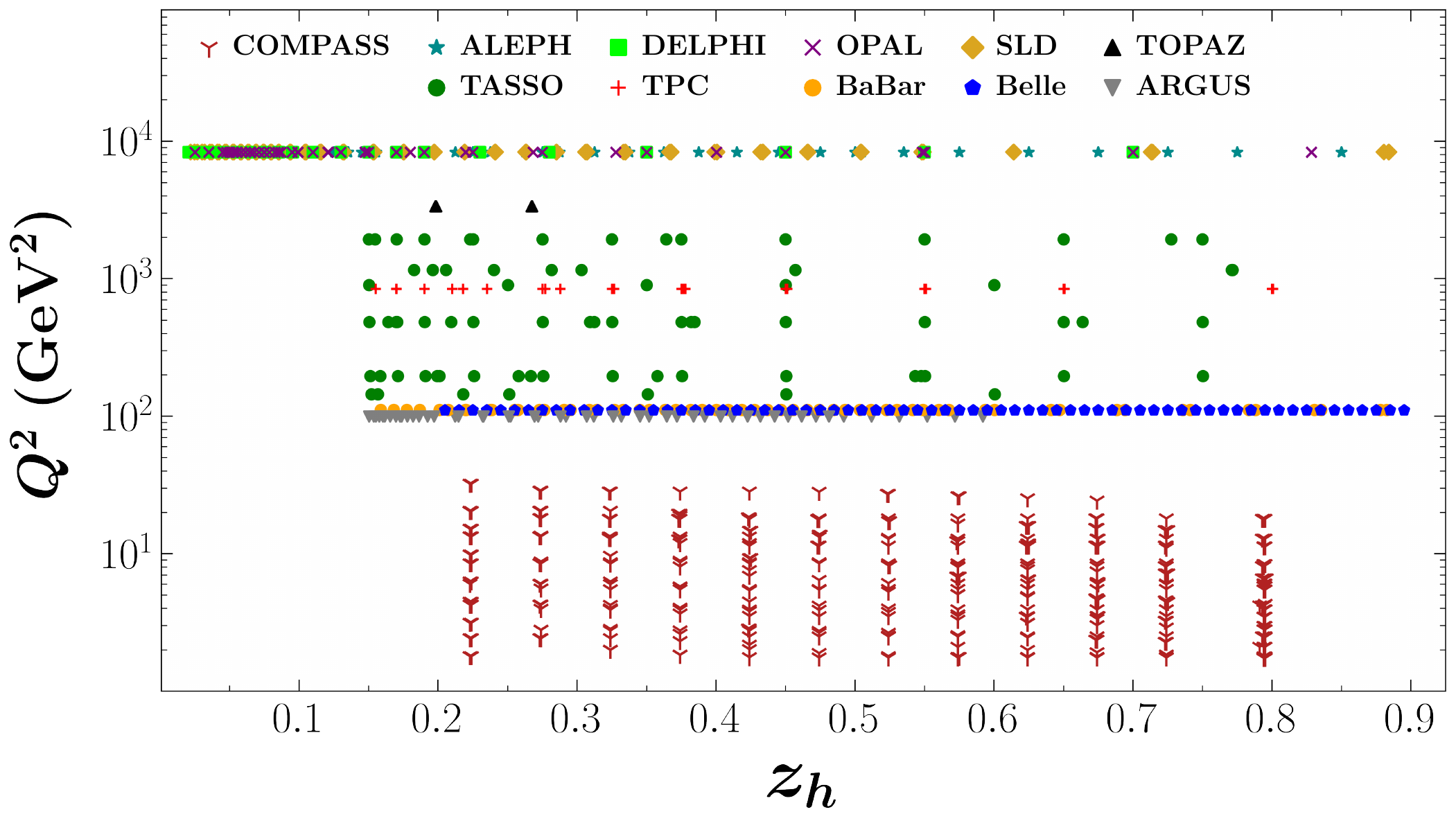}
\caption{Kinematic coverage of datasets used in this analysis, with the scale $Q^2$ versus the Bjorken scaling variable $x_i=x_B$ for DIS/SIDIS, and $x_i=x_1$ or $x_2$ for Drell-Yan/weak boson/jet production data (upper panel), and the fragmentation variable $z_h = p \cdot p_h /p \cdot q$ for SIDIS and $z_h = 2 \, p_h \cdot q/Q^2$ for SIA data (lower panel).}
\label{fig:Q2x_Q2z}
\end{figure}

The kinematic range covered by the data for each type of dataset and experiment is illustrated in Fig.~\ref{fig:Q2x_Q2z}.
Here the scale $Q^2$ is shown versus the Bjorken scaling variable $x_B$ for DIS and SIDIS data, and versus $x_{1,2}$ variables for hadron-induced scattering reactions.
In addition, for SIDIS and SIA data, the scale $Q^2$ is shown versus the fragmentation variables $z_h = p \cdot p_h /p \cdot q$ and $z_h = 2 \, p_h \cdot q/Q^2$, respectively.
The SIDIS data are critical, as these are the only datasets that have sensitivity to both the PDFs and FFs.

\clearpage
\section{Global QCD analysis}
\label{s.results}

Having outlined the theoretical framework and data selection, in this section we present the results of our global QCD analysis.
We will first review the methodological aspects of our analysis, followed by a detailed description of our strategies for constructing posterior distributions for the PDFs and FFs.
The quality of the results will be discussed by examining the agreement between data and theory, with a focus on the impact of SIDIS multiplicities for $\pi^\pm$, $K^\pm$, and unidentified hadron $h^\pm$ production data from COMPASS, as well as $W$+\,charm production cross section data from the LHC, on the magnitude and shape of the strange quark PDF.
We discuss several scenarios, including the baseline results, which exclude SIDIS and $W$+\,charm data; two additional scenarios, where the baseline is supplemented with either SIDIS or $W$+\,charm data; and the final combined result that incorporates all datasets in the analysis. 
We label these analyses as ``baseline'', ``+SIDIS'', ``+$W$-charm'', and ``JAM24'', respectively.

\subsection{Methodology and model calibration}
\label{sub.modelcal}

The reconstruction of PDFs and FFs follows the multi-step Monte Carlo-based approach for Bayesian inference employed in previous JAM global QCD analyses \cite{Cocuzza:2021cbi, Cocuzza:2021rfn, Moffat:2021dji}.
This approach allows us to calibrate our models, quantify their performance across the parameter space, and mitigate overparametrization. 
The key algorithmic aspects of the methodology include the use of data resampling for optimization, and the multi-step strategy developed in Ref.~\cite{JAM19}.

In the current analysis several notable improvements have been introduced.
First, the sea quark PDFs are initially treated as flavor symmetric during the steps prior to incorporating flavor sensitive experimental data. 
Once such data are added, they allow the separation of specific flavors from the total sea.
This incremental approach enables a more reliable determination of PDF uncertainties by gradually introducing flavor asymmetries between different sea quark distributions.

Second, we replace the initial flat prior distributions with Gaussian priors, choosing specific hyperparameters for each type of template parameter. 
The hyperparameters are chosen to span a large range of solutions to ensure the Gaussian priors are not too informative and bias the space of solutions.
We also include corresponding Gaussian penalties of the form $\big( (a_i - \mu_i)/\sigma_i \big)^2$, where $\mu_i$ and $\sigma_i$ are the mean and width hyperparameters, respectively, of the distribution for the parameter $a_i$ was sampled. 
Gaussian penalties are utilized in the optimization procedure across the multi-steps to prevent the parameters from growing indefinitely and accumulating at parameter boundaries.
In addition, the use of Gaussian penalties ensures the self-consistency of the template modeling (for example, the $\gamma$ and $\delta$ parameters in Eq.~(\ref{eq:pdf_temp}) should act as correction terms to the leading $x^\alpha (1-x)^\beta$ term of the template function), and enables the specification of priors in extrapolation regions.

Lastly, when computing any physical observables from the MC parameter samples, we use the mean values as the central results, while uncertainties are estimated using Bayesian credible intervals at the 95\% confidence level. 
Specifically, for any observable such as a PDF, FF, cross section or asymmetry, the edges of the confidence intervals are determined using a nonparametric estimate of the inverse cumulative distribution function~\cite{Hyndman}.

With these improvements we first perform model calibration for the baseline scenario, excluding the SIDIS and $W$+\,charm data.
The multi-step optimization is initiated by first considering only DIS data, using a symmetric sea {\it ansatz},\, $\delta \bar{u} = \delta \bar{d} = \delta s = \delta \bar{s} = 0$ and $S_1 = S_2$.
The obtained posteriors are then used as priors for the next step in which DY data are added, and $\delta \bar{u}$ and $\delta \bar{d}$ allowed to vary away from zero. 
The optimization is continued by incrementally extending the dataset to include $W/Z$ and jet production data in subsequent steps.
An additional step is performed where $S_2$ is varied away from $S_1$ and $\delta \bar{s}$ is varied independently away from $\delta \bar{s}$ to account for possible differences between the light and strange quark PDFs.
The resulting MC replicas then constitute our baseline results.

During the optimization process of the baseline setup, we found that the $\gamma$ and $\delta$ parameters in Eq.~(\ref{eq:pdf_temp}) had little or no effect on the $\delta s$, $\delta \bar{s}$, and $S_2$ distributions. 
Consequently, these parameters were set to zero for these functions. 
However, they were allowed to vary freely for each of the $u_v$, $d_v$, $g$, $\delta \bar{u}$, $\delta \bar{d}$, and $S_1$ distributions. 
This results in a total of 35 free parameters for the PDFs in the baseline analysis, comprising 30 shape parameters and 5 normalization parameters.

The model calibration for FFs is initially performed using only the SIA data. 
For each hadron, two template shape functions were found to be necessary to describe the data for all quark FFs. 
Using the FF relations given in Sec.~\ref{sub.modeling}, this results in 60 free FF parameters for each hadron species (12 normalization parameters and 48 shape parameters).

As mentioned above, we consider different scenarios as extensions of the baseline fit.
These include the addition of SIDIS data, $W$+\,charm data, and both SIDIS and $W$+\,charm data. 
Each scenario is implemented by performing additional steps, starting from the baseline results and incorporating the corresponding datasets.
Importantly, the Gaussian penalties applied during the model calibration stages are removed in the final MC runs for each of the 
scenarios.
The combined full analysis utilized 215 parameters to model the input scale PDFs and FFs, along with an additional 43 free normalization parameters for various datasets, bringing the total number of free parameters to be inferred from the data to 258.

\subsection{Figures of merit}
\label{sub.merit}

As in previous JAM analyses, the optimization of the PDF and FF parameters against the data is performed using nuisance parameters. 
Specifically, a given experimental data point $i$ from a dataset $e$ with value $d_{i,e}$ is compared against the corresponding factorization-based calculation $T_{i,e}$, with additional systematic distortions of the form
\begin{align} 
T_{i,e}\ \to\ \widehat{T}_{i,e}\, \equiv\, \sum_k r^k_e\, \beta^{k}_{i,e} + \frac{T_{i,e}}{N_e},
\label{eq:TThat}
\end{align}
where $\beta^k_{i,e}$ represents the $k$-th quoted source of point-to-point correlated systematic uncertainties. 
The nuisance parameters $r^k_e$ and $N_e$ modify the original theory additively and multiplicatively, respectively, to best describe the data within the quoted systematic uncertainties. 
To avoid overfitting, the nuisance parameters are regulated with Gaussian penalties $\sum_k \left(r^k_e\right)^2$ for the additive shifts and by $(1-N_e)^2/(\delta N_e)^2$ for the multiplicative shift, where $\delta N_e$ is the quoted normalization uncertainty for experiment~$e$.

Following Ref.~\cite{MMHT14}, in previous JAM analyses~\cite{Moffat:2021dji, Cocuzza:2021cbi} a theory-scaled point-by-point correlated systematic uncertainty was used, with $\beta^k_{i,e} \to (T_{i,e}/D_{i,e}) \times \beta^k_{i,e}$ in Eq.~(\ref{eq:TThat}).
Using this convention, we found that the SIDIS data could not be well described without introducing large additive shifts.
Removing this rescaling significantly decreased systematic shifts needed to describe the SIDIS data, while leaving other dataset descriptions unaffected.

The quantity $\widehat{T}_{i,e}$ in Eq.~(\ref{eq:TThat}) represents a modified theory that accounts for systematic biases present in the data, under the assumption that $T_{i,e}$ is the correct underlying law for the reconstructed observable. 
In practice, however, $T_{i,e}$ is only computable within the limits dictated by factorization theorems, the perturbative accuracy of the theoretical computation, and the expressivity of the nonperturbative modeling for PDFs and FFs.
In this context, the deviations of $\widehat{T}_{i,e}$ from $T_{i,e}$ are not solely a measure of experimental biases, but also reflect the computational limitations of $T_{i,e}$.
These include constraints from the theoretical framework and modeling assumptions.
We therefore aim to minimize these deviations by adjusting our model assumptions, as will be discussed in Sec.~\ref{sub.theory} below.

To assess the quality of the global analysis, we utilize two figures of merit: the reduced $\chi^2$ for each dataset, defined as  
\begin{equation} 
    \chi^{2}_{\text{red}} 
    \equiv \frac{1}{N_{\rm dat}}
    \sum_{i,e} 
    \bigg( \frac{d_{i,e} - {\rm E}\,[\widehat{T}_{i,e}]} {\alpha_{i,e}} \bigg)^2,
\label{eq:chi2_short} 
\end{equation}
where E\,[$\cdots$] is the expectation value, and the associated $Z$-score to evaluate the probabilities of the estimated $\chi^2_{{\rm red}}$, taking into account the number of kinematical data points $N_{\rm dat}$ for each dataset.
The quantities $\alpha_{i,e}$ are the quadrature sums of all the quoted uncorrelated point-by-point uncertainties.  
The $Z$-score is defined in terms of the inverse of the normal cumulative distribution function,
\begin{equation} 
Z = \Phi^{-1}(p) \equiv \sqrt{2} \, \text{erf}^{-1}(2p - 1), 
\end{equation}
where the $p$-value is computed according to the $\chi^2 \equiv N_{\rm dat} \chi^2_{\rm red}$ distribution with the number of data points $N_{\rm dat}$ as the degrees of freedom.

\subsection{Data selection}
\label{sub.dataselection}

For the DIS datasets, we apply the kinematic cuts $W^2 > 10$~GeV$^2$ and $Q^2 > m_c^2$ to avoid the nucleon resonance region, and suppress higher twist corrections to the leading twist approximations for the structure functions.
For the SIDIS data, we also examined the effects of various cuts on $W^2_{\rm SIDIS}$ and $z_h$ on the description of the data.
We found that the SIDIS multiplicities can be well described using the kinematic cut $Q^2 > m_c^2$, with no additional cuts required on $W^2_{\rm SIDIS}$ and $z_h$ beyond the kinematic coverage of the COMPASS data, namely, $W^2_{\rm SIDIS} > 7$~GeV$^2$ and $0.2 < z_h < 0.8$.
Note that the computation of $W^2_{\rm SIDIS}$ requires specifying the hadron transverse momentum in the Breit frame, $p_{hT}$~\cite{Boglione:2019nwk}.
Since the SIDIS cross sections receive their largest contributions from regions with small transverse momentum, we set its value to zero when evaluating $W^2_{\rm SIDIS}$.

For other SIDIS data at lower energies, such as those from Jefferson Lab Hall~C~\cite{BHATT2025139485}, additional cuts on $W^2_{\rm SIDIS}$ may be needed to allow one to isolate regions where higher twists and target mass corrections can be neglected and the current and target fragmentation regions can be isolated.
For SIA, to avoid large perturbative effects at small and large values of $z_h$ and allow maximal overlap with the SIDIS kinematics of COMPASS, we restrict the data using the upper cut of $z_h < 0.9$ and lower cuts of $0.02$ and $0.15$ for high-energy and low-energy SIA data, respectively.  An exception is the BaBar $K^\pm$ dataset, for which the cut $z_h > 0.2$ is needed for a good overall fit, as also observed in Ref.~\cite{AbdulKhalek:2022laj}.

\subsection{Data versus theory comparison}
\label{sub.theory}

\begin{table}[t]
\centering
\caption{Reduced $\chi^2_{\text{red}}$ and $Z$-score values for each of the four scenarios considered in this analysis (with the total number of data points for each): baseline~(4326 data points), +SIDIS~(5816), \mbox{+$W$-charm}~(4363), and the full JAM24 analysis~(5853). The $N_{\rm dat}$ values listed in the table for each type of observable correspond to the JAM24 fit.
}
\begin{tabular}{l r c c c c} 
\hline \hline ~& ~& \multicolumn{4}{c}{$\chi^2_{\rm red}$ ($Z$-score)} 
\\ \hline 
~Process 
& $N_{\rm dat}$~~~ 
& baseline 
& ~+SIDIS 
& ~~+$W$-charm 
& ~~\textbf{JAM24}
\\ 
\hline
~DIS                  
&      
&  
&  
&      
\\
$\quad$ fixed target \cite{Whitlow:1991uw, BCDMS:1989ggw, NewMuon:1996fwh, NewMuon:1996uwk}                   
& ~~~1495 ~~
& 1.07 (+1.95) 
& ~~~1.08 (+2.23) 
& ~~~1.07 (+1.92) 
& ~~~$\mathbf{1.08\, (+2.25)}$~
\\
$\quad$ HERA \cite{H1:2015ubc}                                            
& ~~~1185 ~~
& 1.16 (+3.64)  
& ~~~1.18 (+4.22) 
& ~~~1.15 (+3.57)
& ~~~$\mathbf{1.18\, (+4.22)}$~
\\
~Drell-Yan \cite{NuSea:2001idv, SeaQuest:2021zxb}                      
& ~~~205 ~~
& 1.15 (+1.50)   
& ~~~1.17 (+1.69)
& ~~~1.14 (+1.43) 
& ~~~$\mathbf{1.17\, (+1.69)}$~
\\
~$W$-lepton asymmetry
& ~~~70 ~~
& 0.88 ($-0.69$)
& ~~~0.91 ($-0.50$)
& ~~~0.87 ($-0.72$)
& ~~~$\mathbf{0.91\, (-0.52)}$~
\\
~~~~\cite{CMS:2011bet, CMS:2012ivw, CMS:2013pzl, CMS:2016qqr, LHCb:2014liz, LHCb:2015mad, STAR:2020vuq}
& 
& 
&
& 
& 
\\
~$W$ charge asymmetry                              
& ~~~27 ~~
& 0.91 ($-0.24$)
& ~~~1.00 ($+0.08$)
& ~~~0.91 ($-0.25$)
& ~~~$\mathbf{1.00\, (+0.08)}$~
\\
~~~~\cite{D0:2013lql, CDF:2009cjw}                              
& 
& 
&
& 
& 
\\
~$Z$ rapidity \cite{D0:2007djv,CDF:2010vek}                               
& ~~~56 ~~
& 1.18 (+0.98)
& ~~~1.22 (+1.15)
& ~~~1.24 (+1.25)
& ~~~$\mathbf{1.24\, (+1.24)}$~
\\
~Inclusive jets \cite{PhysRevLett.101.062001, PhysRevD.75.092006, PhysRevLett.97.252001}                               
& ~~~198 ~~
& 1.02 ($+0.27$)
& ~~~0.94 ($-0.56$)
& ~~~1.04 ($+0.41$)
& ~~~$\mathbf{0.94\, (-0.54)}$~
\\
~$W$ + charm \cite{ATLAS:2014jkm, CMS:2013wql, CMS:2018dxg}              
& ~~~37 ~~
& ---  
& ---  
& ~~~0.57 ($-2.10$)
& ~~~$\mathbf{0.59\, (-1.99)}$~
\\
~SIDIS                
&     
&      
\\
$\quad$ $\pi^\pm$ \cite{COMPASS:2016xvm}                                      
& ~~~498 ~~
& --- 
& ~~~0.91 ($-1.42$)
& --- 
& ~~~$\mathbf{0.89\, (-1.74)}$~
\\
$\quad$ $K^\pm$ \cite{COMPASS:2016crr}                                      
& ~~~494 ~~
& --- 
& ~~~0.91 ($-1.50$)
& --- 
& ~~~$\mathbf{0.88\, (-1.96)}$~
\\
$\quad$ $h^\pm$ \cite{COMPASS:2016xvm}                                      
& ~~~498 ~~
& --- 
& ~~~0.90 ($-1.65$)
& --- 
& ~~~$\mathbf{0.86\, (-2.22)}$~
\\
~SIA                  
&      
&      
\\
$\quad$ $\pi^\pm$ \cite{ARGUS:1989zdf, BaBar:2013yrg, Belle:2013lfg, BRANDELIK1981357, TASSO:1983cre, TASSO:1988jma, Lu:1986mc, TOPAZ:1994voc, ALEPH:1994cbg, DELPHI:1998cgx, OPAL:1994zan, SLD:2003ogn}          
& ~~~403 ~~
& 0.73 ($-4.30$)
& ~~~0.82 ($-2.77$)
& ~~~0.73 ($-4.30$)
& ~~~$\mathbf{0.82\, (-2.67)}$~
\\
$\quad$ $K^\pm$ \cite{ARGUS:1989zdf, BaBar:2013yrg, Belle:2013lfg, BRANDELIK1981357, TASSO:1983cre, TASSO:1988jma, TPCTwoGamma:1988yjh, TOPAZ:1994voc, ALEPH:1994cbg, DELPHI:1998cgx, OPAL:1994zan, SLD:2003ogn}                                      
& ~~~377 ~~
& 0.65 ($-5.55$)
& ~~~0.78 ($-3.20$) 
& ~~~0.65 ($-5.55$)
& ~~~$\mathbf{0.79\, (-3.04)}$~
\\
$\quad$ $h^\pm$ \cite{BRANDELIK1981357, TASSO:1982bkc, TASSO:1981gag, TASSO:1988jma, TPCTwoGamma:1988yjh, ALEPH:1995njx, DELPHI:1998cgx, OPAL:1998arz, SLD:2003ogn}     
& ~~~310 ~~
& 0.76 ($-3.29$)
& ~~~0.73 ($-3.73$) 
& ~~~0.76 ($-3.29$)
& ~~~$\mathbf{0.73\, (-3.69)}$~
\\
\hline 
~\textbf{Total}       
& ~~~$\mathbf{5853}$ ~~
& $1.00$ ($+0.11$)
& ~~~$1.00$ ($-0.09$)
& ~~~$1.00$ ($-0.06$) 
& ~~~$\mathbf{0.99\, (-0.51)}$~
\\
\hline\hline 
\end{tabular}
\label{tb:chi2}
\end{table}

\begin{figure}[t]
\begin{center}
\includegraphics[scale=0.54]{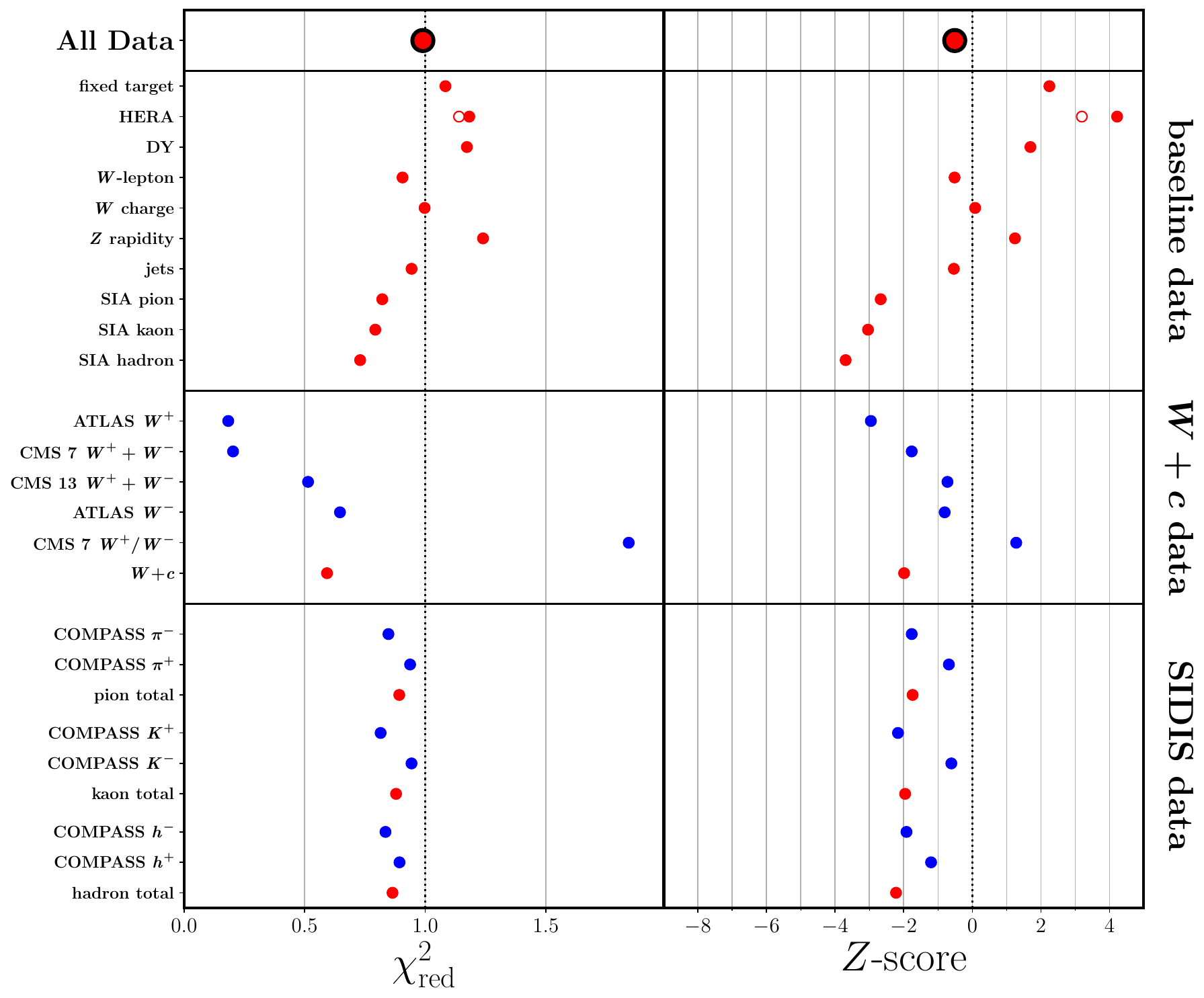}
    \caption{$\chi^2_{\rm res}$ and $Z$-score values of all datasets used in the current analysis, including for the baseline datasets (top panel), $W$+\,charm data from CMS and ATLAS (middle panel), and SIDIS $\pi^\pm$, $K^\pm$, and $h^\pm$ production from COMPASS (bottom panel). The reduced $\chi^2$ for individual datasets and averages over datasets are indicated by the blue and red points, respectively, and an additional open circle shows the reduced $\chi^2$ and $Z$-score of the HERA data using for $Q^2 > 3.5$~GeV$^2$ cut.}
\label{fig:chi2s}
\end{center}
\end{figure}

\begin{table}[t]
\begin{center}
\caption{Comparisons of the reduced $\chi^2_{\rm red}$ and $Z$-score values for HERA inclusive DIS data for the full JAM24 global analysis with the NNPDF4.0~\cite{NNPDF4.0} and CT18~\cite{CT18} fits.\\} 
\begin{tabular}{ l c c c c } 
    \hline\hline
    ~Analysis &
    ~~~~~pQCD accuracy~~~ &
    $Q^2_{\rm cut}$ (GeV$^2$)&
    ~~$N_{\rm dat}$~~ &
    ~~$\chi^2_{\rm red}$ ($Z$-score)~~
    \\ \hline
    ~NNPDF4.0 \cite{NNPDF4.0}&
    NNLO&
    ~3.49&
    1145&
    1.17 (3.89)
    \\ \hline
    ~CT18 \cite{CT18}&
    NNLO&
    4.0&
    1120&
    1.30 (6.51)
    \\ \hline
    ~JAM24&
    NLO&
    3.5&
    1120&
    1.14 (3.19)
    \\ 
    &
    NLO&
    $m_c^2$&
    1185&
    1.18 (4.22)
    \\ \hline\hline
    \end{tabular}
\label{tab:jam-vs-world}
\end{center}
\end{table}

\begin{figure}[t]
    \begin{center}
        \includegraphics[scale=0.4]{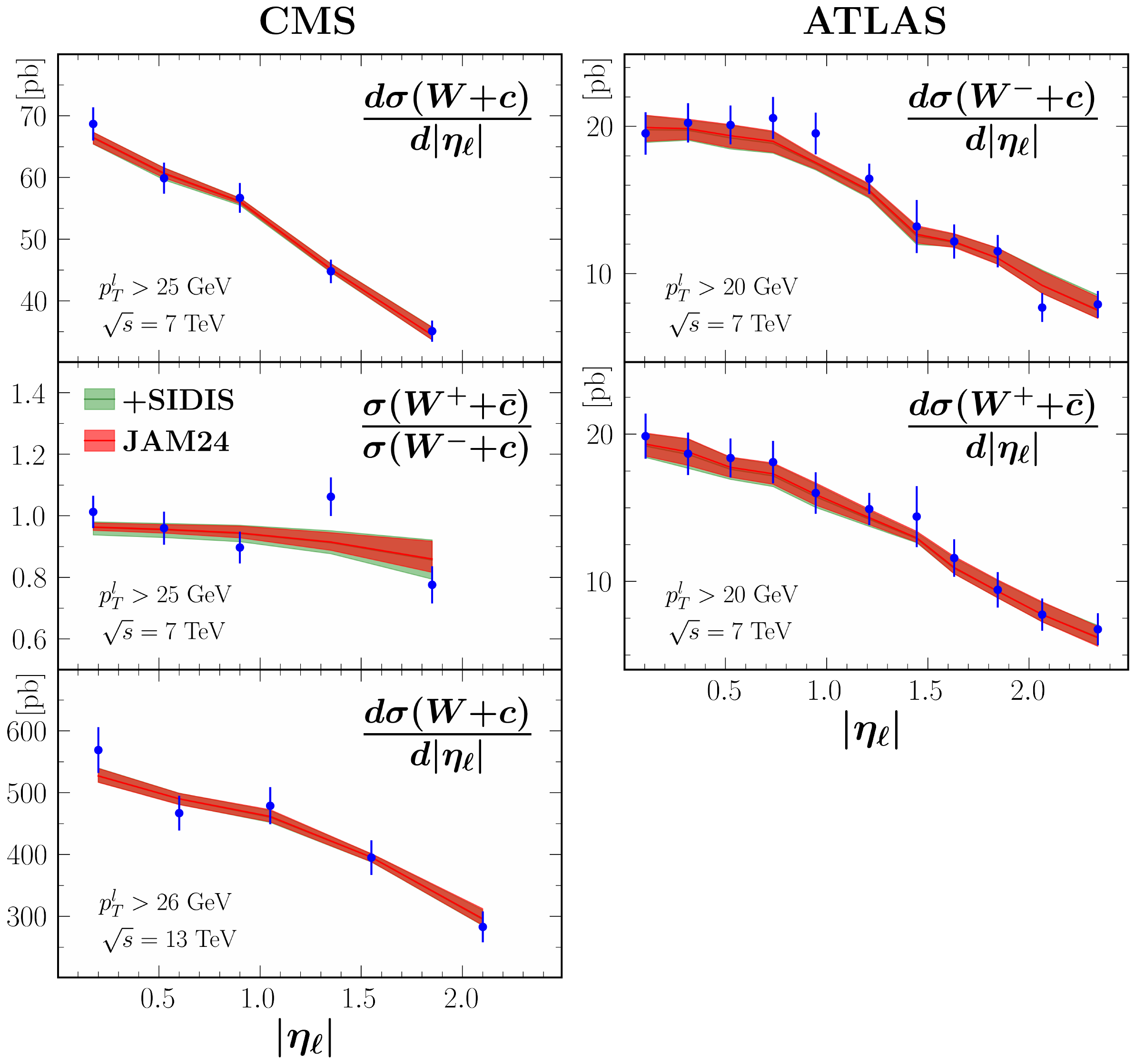}
        \caption{Comparison of CMS and ATLAS $W$+\,charm data with the JAM24 global analysis as a function of lepton pseudorapidity, $|\eta_\ell|$, with 95\% credible interval uncertainty bands.
        (Left): Sum (top) and ratio (middle) at 7~TeV~\cite{CMS:2013wql}, and sum (bottom) at 13~TeV~\cite{CMS:2018dxg} of $W^+ + \bar{c}$ and $W^- + c$ cross sections from CMS. 
        (Right): Differential cross section of $W^- + c$ (top) and $W^+ + \bar{c}$ (middle) from ATLAS at 7~TeV~\cite{ATLAS:2014jkm}. The cuts on the transverse momentum of the final state lepton $p_T^\ell$ are indicated on the panels.}
        \label{fig:w+charm}
    \end{center}
\end{figure}

\begin{figure}[t]
\begin{center}
\includegraphics[scale=0.41]{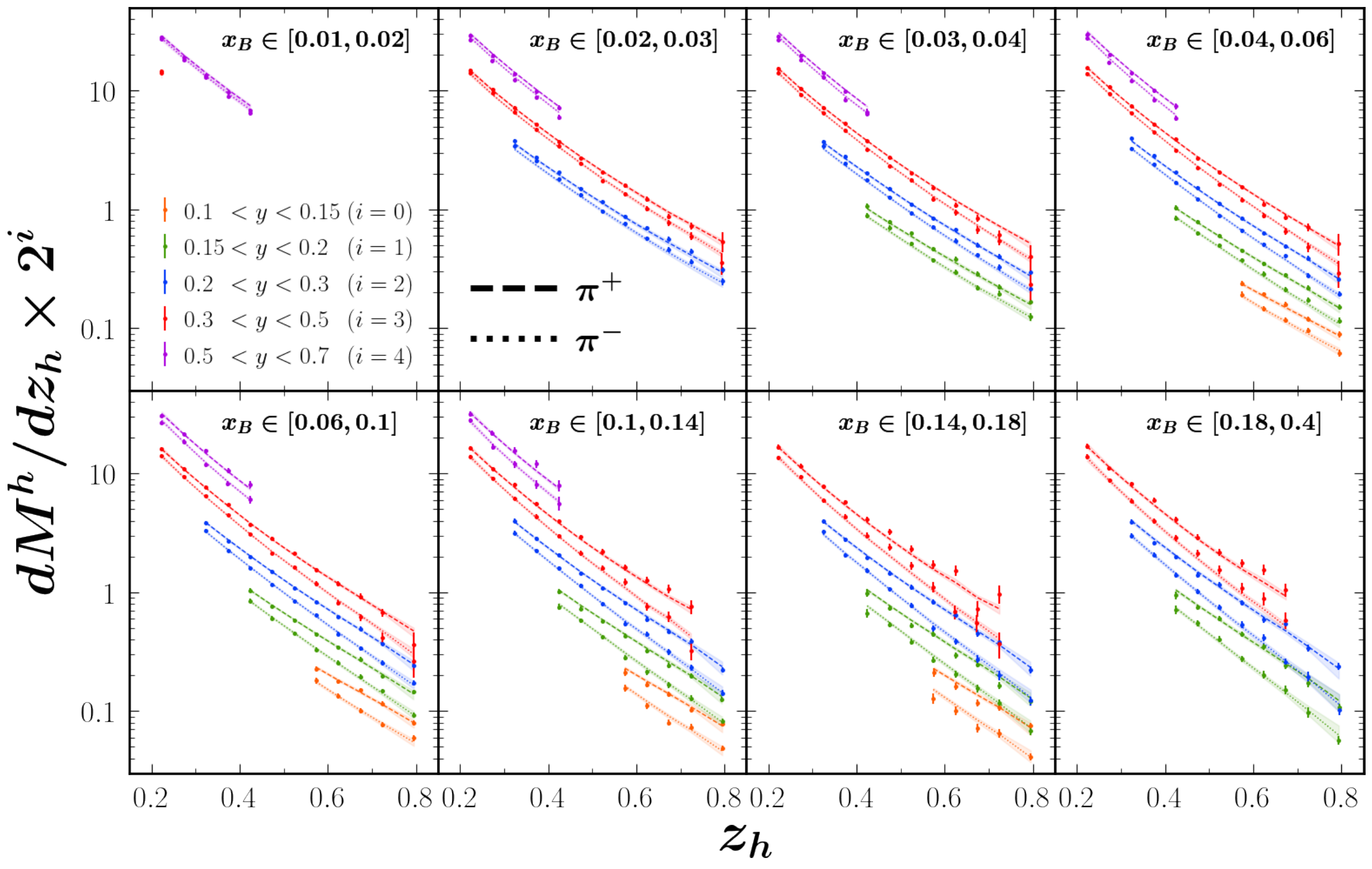}
    \caption{SIDIS multiplicities $\dd M^h/\dd z_h$ from COMPASS \cite{COMPASS:2016xvm} as a function of $z_h$ for $h=\pi^+$ (dashed lines) and $h=\pi^-$ (dotted lines) compared with the JAM24 fit result. The data and curves are scaled by a factor of $2^i$ ($i=0,\ldots, 4$) to more clearly separate and isolate the various $x_B$ and $y$ bins.}
    \label{fig:sidis-p}
\end{center}
\end{figure}

A summary of the figures of merit is presented in Table~\ref{tb:chi2} and Fig.~\ref{fig:chi2s} for the scenarios discussed in this work. 
Starting with the baseline scenario, we find that all datasets considered in the analysis are described quite well.
An exception is the neutral current $e^\pm p$ HERA data at $\sqrt{s} = 318$~GeV.
The reduced $\chi^2$ and $Z$-score values for this dataset are relatively high, with $\chi^2_{\rm red} = 1.35$ and $Z = 4.52$ for $e^+ p$, and $\chi^2_{\rm red} = 1.42$ and $Z = 3.37$ for $e^- p$ scattering, indicating potential difficulties in describing the DIS data across the full range of HERA kinematics.

Since the uncertainties are typically larger at higher $x$, it is likely that the observed tensions occur at small $x$, where higher-order corrections $\sim \log x$ are known to play a more important role.
We have verified this hypothesis by performing the analysis with a larger $Q^2$ cut, $Q^2 > 3.5$~GeV$^2$, on the HERA data to remove data at $x < 3.46 \times 10^{-5}$.
The results, shown in Fig.~{\ref{fig:chi2s}} as open circles, demonstrate that the $Z$-scores can be reduced from 4.22 to as low as 3.19.

To test whether the tensions in the HERA data may be lessened by increasing the perturbative accuracy of our analysis, we compare in Table~\ref{tab:jam-vs-world} the results from the NNLO analyses by NNPDF~\cite{NNPDF4.0} and CT18~\cite{CT18}.
The comparison indicates that our results are similar to those obtained by these groups. 
In particular, we find that applying a larger $Q^2$ cut, as in the NNPDF analysis~\cite{NNPDF4.0}, yields $\chi^2_{\rm red}$ and $Z$-score values very similar to our results.

Moreover, we have verified that our reconstructed PDFs and FFs are not significantly affected by the value of the DIS cut.
However, since our primary objective is to study the reconstruction of the strange quark PDF from SIDIS and $W$+\,charm data --- and because applying a $Q^2 > 3.5$~GeV$^2$ cut would remove a significant amount of SIDIS data --- we retain our nominal $Q^2 > m_c^2$ cut as the final choice for all scenarios.

The inclusion of the SIDIS data in the +SIDIS scenario slightly increases the figures of merit for most PDF-dependent observables, with the most notable cases being the HERA cross sections and $W$ charge asymmetries, which can modify the gluon and light sea quark PDFs, respectively, changing $Z$-scores within 1$\sigma$.  
Interestingly, while the $Z$-scores for the SIA pion and kaon data shift by $1.53\sigma$ and $2.35\sigma$, respectively, those for the unidentified charged hadron SIA data improve by $0.44 \sigma$.  
However, all SIA $\chi^2_{\rm red}$ and $Z$-score values remain acceptable with the inclusion of SIDIS data. 
Moreover, we do not find any significant tensions when combining SIDIS data with the rest of the global dataset considered in our~analysis.

Note also that in Ref.~\cite{Moffat:2021dji} the $h^-$ SIDIS dataset was 
found to have an anomalously low $\chi_{\rm red}^2$.  We have verified that this is in fact to an overestimation of the experimental uncorrelated uncertainties, and the corrected results in the current analysis give a $\chi_{\rm red}^2$ and $Z$-score comparable to other SIDIS datasets.

The figures of merit in the +$W$-charm scenario are largely compatible with the baseline results.
We find that most dataset descriptions improve, with the exception of the $Z$ rapidity data from CDF and D\O, possibly due to sea quark sensitivity in the low-$x$ region.
For the $W$+\,charm datasets, both the ATLAS and CMS cross sections are well described, with $\chi^2_{\rm red} < 1$, apart from the CMS $W$+\,charm ratio measurement, which has a $\chi^2_{\rm red} = 1.76$.
However, the relatively low $Z$-score of 1.19 does not indicate any anomaly.

\begin{figure}[t]
\begin{center}   
\includegraphics[scale=0.41]{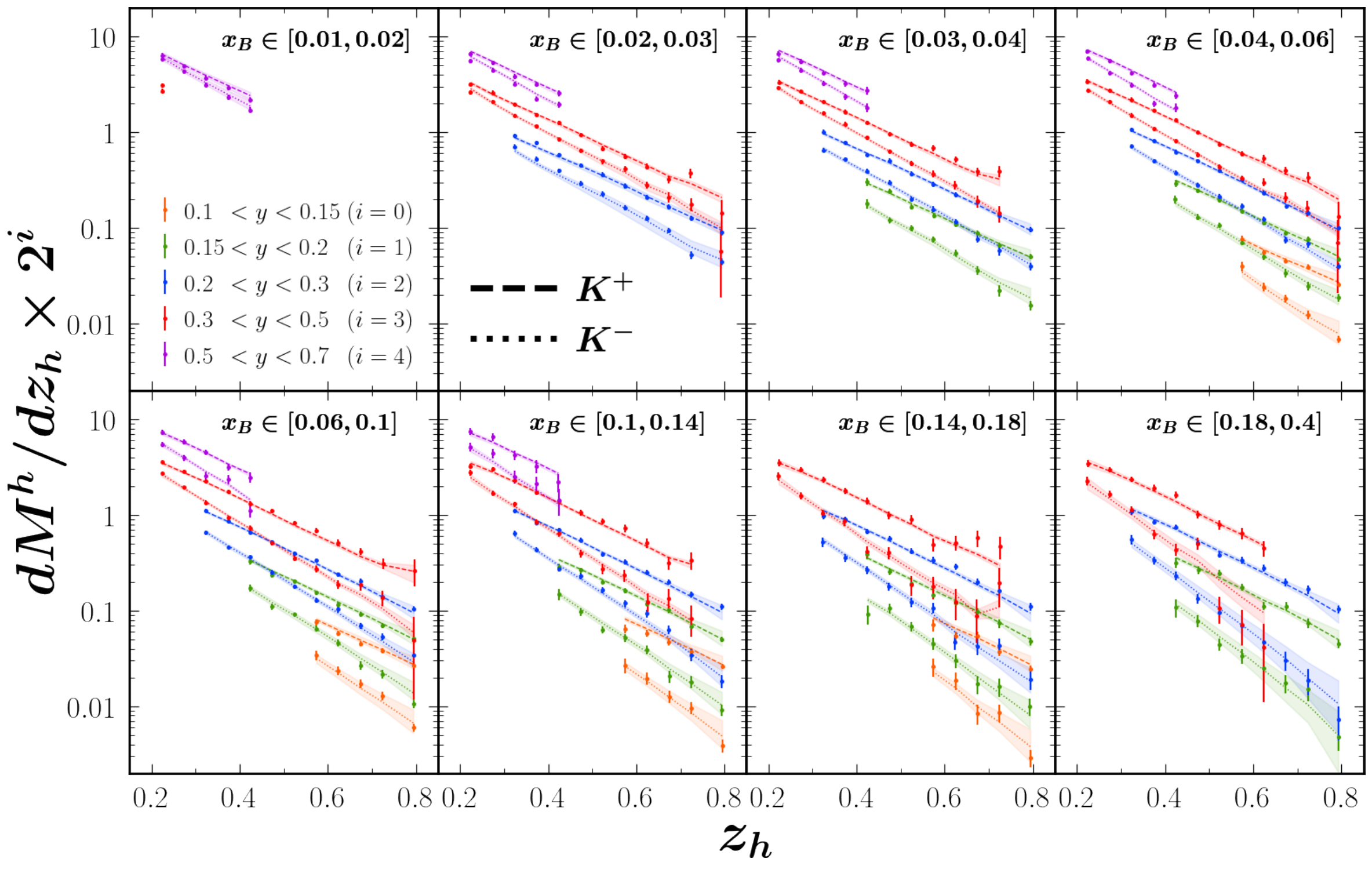}
    \caption{As in Fig.~\ref{fig:sidis-p}, but for $K^\pm$ data.}
\label{fig:sidis-k}
\end{center}
\end{figure}

In the final JAM24 scenario, where both SIDIS and $W$+\,charm data are fitted along with the baseline datasets, the figures of merit are most similar to the +SIDIS fit.
For SIDIS data, we find a decrease in $\chi^2_{\rm red}$ compared to the +SIDIS fit, while the $\chi^2_{\rm red}$ for the SIA and $W$+charm data increase marginally compared to the +SIDIS and +$W$-charm scenarios, respectively.
Focusing on the main datasets of interest for the current analysis, a detailed comparison between theory and data for the ATLAS and CMS $W$+\,charm production cross sections, differential in the lepton pseudorapidity $|\eta_{\ell}|$, is shown in Fig.~\ref{fig:w+charm} for various values of the lepton transverse momentum lower cutoff, $p_T^\ell$. 
Remarkably, we find a consistent agreement between the description of the $W$+\,charm data in the +SIDIS and JAM24 fits, indicating no tension between the observables.
For all of the CMS $W$+\,charm summed cross sections, as well as the ATLAS $W^+$+\,$\bar{c}$ data, excellent overall agreement is seen for both the 7~TeV and 13~TeV data.
For the ATLAS $W^-$+\,$c$ data, the theoretical result slightly underestimates the measurements at $|\eta_\ell| \sim 0.8$, while for the CMS $(W^+$+\,$\bar c)/(W^-$+\,$c)$ ratio the description at $|\eta_\ell| = 1.35$ differs by $2.35\sigma$, leading to a larger $\chi^2_{\rm red}$ than for the other datasets.

Both the ATLAS and CMS data indicate a slightly larger $W^-$+\,$c$ cross section compared with that for $W^+$+\,$\bar{c}$.
In principle, this could suggest a positive strange-antistrange asymmetry, $s - \bar{s}$, arising from a small preference of the $g s \to W^- + c$ channel versus the $g \bar{s} \to W^+ + \bar{c}$ channel.
However, this conclusion is still somewhat limited by the current experimental uncertainties, and additional, higher precision data would be needed to make more a definitive statement.

\begin{figure}[t]
\begin{center}   
\includegraphics[scale=0.41]{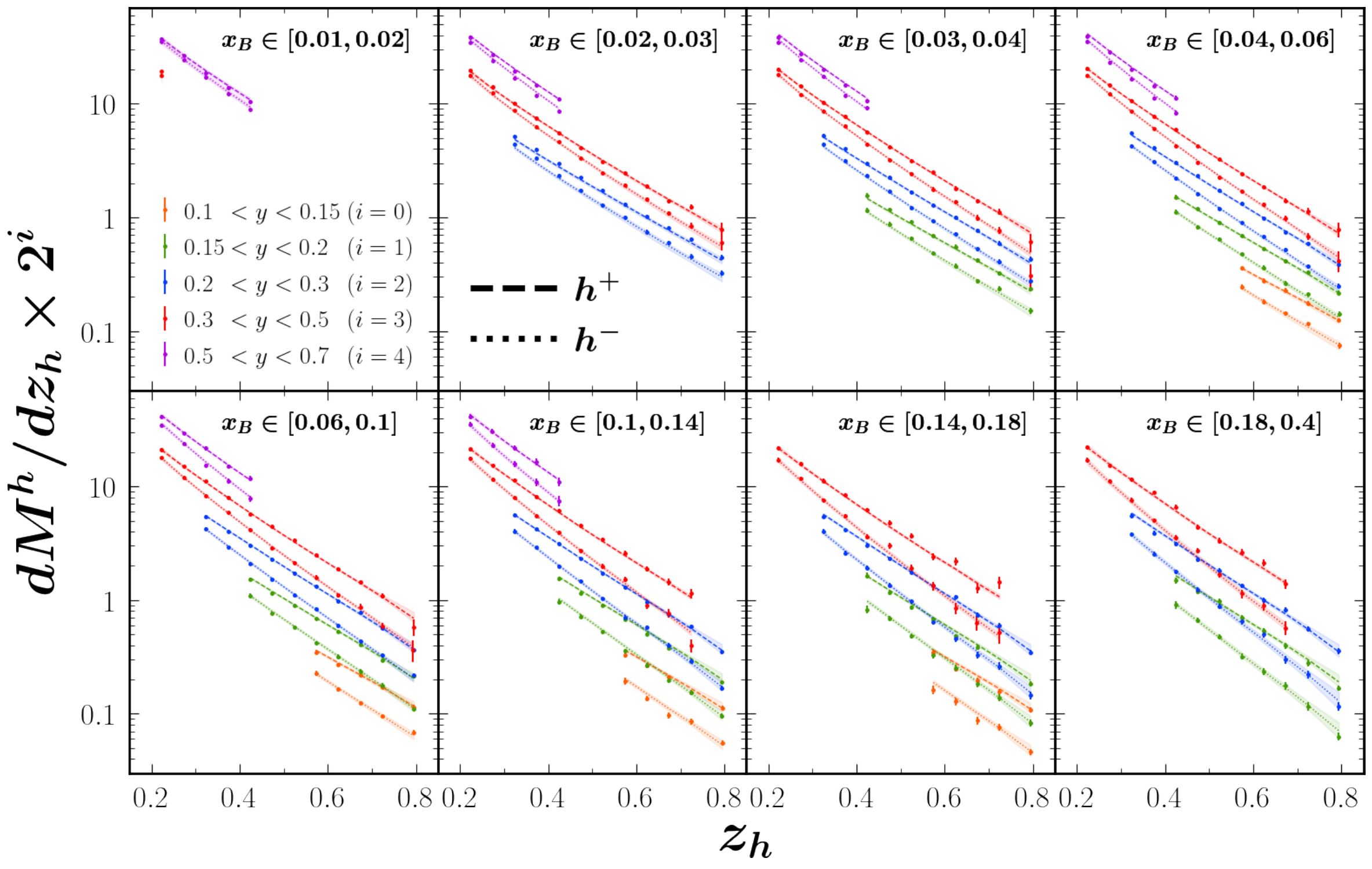}
    \caption{As in Fig.~\ref{fig:sidis-p}, but for unidentified hadrons $h^\pm$.}
\label{fig:sidis-h}
\end{center}
\end{figure}

Turning now to the SIDIS data from COMPASS, in Figs.~\ref{fig:sidis-p}, \ref{fig:sidis-k}, and \ref{fig:sidis-h} we show the $z_h$~dependence of the $\pi^\pm$, $K^\pm$, and $h^\pm$ multiplicities, respectively, defined as ratios of cross sections for SIDIS and inclusive DIS at the same values of $x_B$ and $Q^2$,
\begin{equation}
    \frac{\dd M^h}{\dd z_h}
    = \frac{\dd \sigma^h/\dd x_B \, \dd z_h \, \dd Q^2}
           {\dd \sigma^{{\mbox{\rm \tiny DIS}}}/\dd x_B \, \dd Q^2}.
    \label{eq:mult}
\end{equation}
For all pions and unidentified hadrons, better agreement between theory and experimental data is found for negatively charged hadrons than for positively charged hadrons, while the opposite is found for kaon production.
This can be observed in both the high-$x_B$ and \mbox{low-$Q^2$} (low-$y$) regions, where the differences between the positively and negatively charged hadron multiplicities become larger, especially for kaons.
In general, a good description of the COMPASS data is obtained across most kinematic regions and bins, with poorer agreement in the high-$z_h$ region and the low-$Q^2$ region, particularly for the high-$x_B$ bins and some of the lower-$x_B$ bins.
Better agreement may be obtained by including hadron mass and other power corrections~\cite{Accardi:2009md, Guerrero:2015wha, Guerrero:2017yvf}, which are known to be more important at large $x_B$ and large~$z_h$, as one approaches the exclusive limit.

\begin{figure}[t]
\begin{center}   
\includegraphics[scale=0.46]{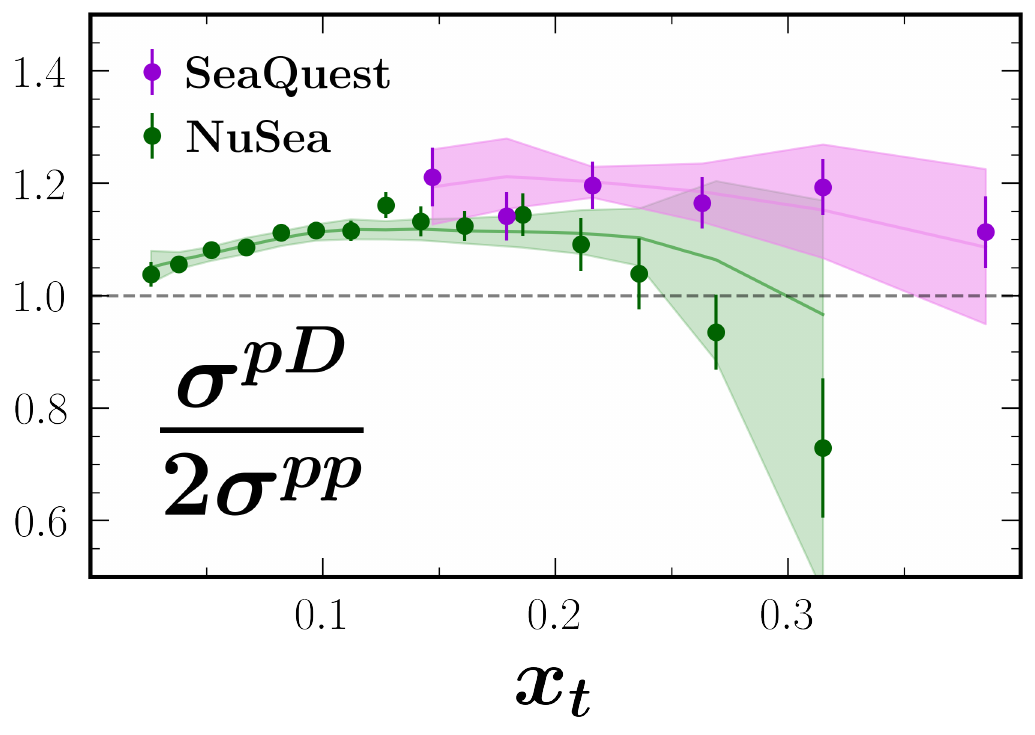}
\includegraphics[scale=0.46]{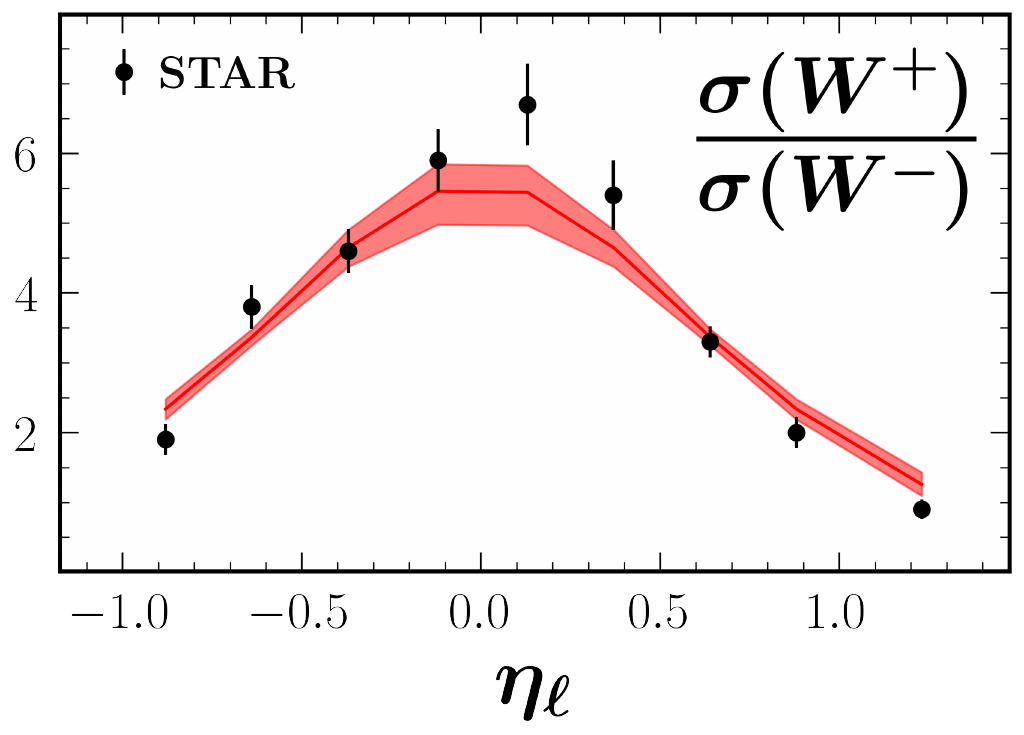}
    \caption{
    (Left) Ratio of $pD$ to $pp$ Drell-Yan cross sections versus the target momentum variable $x_t$ from SeaQuest (purple circles) for $0.48 < x_b < 0.69$ and NuSea (green circles) for $0.32 < x_b < 0.56$, compared to the JAM24 fit result (colored bands). 
    (Right) $W$-lepton ratio cross sections for $pp$ collisions from STAR at $\sqrt{s} = 510$~GeV and $p^\ell_T > 25$~GeV as a function of pseduorapidity $\eta_\ell$ compared to the JAM24 fit (red band).}
\label{fig:dy_RHIC}
\end{center}
\end{figure}

For completeness, we also present in Fig.~\ref{fig:dy_RHIC} the data and theory comparisons for DY datasets from the NuSea and SeaQuest experiments, as well as the $W$-lepton cross section ratios from STAR, demonstrating an excellent description of these data within the JAM24 scenario.
While these datasets are not directly sensitive to the strange quark PDF, they do provide constraints on the light sea sector, particularly the $\bar{d}-\bar{u}$ asymmetry.
Changes in the light sea quark sector can, however, induce modifications in the strange quark PDF through QCD evolution and the overall description of the global dataset.

Lastly, we address the impact of adopting different small-$x$ asymptotic behaviors for the light sea and strange quark PDFs, modeled using the two independent functions, $S_1$ and $S_2$, in Eq.~(\ref{eq:pdf_shapes}).
During the model calibration, we observed differences in the data by contrasting the results for $S_1 \neq S_2$ and $S_1 = S_2$. 
While both cases provide good descriptions of the data, we found that $S_1 = S_2$ gives a slightly poorer description of the data requiring the need to include larger systematic shifts to the theory.
We therefore choose the more general $S_1 \neq S_2$ condition in all the scenarios explored in this analysis.

\subsection{Reconstructed PDFs and FFs}
\label{sub.pdf}

\begin{figure}[t]
\begin{center}   
\includegraphics[scale=0.43]{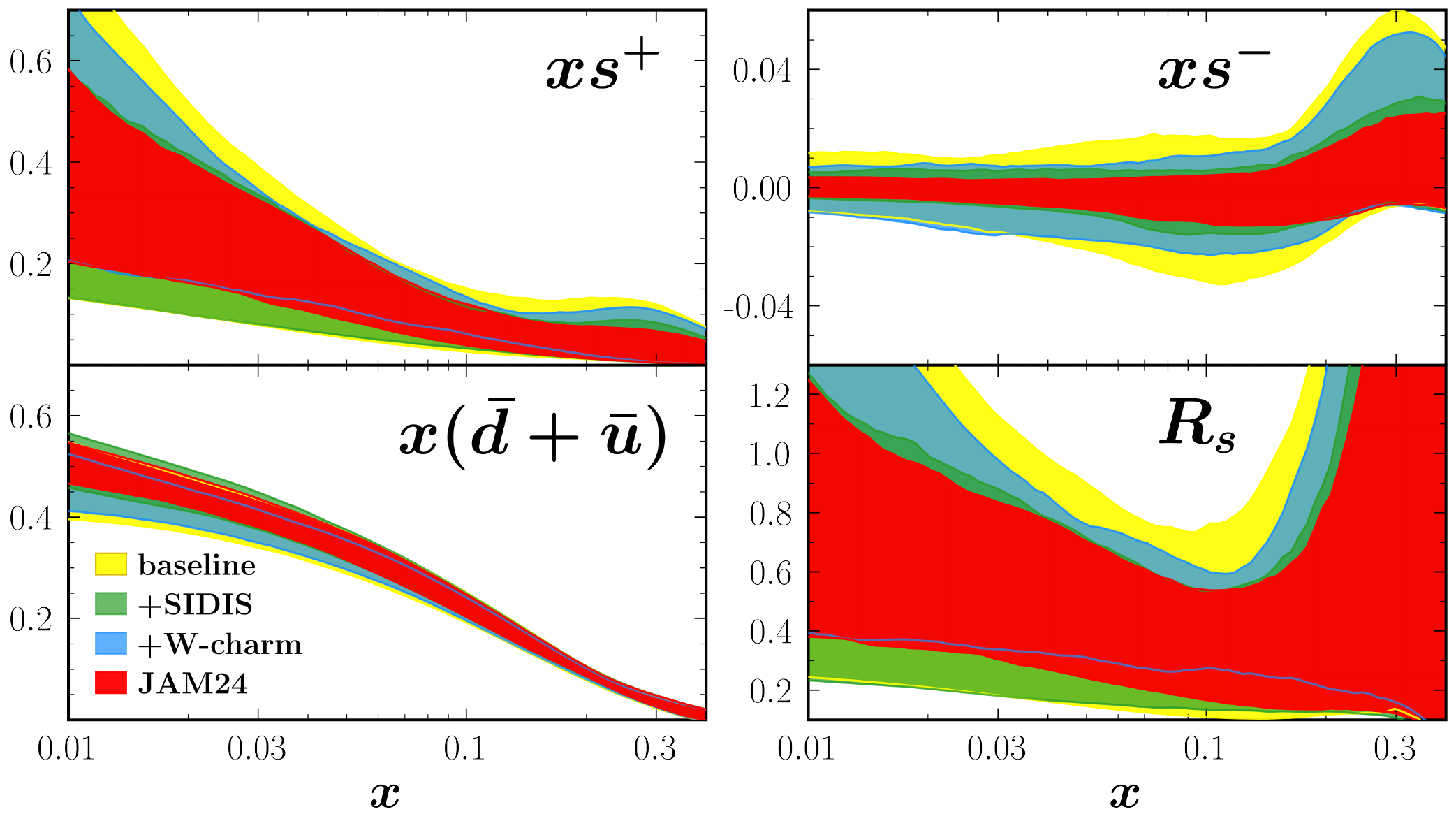}
    \caption{Impact of the individual SIDIS (green bands) and $W$+\,charm (blue bands) data on the PDFs $xs^+$, $xs^-$, $x(\bar d+\bar u)$ and ratio $R_s$ compared with the baseline fit (yellow bands) and the full JAM24 fit (red bands) that includes all of the data. Uncertainty bands represent a 95\% credible interval at the scale $Q^2 = 4$~GeV$^2$.}
\label{fig:combined_pdfs}
\end{center}
\end{figure}

Having established the agreement between our fits and the data, we now present the reconstructed PDFs and FFs in the various scenarios, focusing in particular on understanding the impact of the SIDIS and $W$+\,charm data on both the magnitude and constraints of the strange quark PDF.
The effects of the additional datasets beyond the baseline on the strange and light antiquark distributions are illustrated in Fig.~\ref{fig:combined_pdfs}, where results for the initial baseline, +SIDIS, +$W$-charm, and the final JAM24 fit that incorporates both the SIDIS and $W$+\,charm datasets, are compared at a common scale of $Q^2 = 4$~GeV$^2$.

The inclusion of either the SIDIS or $W$+\,charm datasets is clearly observed to decrease the uncertainties in the $s^+ = s + \bar{s}$ distribution for all $x$.  
Both datasets decrease the upper limit of uncertainty in $s^+$ compared to the baseline analysis, with SIDIS data constraining the upper limit more than $W$+\,charm data; however, the $W$+\,charm data decrease the lower uncertainty, whereas SIDIS data give results similar to those for the baseline fit.  
The net strange quark content between all scenarios is within uncertainties, but with SIDIS data having a slight preference for lower strangeness, and $W$+\,charm data preferring larger strangeness.  
For the light sea quark sector, there is agreement between scenarios for all $x$ values within uncertainties.

The ratio of strange to nonstrange sea quark PDFs follows a similar behavior as the $s^+$ PDF, with the $W$+\,charm data raising the lower uncertainty bound and expanding the upper uncertainty bound, while the SIDIS data align with the baseline fit at the lower uncertainty bound but reduce the upper uncertainty compared to the $W$+\,charm data.
All scenarios produce a monotonically decreasing $R_s$ in the range $0.01 < x < 0.3$, while outside of this range the uncertainties become too large to draw clear conclusions.
This suggests compatibility with an SU(3) symmetric sea for $x \lesssim 0.02$, albeit within sizeable uncertainties, but a suppressed strange quark sea at higher $x$, in contrast to the earlier finding by the ATLAS Collaboration~\cite{ATLAS:2014jkm}.
This discrepancy may arise due to dataset limitations or insufficient parametrization flexibility in that analysis.
Interestingly, the apparent rise in $R_s$ at $x \gtrsim 0.2$ is similar to that observed in Ref.~\cite{Faura:2020oom}, which included also neutrino-nucleus cross sections but not SIDIS data.
The final results of the full JAM24 fit favor a small strange quark suppression consistent with the +SIDIS scenario, which reflects the larger quantity of SIDIS data (1490 points) compared to the $W$+\,charm data (37 points).

While both the SIDIS and $W$+\,charm data reduce the uncertainty on $s^+$, particularly at $x \lesssim 0.1$, future data with greater sensitivity to the strange quark will be needed to further reduce the uncertainty in this region.
Such data could come from experiments at Jefferson Lab, involving parity-violating inclusive DIS measurements and high-precision SIDIS multiplicity data from Halls~B and C.
For the $s^- = s - \bar{s}$ difference, we find a notable consistency for all scenarios, which show a magnitude consistent with zero for all $x$, but a slight rise at $x \sim 0.3$. 
The datasets with sensitivity to strangeness show a reduction of uncertainty, with SIDIS data constraining the upper limit more than the $W$+\,charm data at high $x$.
Although QCD does not require the strange and antistrange PDFs to be identical, the current datasets provide no clear indication of a nonzero $s^-$ asymmetry within the present uncertainties.

\begin{figure}[t]
\begin{center}   
\includegraphics[scale=0.44]{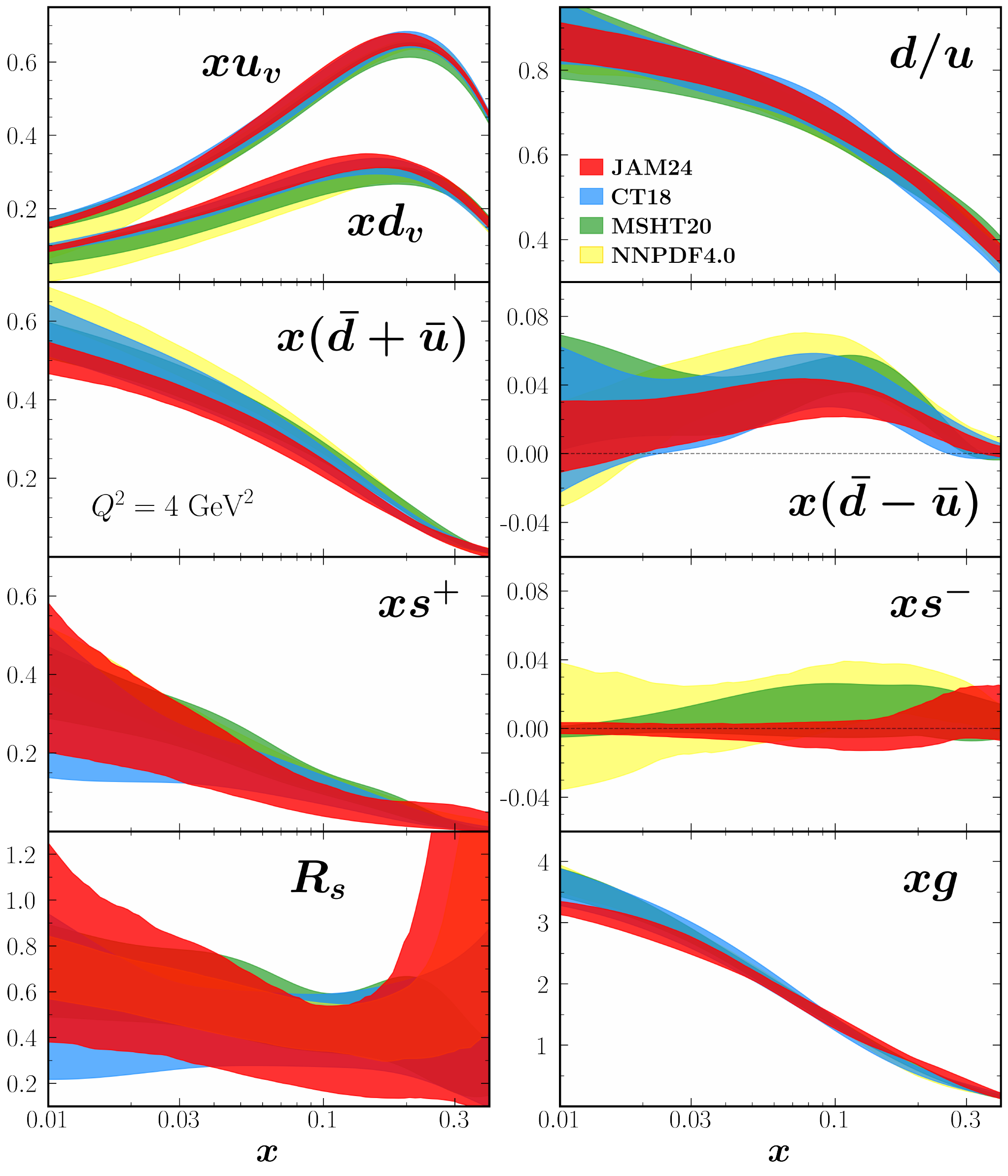}
    \caption{Comparison of various PDFs from the present JAM24 global analysis (red bands) with results from the CT18~\cite{CT18} (blue), MSHT20~\cite{Bailey:2020ooq} (green), and NNPDF4.0~\cite{NNPDF4.0} (yellow) NLO parametrizations at the scale $Q^2 = 4$~GeV$^2$. Bands represent a 95\% credible interval for JAM24 and NNPDF4.0, a 90\% C.L. for CT18, and a 68\% C.L. for MSHT20.}
\label{fig:lhapdfs}
\end{center}
\end{figure}

\begin{figure}[t]
\begin{center}   
\includegraphics[scale=0.36]{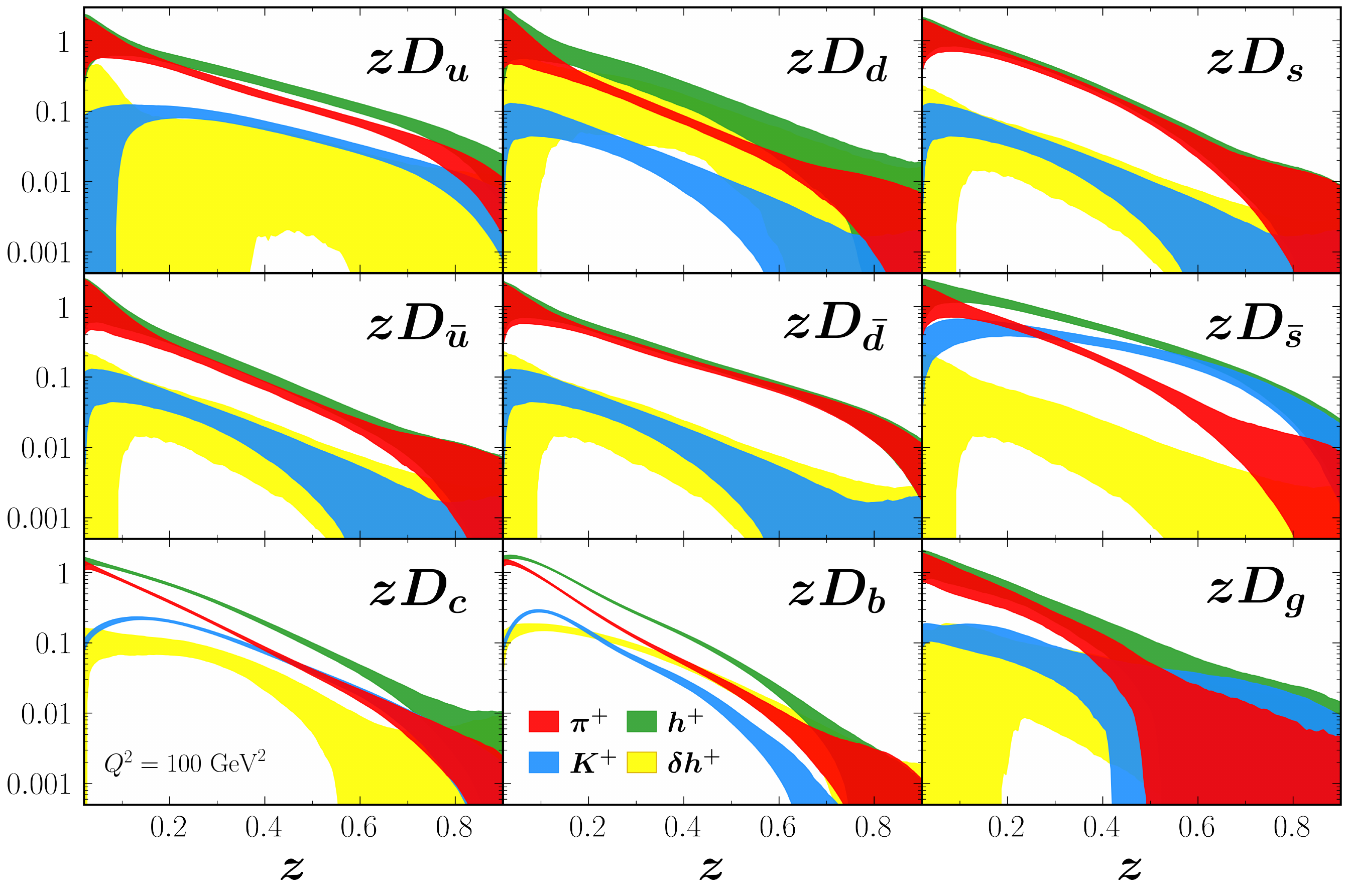}
    \caption{Parton to hadron fragmentation functions $z D_f$ versus $z$ for the production of charged pions $\pi^+$ (red bands), kaons $K^+$ (blue), unidentified hadrons $h^+$ (green), and residual hadrons $\delta h^+$ (yellow) for 95\% credible interval at $Q^2 = 100$~GeV$^2$.}
\label{fig:FFs}
\end{center}   
\end{figure}

The results of our full global analysis for the PDFs are illustrated in Fig.~\ref{fig:lhapdfs}, where we show each of the fitted PDFs at $Q^2 = 4$~GeV$^2$.
For comparison, we also show the results from other NLO parametrizations, including the CT18~\cite{CT18}, MSHT20~\cite{Bailey:2020ooq} and NNPDF4.0~\cite{NNPDF4.0}, with uncertainty bands representing 90\% confidence level (C.L.), 68\% C.L., and 95\% credible interval, respectively.
In these comparisons we focus on the kinematic region of parton momentum fractions $0.01 \lesssim x \lesssim 0.4$ where the $W$+\,charm and SIDIS datasets have the greatest impact on the strange quark PDF.

\begin{figure}[t]
\begin{center}   
\includegraphics[scale=0.5]{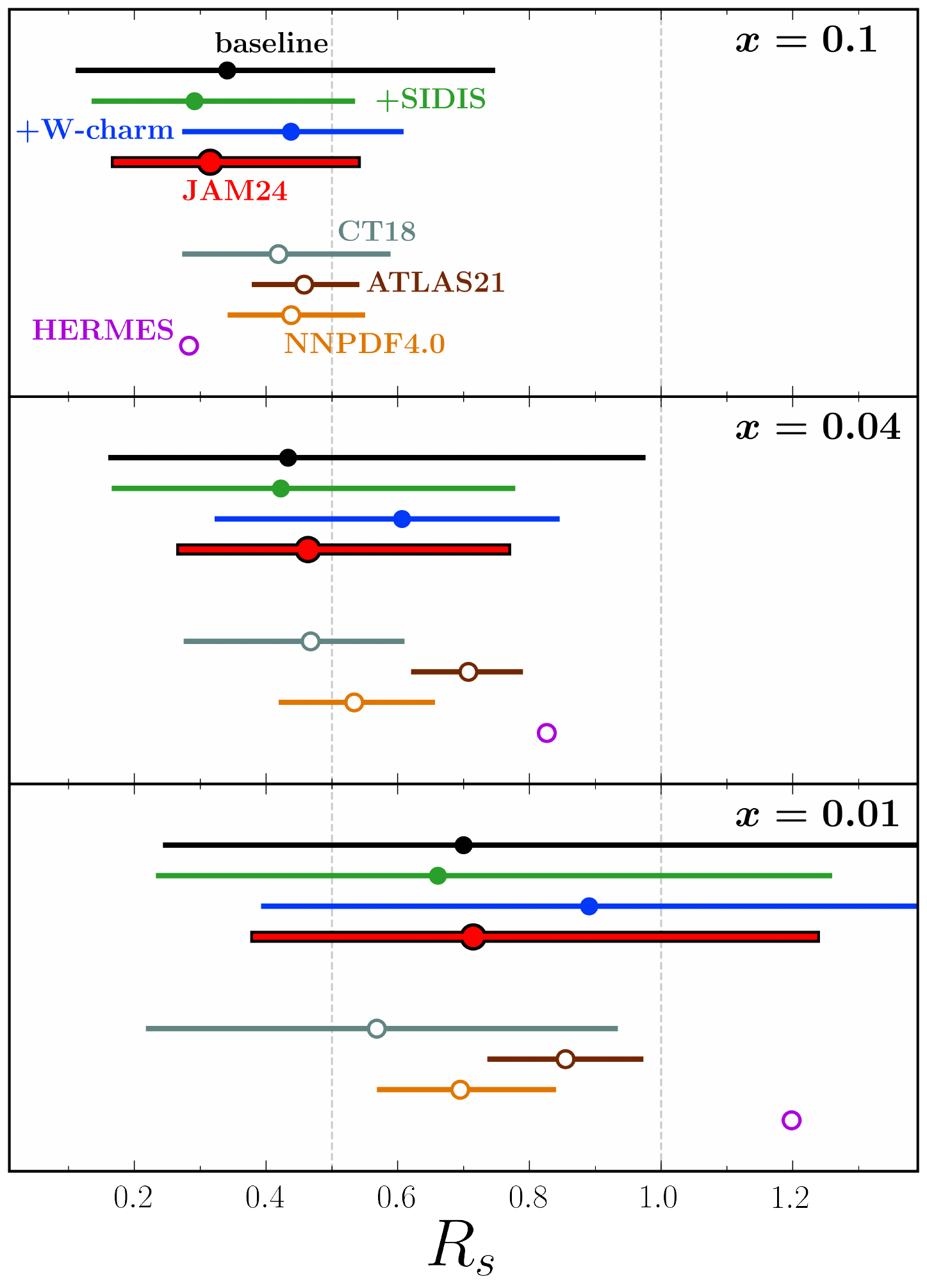}
\caption{Strange to nonstrange quark PDF ratio $R_s$ at $x=0.1$ (top), $x=0.04$ (middle), and $x=0.01$ (bottom) at $Q^2=2.5$~GeV$^2$ for HERMES data~\cite{HERMES:2008pug} and $Q^2=4$~GeV$^2$ for all other fits~\cite{CT18,ATLAS:2021vod, NNPDF4.0}. The main JAM24 results (red circles) are compared with various other scenarios and analyses.  Uncertainties shown represent a 95\% credible interval for baseline, +SIDIS, +$W$-charm, JAM24, and NNPDF4.0, a 90\% confidence level for CT18, and a tolerance $T=1$ for ATLAS21.}
\label{fig:Rs_pdfs}
\end{center}
\end{figure}

For the valence $u$- and $d$-quark distributions, as well as the $d/u$ PDF ratio, our results are in overall agreement with those from other groups.
The total light antiquark distribution $\bar{d}+\bar{u}$ is slightly below the NNPDF4.0 result, but is in agreement with other fits for $x \lesssim 0.03$, and is consistent for higher $x$, while the $\bar{d}-\bar{u}$ difference sits at the lower end of the other PDF sets in the intermediate-$x$ range.
The total strange distribution $s^+$ is in general agreement with other fits at low $x$, but is slightly lower compared to NNPDF4.0 and MSHT20 in the $0.1 \lesssim x \lesssim 0.2$ range.
This is also seen with the strange to nonstrange sea quark ratio, $R_s$, which is $0.72^{+0.52}_{-0.34}$ at $x=0.01$, but falls to $0.34^{+0.47}_{-0.21}$ at $x=0.2$.  The observed differences in uncertainty bands are likely attributed to methodological choices in uncertainty estimation rather than differences in data selection.

For the FFs extracted in our analysis, in Fig.~\ref{fig:FFs} we show the $z$ dependence for positively charged $\pi^+$, $K^+$, unidentified hadrons $h^+$, and residual hadrons $\delta h^+$ defined in Eq.~({\ref{eq:had_def}}) at a scale $Q^2 = 100$~GeV$^2$, for a 95\% credible interval.
Since the pion is the lightest hadron, as expected we find the magnitude of the $\pi^+$ FFs is generally larger than those for kaons and other hadrons for most quark flavors.
An exception to this is the FF for $\bar{s} \to K^+$, where the $\bar{s}$ flavor is favored by $K^+$ but unfavored by $\pi^+$.
Additionally, the $\pi^+$ FFs are comparable to the $c \to K^{+}$ and the $b \to \delta h^+$ FFs in the intermediate-$z$ region, and $g \to \pi^+$ comparable to other hadrons in the high-$z$ sector.
These findings are similar to those found in previous JAM analyses~\cite{JAM19,
Moffat:2021dji} and by other groups~\cite{HKNS}.

A summary of our findings for the strange to nonstrange ratio $R_s$ is shown in Fig.~{\ref{fig:Rs_pdfs}} for $x=0.01$, $0.04$, and $0.1$ at $Q^2=4$~GeV$^2$.  Here we compare the $R_s$ values from the all fits considered in this analysis, namely, the baseline +SIDIS, +$W$-charm, and the full JAM24 analysis, with results from the CT18~\cite{CT18}, ATLAS21~\cite{ATLAS:2021vod}, and NNPDF4.0~\cite{NNPDF4.0} PDF analyses at the same $x$ and $Q^2$, and with the HERMES extraction at $Q^2=2.5$~GeV$^2$~\cite{HERMES:2008pug}.
For the final JAM24 result, we find a strange to nonstrange sea quark ratio of $R_s = 0.72^{+0.52}_{-0.34}$ at $x=0.01$ to be compatible with other fits in this analysis and results from other groups.
In the intermediate-$x$ region, we find a more suppressed strange-quark PDF, with $R_s = 0.46^{+0.30}_{-0.20}$ at $x=0.04$ and $R_s = 0.32^{+0.23}_{-0.15}$ at a higher $x=0.1$.
This is again compatible with most other fits at these kinematics and is most similar in magnitude and uncertainty to the +SIDIS fit.

\section{Conclusions}
\label{s.conclusions}

How ``strange'' is the proton has been a question that has perplexed nuclear and particle physicists for decades, attracting considerable experimental and theoretical attention in the quest to understand the detailed structure of the proton's sea quarks, and in particular the size of the strange sea relative to the light quark sea.
In this study we have for the first time performed a comprehensive global QCD analysis of the effect on the strange quark PDF in the proton from simultaneously including $W$+\,charm production data in $pp$ collisions at the LHC and hadron production data in semi-inclusive muon-deuterium scattering from COMPASS, both of which are expected to have sensitivity to the $s$ and $\bar s$ quark distributions.
The SIDIS data in particular require a simultaneous fit to both unpolarized PDFs and parton to hadron FFs.

An excellent overall reduced $\chi^2$ value of $\chi^2_{\rm red} = 0.99$ is obtained for the global fit, which includes datasets from inclusive DIS, Drell-Yan lepton-pair production, weak boson and jet production, SIDIS, and $e^+ e^-$ SIA reactions, comprising nearly 6\,000 data points.
It is the first time that such a large body of data, constraining both unpolarized PDFs and FFs, has been successfully described within a collinear QCD factorization framework.
We find $\chi^2_{\rm red} \lesssim 1$ for both the LHC $W$+\,charm datasets, and the COMPASS SIDIS data on $\pi^\pm$, $K^\pm$, and unidentified $h^\pm$ leptoproduction.
In comparison to a baseline global fit that does not include these datasets, we find that the combined SIDIS and $W$+\,charm datasets are consistent in magnitude but reduce the uncertainty bounds for all $x$. 
Without the $W$+\,charm data, a broader range of suppressed $s^+$ PDFs are allowed for $x \lesssim 0.1$, while without SIDIS data a more enhanced range of $s^+$ PDFs is permitted for all $x$.

For the combined JAM24 fit, the SIDIS data have a stronger pull on $s^+$ at high $x$, while both the SIDIS and $W$+\,charm data have similar pulls at lower $x$ values.
Overall, the strange to nonstrange sea quark distribution ratio at $Q^2=4$~GeV$^2$ is determined to be $R_s = \{ 0.72^{+0.52}_{-0.34}, 0.46^{+0.30}_{-0.20}, 0.32^{+0.22}_{-0.15} \}$ for $x = \{ 0.01, 0.04, 0.1 \}$.
We conclude therefore that at $x \approx 0.01$, the global dataset is compatible with an SU(3) flavor symmetric sea, but indicates strong SU(3) symmetry breaking in the sea for $x \gtrsim 0.02$.
Our analysis also places more stringent constraints on the strange--antistrange PDF asymmetry, $s-\bar s$, with a significant reduction relative to the baseline fit, but still compatible with zero over the entire $x$ range studied.

In the future our analysis can be extended in several ways.
Firstly, we plan to include inclusive $pp \to h X$ data to better constrain the FFs~\cite{JAMFF}, supplementing the constraints from SIA and SIDIS.
We will also explore the possibility of using new lattice QCD simulations of PDF moments and pseudo-Ioffe time distributions to provide complementary constraints on the $s$ and $\bar s$ distributions.
Upcoming data that will help with the reconstruction of the $s$ and $\bar s$ PDFs will include parity-violating DIS, which gives access to a new combination of $u^+$, $d^+$ and $s^+$ PDFs, as well as SIDIS for pion and kaon production at Jefferson Lab, with a 12~GeV and possibly a 22~GeV electron beam, as well as the Electron-Ion Collider.
Finally, a definitive analysis will also include neutrino-nucleus DIS data, which historically been used to constrain the strange-quark PDF, with a systematic treatment of nuclear and hadronization uncertainties.

\begin{acknowledgments}
We thank Rabah Abdul Khalek, Patrick Barry, Chris Cocuzza, Amanda Cooper-Sarkar, Emanuele Nocera, and Juan Rojo for helpful discussions and communications. 
This work was supported by the DOE contract No.~DE-AC05-06OR23177, under which Jefferson Science Associates, LLC operates Jefferson Lab. T.A. acknowledges support from JSA/JLab Graduate Fellowship Program.  
The work of N.S. was supported by the DOE, Office of Science, Office of Nuclear Physics in the Early Career Program.
\end{acknowledgments}

\clearpage
\appendix
\section{Fit Summary}
\label{summary}

In this appendix we collect tables summarizing figues of merit for all datasets used in the JAM24 analysis.  The values presented are determined from the expectation values of the modified theory given in Eq.~({\ref{eq:TThat}}). \\ \\

\begin{table*}[h]
\centering
\caption{DIS datasets used in the present JAM24 analysis.  Along with each dataset is the corresponding observable measured, number of data points after/before kinematic cuts, number of correlated systematic uncertainties, reduced $\chi^2$ and $Z$-score, correlated~$\chi^2$, normalized $\chi^2$, and the fitted normalization.\\}
\begin{tabular}{l c | c c | c r c c}
\hhline{========}
~dataset & observable~ & 
$N_{\rm dat}$ &
~$N_{\rm cor}$~ &
~$\chi^2_{\rm red}$ ($Z$-score) &
$\chi^2_{\rm cor}$~~ &
$\chi^2_{\rm norm}$ &
fitted norm. \\
\hline
~SLAC               \cite{Whitlow:1991uw}        & $F_2^p$               & ~218/661 & ~---~ & 0.87 ($-1.38$) & ---~~~  & 0.55  & 0.984(9)   \\
~SLAC               \cite{Whitlow:1991uw}        & $F_2^d$               & ~228/692 & ~---~ & 0.65 ($-4.25$) & ---~~~  & 2.04  & 0.976(9)   \\
~BCDMS              \cite{BCDMS:1989ggw}         & $F_2^p$               & ~348/351 & ~5~   & 1.14 ($+1.74$) & 13.44~  & 1.02  & 0.970(6)   \\
~BCDMS              \cite{BCDMS:1989ggw}         & $F_2^d$               & ~254/254 & ~5~   & 1.07 ($+0.84$) & 7.99~   & 0.21  & 0.986(8)   \\
~NMC                \cite{NewMuon:1996fwh}       & $F_2^p$               & ~273/292 & ~11~  & 1.67 ($+6.62$) & 8.35~   & 0.71  & ~1.017(10) \\
~NMC                \cite{NewMuon:1996uwk}       & $F_2^d/F_2^p$         & ~174/260 & ~5~   & 0.90 ($-0.90$) & 3.36~   & 0.08  & 0.997(6)   \\
\hline
~HERA NC $e^+$ (1) \cite{H1:2015ubc}           & $\sigma^p_{\rm red}$  &
~402/485 & ~169~   & 1.42 ($+5.27$) & 120.65~  & --- & ---  \\
~HERA NC $e^+$ (2) \cite{H1:2015ubc}           & $\sigma^p_{\rm red}$  & 
~75/112  & ~169~   & 1.03 ($+0.22$) & 13.77~   & --- & ---  \\
~HERA NC $e^+$ (3) \cite{H1:2015ubc}           & $\sigma^p_{\rm red}$  & 
~259/260 & ~169~   & 0.89 ($-1.25$) & 14.13~   & --- & ---  \\
~HERA NC $e^+$ (4) \cite{H1:2015ubc}           & $\sigma^p_{\rm red}$  & 
~209/209 & ~169~   & 0.98 ($-0.19$) & 15.74~   & --- & ---  \\
~HERA NC $e^-$     \cite{H1:2015ubc}           & $\sigma^p_{\rm red}$  & 
~159/159 & ~169~   & 1.45 ($+3.58$) & 25.58~   & --- & ---  \\
~HERA CC $e^+$     \cite{H1:2015ubc}           & $\sigma^p_{\rm red}$  & 
~39/39   & ~169~   & 1.17 ($+0.80$) & 2.78~    & --- & ---  \\
~HERA CC $e^-$     \cite{H1:2015ubc}           & $\sigma^p_{\rm red}$  & 
~42/42   & ~169~   & 1.02 ($+0.15$) & 5.91~    & --- & ---  \\
\hhline{========}
\end{tabular}
\label{tb:stats DIS}
\end{table*}

\clearpage
\begin{table*}
\centering
\caption{As in Table~\ref{tb:stats DIS}, but for Drell-Yan data.}
{
\begin{tabular}{l c | c c | c c c c}
\hhline{========}
dataset             & 
observable          & 
$N_{\rm dat}$    & 
$N_{\rm cor}$       & 
~$\chi^{2}_{\rm red}$ ($Z$-score) &
~$\chi^2_{\rm cor}$  &
~$\chi^2_{\rm norm}$ &
fitted norm. 
\\ \hline
E866 (NuSea) \cite{NuSea:2001idv}       &
$M^3 \dd^2 \sigma/\dd M \, \dd x_F$~   &
184/184                                 &
---                                     &
1.20 ($+1.88$)                          &
---                                     & 
0.31                                    &
0.961(28) 
\\
E866 (NuSea) \cite{NuSea:2001idv}       &
$\dd \sigma^{p/d}/2 \, \dd \sigma^{pp}$ & 
15/15                                   & 
---                                     & 
1.00 ($+0.12$)                          & 
---                                     &
1.07                                    &
0.990(13) 
\\
E906 (SeaQuest) \cite{SeaQuest:2021zxb} &
$\dd \sigma^{p/d}/2 \, \dd \sigma^{pp}$ &
6/6                                     & 
1                                       &
0.65 ($-0.51$)                          &
0.07                                    &
0.36                                    & 
1.012(10) 
\\
\hhline{========}
\end{tabular}}
\label{tb:stas DY}
\end{table*}

\begin{table*}
\centering
\caption{As in Table~\ref{tb:stats DIS}, but for $W$-lepton asymmetry data.}
{
\begin{tabular}{l c | c c | c c }
\hhline{======}
dataset         & 
observable      & 
~$N_{\rm dat}$  & 
~$N_{\rm cor}$~   &
~$\chi^2_{\rm red}$ ($Z$-score)~ & 
~$\chi^2_{\rm cor}$~ 
\\
\hline
CMS $8~$TeV \cite{CMS:2016qqr}          &
$A_\mu$                                 & 
~11/11           & 
6               & 
0.36 ($-1.91$)  & 
0.22            
\\
CMS $7~$TeV \cite{CMS:2013pzl}          &
$A_\mu$                                 & 
~11/11           &
12              & 
1.44 ($+1.05$)  &
0.39            
\\
CMS $7~$TeV \cite{CMS:2012ivw}          & 
$A_e$                                   & 
~11/11           & 
2               & 
0.83 ($-0.27$)  & 
0.32            
\\
CMS $7~$TeV \cite{CMS:2011bet}          & 
$A_e$                                   & 
~6/6             & 
2               &
0.79 ($-0.19$)  & 
0.36            
\\
CMS $7~$TeV \cite{CMS:2011bet}          & 
$A_\mu$                                 & 
~6/6             & 
2               & 
0.05 ($-3.34$)  & 
0.43            
\\
LHCb $7~$TeV \cite{LHCb:2014liz}        & 
$A_\mu$                                 & 
~8/8             &  
---             &
0.48 ($-1.12$)  &
---  
\\
LHCb $8~$TeV \cite{LHCb:2015mad}        & 
$A_\mu$                                 & 
~8/8             & 
1               & 
0.39 ($-1.43$)  &
0.03            
\\
STAR \cite{STAR:2020vuq}                &
$\sigma^{W^+}/\sigma^{W^-}$             &
~9/9             &
---             & 
2.49 ($+2.42$)  & 
---             
\\
\hhline{======}
\end{tabular}}
\label{tb:stats WZrv}
\end{table*}

\begin{table*}
\centering
\caption{As in Table~\ref{tb:stats DIS}, but for $W$ charge asymmetry data.}
{
\begin{tabular}{l c | c c | c  c c}
\hhline{======}
~dataset &
observable~ &
~$N_{\rm dat}$~ & 
~$N_{\rm cor}$~ & 
~$\chi^{2}_{\rm red}$ ($Z$-score)~ &
~$\chi^2_{\rm cor}$~
\\ \hline
~CDF \cite{CDF:2009cjw} & $A_W$ &
13/13  & 5   & 1.25 ($+0.73$) & 2.80 
\\
~D\O~\cite{D0:2013lql}  & $A_W$ & 
14/14  & --- & 0.76 ($-0.57$) & ---  
\\
\hhline{======}
\end{tabular}}
\label{tb:stats Wasym}
\end{table*}

\begin{table*}
\centering
\caption{As in Table~\ref{tb:stats DIS}, but for $Z$ rapidity data.}
{
\begin{tabular}{l c | c c | c c c c}
\hhline{========}
~dataset & 
observable & 
~$N_{\rm dat}$~ &
~$N_{\rm cor}$~ & 
~$\chi^{2}_{\rm red}$ ($Z$-score)~ & 
~$\chi^2_{\rm cor}$~ & 
~$\chi^2_{\rm norm}$~ & 
~fitted norm.~ \\
\hline
~CDF \cite{CDF:2010vek} &
$\dd \sigma/\dd y$ &
28/28  & 1 & 1.35 ($+1.28$) & 2.32 & 1.11 & 0.937(12)
\\
~D\O~\cite{D0:2007djv}  &
~$\sigma^{-1} \dd \sigma/\, \dd y$~ &
28/28  & 1 & 1.13 ($+0.54$) & 0.41 & ---  & ---       
\\
\hhline{========}
\end{tabular}}
\label{tb:stats Zrap}
\end{table*}

\begin{table*}
\centering
\caption{As in Table~\ref{tb:stats DIS}, but for inclusive jet production data.}
{
\begin{tabular}{l c | c c | c c c c}
\hhline{========}
Dataset & 
Observable & 
~$N_{\rm dat}$~ & 
~$N_{\rm cor}$~ & 
~$\chi^{2}_{\rm red}$ ($Z$-score) &
$\chi^2_{\rm cor}$ & 
$\chi^2_{\rm norm}$ & 
fitted norm.
\\ \hline
D\O~\cite{PhysRevLett.101.062001}    & 
$\dd^2\sigma/\dd\eta\, \dd p_T$ & 
110/110 & 22    & 0.91 ($-0.62$) & 17.71  & 4.46  & 1.137(34) 
\\
CDF \cite{PhysRevD.75.092006}    & $\dd^2 \sigma/\dd \eta\, \dd p_T$ &
76/76   & 20    & 0.91 ($-0.54$) & 15.37  & 0.93  & 1.056(37) 
\\
STAR \cite{PhysRevLett.97.252001}     & ~$\dd^2\sigma/2\pi\, \dd\eta\, \dd p_T$~ &
3/5     & 1     & 0.11 ($-1.66$) & ~0.42   & 0.03  & \!\!1.013(3)  
\\
STAR \cite{PhysRevLett.97.252001}     & ~$\dd^2\sigma/2\pi\, \dd\eta\, \dd p_T$~ &
9/9     & 1     & 1.91 ($+1.68$) & ~0.23   & 1.61  & 1.101(16) 
\\
\hhline{========}
\end{tabular}}
\label{tb:stats jet}
\end{table*}

\begin{table*}
\centering
\caption{As in Table~\ref{tb:stats DIS}, but for $W$+\,charm production data.}
{
\begin{tabular}{l c | c c | c c c c}
\hhline{========}
dataset & 
observable & 
$N_{\rm dat}$ & 
~$N_{\rm cor}$~ & 
~$\chi^{2}_{\rm red}$ ($Z$-score)~ &
~$\chi^2_{\rm cor}$~ & 
~$\chi^2_{\rm norm}$~ & 
fitted norm. 
\\ \hline
ATLAS 7 TeV \cite{ATLAS:2014jkm} &
$\dd \sigma^{W^- + c}/ \dd |\eta|$ &
~11/11 & 86  & 0.65 ($-0.81$) & 6.88 & 0.08 & 0.995(55) 
\\
ATLAS 7 TeV \cite{ATLAS:2014jkm} & 
$\dd \sigma^{W^- + \bar{c}}/ \dd |\eta|$ &
~11/11 & 86  & 0.18 ($-2.96$) & 4.21 & 0.10 & 0.994(54)
\\
CMS 7 TeV \cite{CMS:2013wql} &
$\dd \sigma^{W+c} / \dd |\eta|$ &
~5/5   & 5   & 0.20 ($-1.77$) & 3.14 & ---  & ---  
\\
CMS 7 TeV \cite{CMS:2013wql} &
$\sigma^{W^+ +\bar{c}} /\sigma^{W^- +c}$ &
~5/5   & --- & 1.84 ($+1.28$) & ---  & ---  & ---
\\
CMS 13 TeV \cite{CMS:2018dxg} &
$\dd \sigma^{W+c} / \dd |\eta|$ &
~5/5   & 12  & 0.51 ($-0.73$) & 0.93 & 0.00 & 1.000(58)
\\
\hhline{========}
\end{tabular}}
\label{tb:stats wc}
\end{table*}

\begin{table*}
\centering
\caption{As in Table~\ref{tb:stats DIS}, but for SIDIS data.}
{
\begin{tabular}{l c | c c | c c }
\hhline{======}
~dataset & 
observable & 
$N_{\rm dat}$ & 
$N_{\rm cor}$~ & 
~$\chi^{2}_{\rm red}$ ($Z$-score)~ & 
~$\chi^2_{\rm cor}$~
\\ \hline
~COMPASS \cite{COMPASS:2016xvm}~ & 
$\dd M^{\pi^+}/\dd z_h$~ &
~249/311 & 1     & 0.94 ($-0.68$) & 2.41        
\\
~COMPASS \cite{COMPASS:2016xvm}~ &
$\dd M^{\pi^-}/\dd z_h$~ &
~249/311 & 1     & 0.85 ($-1.77$) & 4.97        
\\
~COMPASS \cite{COMPASS:2016crr}~ &
$\dd M^{K^+}/\dd z_h$~ &
~247/309 & 1     & 0.82 ($-2.17$) & 4.21        
\\
~COMPASS \cite{COMPASS:2016crr}~ &
$\dd M^{K^-}/\dd z_h$~ &
~247/309 & 1     & 0.94 ($-0.61$) & 6.50        
\\
~COMPASS \cite{COMPASS:2016xvm}~ &
$\dd M^{h^+}/\dd z_h$~ &
~249/311 & 1     & 0.89 ($-1.20$) & 0.74        
\\
~COMPASS \cite{COMPASS:2016xvm}~ &
$\dd M^{h^-}/\dd z_h$~ &
~249/311 & 1     & 0.84 ($-1.92$) & 2.49        
\\
\hhline{======}
\end{tabular}}
\label{tb:stats SIDIS}
\end{table*}

\begin{table*}
\centering
\caption{As in Table~\ref{tb:stats DIS}, but for $e^+ e^-$ SIA data with $\pi^\pm$ production. Note that an asterisk ${(*)}$ indicates that a normalization is fitted without penalty for datasets not providing normalized measurements, a dagger ${(\dagger)}$ indicates datasets with shared normalizations, and a star ${(\star)}$ indicates that a covariance matrix was provided for systematic uncertainties.}
{
\begin{tabular}{l c | c c | c c c l}
\hhline{========}
~dataset &
observable &
~$N_{\rm dat}$~ & 
~$N_{\rm cor}$~ & 
~$\chi^{2}_{\rm red}$ ($Z$-score)~ &
~$\chi^2_{\rm cor}$~ &
$\chi^2_{\rm norm}$ & 
fitted norm. \\
\hline
~Belle \cite{Belle:2013lfg} &
$\dd \sigma/ \dd z$ &
~70/78 & ---             & 0.19 ($-7.71$) & ---   & ---   & ~~0.624(9)$^*$
\\
~TASSO 12 GeV \cite{BRANDELIK1981357} &
$\dd \sigma/ \dd z$ &
~1/5   & ---             & 0.15 ($-0.51$) & ---   & ---   & ~~2.7(3.1)$^*$ 
\\
~TASSO 14 GeV \cite{TASSO:1983cre} &
$\dd \sigma/ \dd z$ &
~5/11  & ---             & 2.58 ($+1.98$) & ---   & ---   & ~~0.956(49)$^*$
\\
~TASSO 22 GeV \cite{TASSO:1983cre} &
$\dd \sigma/ \dd z$ &
~5/13  & ---             & 0.57 ($-0.59$) & ---   & ---   & ~~1.075(72)$^*$
\\
~TASSO 34 GeV \cite{TASSO:1988jma} &
$\sigma^{-1} \, \dd \sigma/\dd z_p$ &
~6/16  & ---             & 1.60 ($+1.07$) & ---   & 0.65  & ~~0.952(24)
\\
~TASSO 44 GeV \cite{TASSO:1988jma} &
$\sigma^{-1} \, \dd \sigma/\dd z_p$ &
~4/12  & ---             & 1.87 ($+1.22$) & ---   & 0.52  & ~~0.957(28)
\\
~TPC(c) \cite{Lu:1986mc} &
~$(\sigma \beta)^{-1} \, \dd \sigma/\dd z$ (c)~ &
~4/15  & ---             & 0.59 ($-0.43$) & ---   & ---   & ~~~~~--- 
\\
~TPC(b) \cite{Lu:1986mc} &
$(\sigma \beta)^{-1} \, \dd \sigma/\dd z$ (b) &
~4/15  & ---             & 1.85 ($+1.19$) & ---   & ---   & ~~~~~---
\\
~TOPAZ \cite{TOPAZ:1994voc} &
$\sigma^{-1} \, \dd \sigma/\dd \xi$ &
~2/17  & ---             & 0.54 ($-0.21$) & ---   & ---   & ~~~~~---
\\
~SLD \cite{SLD:2003ogn} &
$\sigma^{-1} \, \dd \sigma/\dd z_p$ &
~34/40 & ---             & 0.83 ($-0.65$) & ---   & 0.07  & ~~$1.003(3)^\dagger$ 
\\
~SLD(uds) \cite{SLD:2003ogn} &
~$\sigma^{-1} \, \dd \sigma/\dd z_p$ (uds)~ &
~34/40 & ---             & 0.77 ($-0.94$) & ---   & ---   & ~~$1.003(3)^\dagger$ 
\\
~SLD(c) \cite{SLD:2003ogn} &
~$\sigma^{-1} \, \dd \sigma/\dd z_p$ (c)~ &
~34/40 & ---             & 1.07 ($+0.36$) & ---   & ---   & ~~$1.003(3)^\dagger$ 
\\
~SLD(b) \cite{SLD:2003ogn} &
~$\sigma^{-1} \, \dd \sigma/\dd z_p$ (b)~ &
~34/40 & ---             & 0.60 ($-1.84$) & ---   & ---   & ~~$1.003(3)^\dagger$ 
\\
~ALEPH \cite{ALEPH:1994cbg} &
$\sigma^{-1} \, \dd \sigma/\dd z_p$ &
~23/39 & ---             & 0.54 ($-1.82$) & ---   & 0.11  & ~~0.990(3)
\\
~OPAL \cite{OPAL:1994zan} &
$\sigma^{-1} \, \dd \sigma/\dd p$ &
~24/51 & ---             & 1.73 ($+2.18$) & ---   & ---   & ~~~~~---
\\
~DELPHI \cite{DELPHI:1998cgx} &
$\sigma^{-1} \, \dd \sigma/\dd z_p$ &
~21/23 & ---             & 0.99 ($+0.07$) & ---   & ---   & ~~~~~---
\\
~DELPHI(uds) \cite{DELPHI:1998cgx} &
~$\sigma^{-1} \, \dd \sigma/\dd z_p$ (uds)~ &
~21/23 & ---             & 0.89 ($-0.25$) & ---   & ---   & ~~~~~---
\\
~DELPHI(b) \cite{DELPHI:1998cgx} &
~$\sigma^{-1} \, \dd \sigma/\dd z_p$ (b)~ &
~21/23 & ---             & 0.77 ($-0.71$) & ---   & ---   & ~~~~~---
\\
~BaBar (prompt) \cite{BaBar:2013yrg} &
$\sigma^{-1} \, \dd \sigma/\dd p$ &
~31/45 & 1\large$^\star$ & 0.34 ($-3.54$) & 50.71 & 40.10 & ~~0.938(5)
\\
~ARGUS \cite{ARGUS:1989zdf} &
$\sigma^{-1} \, \dd \sigma/\dd p$ &
~25/52 & ---             & 1.72 ($+2.19$) & ---   & 11.50 & ~~0.939(10)
\\
\hhline{========}
\end{tabular}}
\label{tb:stats sia-p}
\end{table*}

\begin{table*}
\centering
\caption{As in Table~\ref{tb:stats sia-p}, but for $e^+ e^-$ SIA data with $K^\pm$ production.} 
{
\begin{tabular}{l c | c c | c c c l}
\hhline{========}
~dataset & 
observable & 
$N_{\rm dat}$ & 
~$N_{\rm cor}$~ &
~$\chi^{2}_{\rm red}$ ($Z$-score)~ &
~$\chi^2_{\rm cor}$~ & 
~$\chi^2_{\rm norm}$~ & 
\!\!fitted norm. 
\\ \hline
~Belle \cite{Belle:2013lfg} &
$\dd \sigma/\dd z$ &
~70/78 & ---             & 0.06 ($-11.54$) & ---   & ---   & 0.852(10)*        
\\
~TASSO 12 GeV \cite{BRANDELIK1981357} &
$\dd \sigma/\dd z$ &
~1/3   & ---             & 0.26 ($-0.28$)  & ---   & ---   & 0.8(2.6)*
\\
~TASSO 14 GeV \cite{TASSO:1983cre} &
$\dd \sigma/\dd z$ &
~3/9   & ---             & 1.00 ($+0.28$)  & ---   & ---   & 0.860(131)*       
\\
~TASSO 22 GeV \cite{TASSO:1983cre} &
$\dd \sigma/\dd z$ &
~3/10  & ---             & 0.12 ($-1.61$)  & ---   & ---   & 1.180(227)*       
\\
~TASSO 34 GeV \cite{TASSO:1988jma} &
$\sigma^{-1} \, \dd \sigma/\dd z_p$ &
~3/11  & ---             & 0.34 ($-0.84$)  & ---   & 0.17  & 0.975(22)
\\
~TPC \cite{TPCTwoGamma:1988yjh} &
$\sigma^{-1} \, \dd \sigma/\dd z_p$ &
~7/21  & ---             & 2.15 ($+1.81$)  & ---   & ---   & ---
\\
~TOPAZ \cite{TOPAZ:1994voc} &
$\sigma^{-1} \, \dd \sigma/\dd \xi$ &
~2/12  & ---             & 0.12 ($-1.22$)  & ---   & ---   & ---
\\
~SLD \cite{SLD:2003ogn} &
$\sigma^{-1} \, \dd \sigma/\dd z_p$ &
~35/36 & ---             & 0.45 ($-2.90$)  & ---   & 0.30  & $0.995(6)\dagger$ 
\\
~SLD(uds) \cite{SLD:2003ogn} &
$\sigma^{-1} \, \dd \sigma/\dd z_p$ (uds) &
~35/36 & ---             & 1.80 ($+2.81$)  & ---   & ---   & $0.995(6)\dagger$
\\ 
~SLD(c) \cite{SLD:2003ogn} &
$\sigma^{-1} \, \dd \sigma/\dd z_p$ (c) &
~35/36 & ---             & 1.43 ($+1.68$)  & ---   & ---   & $0.995(6)\dagger$
\\ 
~SLD(b) \cite{SLD:2003ogn} &
$\sigma^{-1} \, \dd \sigma/\dd z_p$ (b) &
~35/36 & ---             & 1.15 ($+0.68$)  & ---   & ---   & $0.995(6)\dagger$
\\
~ALEPH \cite{ALEPH:1994cbg} &
$\sigma^{-1} \, \dd \sigma/\dd z_p$ &
~19/29 & ---             & 0.92 ($-0.13$)  & ---   & 0.51  & 1.021(12)
\\
~OPAL \cite{OPAL:1994zan} &
$\sigma^{-1} \, \dd \sigma/\dd p$ &
~10/33 & ---             & 0.35 ($-1.83$)  & ---   & ---   & ~~~~---
\\
~DELPHI \cite{DELPHI:1998cgx} &
$\sigma^{-1} \, \dd \sigma/\dd z_p$ &
~22/23 & ---             & 0.99 ($+0.06$)  & ---   & ---   & ~~~~---
\\
~DELPHI(uds) \cite{DELPHI:1998cgx} &
$\sigma^{-1} \, \dd \sigma/\dd z_p$ (uds) &
~22/23 & ---             & 1.15 ($+0.58$)  & ---   & ---   & ~~~~---
\\
~DELPHI(b) \cite{DELPHI:1998cgx} &
$\sigma^{-1} \, \dd \sigma/\dd z_p$ (b) &
~22/23 & ---             & 0.61 ($-1.41$)  & ---   & ---   & ~~~~---
\\
~BaBar (prompt) \cite{BaBar:2013yrg} &
$\sigma^{-1} \, \dd \sigma/\dd p$ &
~28/45 & 1\large$^\star$ & 0.12 ($-5.74$)  & 53.77 & 27.81 & 0.948(4)
\\
~ARGUS \cite{ARGUS:1989zdf} &
$\sigma^{-1} \, \dd \sigma/\dd p$ &
~25/42 & ---             & 0.84 ($-0.52$)  & ---   & 1.10  & 0.981(7)
\\
\hhline{========}
\end{tabular}}
\label{tb:stats sia-k}
\end{table*}

\begin{table*}
\centering
\caption{As in Table~\ref{tb:stats sia-p}, but for $e^+ e^-$ SIA data with unidentified $h^\pm$ hadron production.} 
{
\begin{tabular}{l c | c | c c c}
\hhline{======}
~dataset & 
observable & 
$N_{\rm dat}$ & 
~$\chi^{2}_{\rm red}$ ($Z$-score)~ & 
$\chi^2_{\rm norm}$ &
fitted norm.
\\ \hline
~TASSO 12 GeV \cite{BRANDELIK1981357} &
$\sigma^{-1} \, \dd \sigma/\dd z_p$ &
~5/7~   & 0.81 ($-0.11$) & 2.75 & 0.925(22)
\\
~TASSO 14 GeV \cite{TASSO:1982bkc} &
$\sigma^{-1} \, \dd \sigma/\dd z_p$ &
~11/20~ & 0.92 ($-0.05$) & 2.88 & 0.924(17)
\\
~TASSO 22 GeV \cite{TASSO:1982bkc} &
$\sigma^{-1} \, \dd \sigma/\dd z_p$ &
~11/20~ & 1.41 ($+0.99$) & 2.36 & 0.931(17)
\\
~TASSO 30 GeV \cite{TASSO:1981gag} &
$\sigma^{-1} \, \dd \sigma/\dd z_p$ &
~5/7~   & 1.86 ($+1.29$) & 0.05 & 1.010(18)
\\
~TASSO 44 GeV \cite{TASSO:1988jma} &
$\sigma^{-1} \, \dd \sigma/\dd z_p$ &
~11/20~ & 1.59 ($+1.31$) & 2.32 & 0.932(15)
\\
~TPC \cite{TPCTwoGamma:1988yjh} &
$\sigma^{-1} \, \dd \sigma/\dd z_p$ &
~12/34~ & 1.65 ($+1.47$) & ---  & ---
\\
~SLD \cite{SLD:2003ogn} &
$\sigma^{-1} \, \dd \sigma/\dd z_p$ &
~34/40~ & 0.68 ($-1.39$) & 1.24 & $1.011(2)\dagger$ 
\\
~SLD(uds) \cite{SLD:2003ogn} &
$\sigma^{-1} \, \dd \sigma/\dd z_p$ (uds) &
~34/40~ & 0.68 ($-1.44$) & ---  & $1.011(2)\dagger$ 
\\
~SLD(c) \cite{SLD:2003ogn} &
$\sigma^{-1} \, \dd \sigma/\dd z_p$ (c) &
~34/40~ & 0.86 ($-0.51$) & ---  & $1.011(2)\dagger$ 
\\
~SLD(b) \cite{SLD:2003ogn} &
$\sigma^{-1} \, \dd \sigma/\dd z_p$ (b) &
~34/40~ & 0.67 ($-1.46$) & ---  & $1.011(2)\dagger$
\\
~ALEPH \cite{ALEPH:1995njx}       &
$\sigma^{-1} \, \dd \sigma/\dd z_p$ &
~37/46~ & 0.44 ($-3.04$) & 2.01 & 1.014(3)\,~ 
\\
~OPAL \cite{OPAL:1998arz} &
$\sigma^{-1} \, \dd \sigma/\dd z$ &
~19/22~ & 0.73 ($-0.82$) & ---  & --- 
\\
~DELPHI \cite{DELPHI:1998cgx} &
$\sigma^{-1} \, \dd \sigma/\dd z_p$ &
~21/27~ & 0.28 ($-3.32$) & ---  & --- 
\\
~DELPHI(uds) \cite{DELPHI:1998cgx} &
$\sigma^{-1} \, \dd \sigma/\dd p$ &
~21/27~ & 0.34 ($-2.82$) & ---  & ---
\\
~DELPHI(b) \cite{DELPHI:1998cgx} &
~$\sigma^{-1} \, \dd \sigma/\dd z_p$ &
~21/27~ & 0.42 ($-2.37$) & ---  & ---               
\\
\hline
\hhline{======}
\end{tabular}}
\label{tb:stats sia-h}
\end{table*}

\clearpage
\bibliography{bibliography}

\begin{thebibliography}{88}%
\makeatletter
\providecommand \@ifxundefined [1]{%
 \@ifx{#1\undefined}
}%
\providecommand \@ifnum [1]{%
 \ifnum #1\expandafter \@firstoftwo
 \else \expandafter \@secondoftwo
 \fi
}%
\providecommand \@ifx [1]{%
 \ifx #1\expandafter \@firstoftwo
 \else \expandafter \@secondoftwo
 \fi
}%
\providecommand \natexlab [1]{#1}%
\providecommand \enquote  [1]{``#1''}%
\providecommand \bibnamefont  [1]{#1}%
\providecommand \bibfnamefont [1]{#1}%
\providecommand \citenamefont [1]{#1}%
\providecommand \href@noop [0]{\@secondoftwo}%
\providecommand \href [0]{\begingroup \@sanitize@url \@href}%
\providecommand \@href[1]{\@@startlink{#1}\@@href}%
\providecommand \@@href[1]{\endgroup#1\@@endlink}%
\providecommand \@sanitize@url [0]{\catcode `\\12\catcode `\$12\catcode
  `\&12\catcode `\#12\catcode `\^12\catcode `\_12\catcode `\%12\relax}%
\providecommand \@@startlink[1]{}%
\providecommand \@@endlink[0]{}%
\providecommand \url  [0]{\begingroup\@sanitize@url \@url }%
\providecommand \@url [1]{\endgroup\@href {#1}{\urlprefix }}%
\providecommand \urlprefix  [0]{URL }%
\providecommand \Eprint [0]{\href }%
\providecommand \doibase [0]{http://dx.doi.org/}%
\providecommand \selectlanguage [0]{\@gobble}%
\providecommand \bibinfo  [0]{\@secondoftwo}%
\providecommand \bibfield  [0]{\@secondoftwo}%
\providecommand \translation [1]{[#1]}%
\providecommand \BibitemOpen [0]{}%
\providecommand \bibitemStop [0]{}%
\providecommand \bibitemNoStop [0]{.\EOS\space}%
\providecommand \EOS [0]{\spacefactor3000\relax}%
\providecommand \BibitemShut  [1]{\csname bibitem#1\endcsname}%
\let\auto@bib@innerbib\@empty
\bibitem [{\citenamefont {Thomas}(1983)}]{Thomas:1983fh}%
  \BibitemOpen
  \bibfield  {author} {\bibinfo {author} {\bibfnamefont {A.~W.}\ \bibnamefont
  {Thomas}},\ }\href {\doibase 10.1016/0370-2693(83)90026-6} {\bibfield
  {journal} {\bibinfo  {journal} {Phys. Lett. B}\ }\textbf {\bibinfo {volume}
  {126}},\ \bibinfo {pages} {97} (\bibinfo {year} {1983})}\BibitemShut
  {NoStop}%
\bibitem [{\citenamefont {Towell}\ \emph {et~al.}(2001)\citenamefont {Towell}
  \emph {et~al.}}]{NuSea:2001idv}%
  \BibitemOpen
  \bibfield  {author} {\bibinfo {author} {\bibfnamefont {R.~S.}\ \bibnamefont
  {Towell}} \emph {et~al.},\ }\href {\doibase 10.1103/PhysRevD.64.052002}
  {\bibfield  {journal} {\bibinfo  {journal} {Phys. Rev. D}\ }\textbf {\bibinfo
  {volume} {64}},\ \bibinfo {pages} {052002} (\bibinfo {year}
  {2001})}\BibitemShut {NoStop}%
\bibitem [{\citenamefont {Dove}\ \emph {et~al.}(2021)\citenamefont {Dove} \emph
  {et~al.}}]{SeaQuest:2021zxb}%
  \BibitemOpen
  \bibfield  {author} {\bibinfo {author} {\bibfnamefont {J.}~\bibnamefont
  {Dove}} \emph {et~al.},\ }\href {\doibase 10.1038/s41586-022-04707-z}
  {\bibfield  {journal} {\bibinfo  {journal} {Nature}\ }\textbf {\bibinfo
  {volume} {590}},\ \bibinfo {pages} {561} (\bibinfo {year} {2021})},\ \bibinfo
  {note} {[Erratum: Nature {\bf 604}, E26 (2022)]}\BibitemShut {NoStop}%
\bibitem [{\citenamefont {Cocuzza}\ \emph
  {et~al.}(2021{\natexlab{a}})\citenamefont {Cocuzza}, \citenamefont
  {Melnitchouk}, \citenamefont {Metz},\ and\ \citenamefont
  {Sato}}]{Cocuzza:2021cbi}%
  \BibitemOpen
  \bibfield  {author} {\bibinfo {author} {\bibfnamefont {C.}~\bibnamefont
  {Cocuzza}}, \bibinfo {author} {\bibfnamefont {W.}~\bibnamefont
  {Melnitchouk}}, \bibinfo {author} {\bibfnamefont {A.}~\bibnamefont {Metz}}, \
  and\ \bibinfo {author} {\bibfnamefont {N.}~\bibnamefont {Sato}},\ }\href
  {\doibase 10.1103/PhysRevD.104.074031} {\bibfield  {journal} {\bibinfo
  {journal} {Phys. Rev. D}\ }\textbf {\bibinfo {volume} {104}},\ \bibinfo
  {pages} {074031} (\bibinfo {year} {2021}{\natexlab{a}})}\BibitemShut
  {NoStop}%
\bibitem [{\citenamefont {Signal}\ and\ \citenamefont
  {Thomas}(1987)}]{Signal:1987gz}%
  \BibitemOpen
  \bibfield  {author} {\bibinfo {author} {\bibfnamefont {A.~I.}\ \bibnamefont
  {Signal}}\ and\ \bibinfo {author} {\bibfnamefont {A.~W.}\ \bibnamefont
  {Thomas}},\ }\href {\doibase 10.1016/0370-2693(87)91348-7} {\bibfield
  {journal} {\bibinfo  {journal} {Phys. Lett. B}\ }\textbf {\bibinfo {volume}
  {191}},\ \bibinfo {pages} {205} (\bibinfo {year} {1987})}\BibitemShut
  {NoStop}%
\bibitem [{\citenamefont {Ji}\ and\ \citenamefont {Tang}(1995)}]{Ji:1995rd}%
  \BibitemOpen
  \bibfield  {author} {\bibinfo {author} {\bibfnamefont {X.}~\bibnamefont
  {Ji}}\ and\ \bibinfo {author} {\bibfnamefont {J.}~\bibnamefont {Tang}},\
  }\href {\doibase 10.1016/0370-2693(95)01153-H} {\bibfield  {journal}
  {\bibinfo  {journal} {Phys. Lett. B}\ }\textbf {\bibinfo {volume} {362}},\
  \bibinfo {pages} {182} (\bibinfo {year} {1995})}\BibitemShut {NoStop}%
\bibitem [{\citenamefont {Melnitchouk}\ and\ \citenamefont
  {Malheiro}(1997)}]{Melnitchouk:1996fj}%
  \BibitemOpen
  \bibfield  {author} {\bibinfo {author} {\bibfnamefont {W.}~\bibnamefont
  {Melnitchouk}}\ and\ \bibinfo {author} {\bibfnamefont {M.}~\bibnamefont
  {Malheiro}},\ }\href {\doibase 10.1103/PhysRevC.55.431} {\bibfield  {journal}
  {\bibinfo  {journal} {Phys. Rev. C}\ }\textbf {\bibinfo {volume} {55}},\
  \bibinfo {pages} {431} (\bibinfo {year} {1997})}\BibitemShut {NoStop}%
\bibitem [{\citenamefont {Bazarko}\ \emph {et~al.}(1995)\citenamefont {Bazarko}
  \emph {et~al.}}]{CCFR:1994ikl}%
  \BibitemOpen
  \bibfield  {author} {\bibinfo {author} {\bibfnamefont {A.~O.}\ \bibnamefont
  {Bazarko}} \emph {et~al.},\ }\href {\doibase 10.1007/BF01571875} {\bibfield
  {journal} {\bibinfo  {journal} {Z. Phys. C}\ }\textbf {\bibinfo {volume}
  {65}},\ \bibinfo {pages} {189} (\bibinfo {year} {1995})}\BibitemShut
  {NoStop}%
\bibitem [{\citenamefont {Mason}\ \emph {et~al.}(2007)\citenamefont {Mason}
  \emph {et~al.}}]{NuTeV:2007uwm}%
  \BibitemOpen
  \bibfield  {author} {\bibinfo {author} {\bibfnamefont {D.}~\bibnamefont
  {Mason}} \emph {et~al.},\ }\href {\doibase 10.1103/PhysRevLett.99.192001}
  {\bibfield  {journal} {\bibinfo  {journal} {Phys. Rev. Lett.}\ }\textbf
  {\bibinfo {volume} {99}},\ \bibinfo {pages} {192001} (\bibinfo {year}
  {2007})}\BibitemShut {NoStop}%
\bibitem [{\citenamefont {Kayis-Topaksu}\ \emph {et~al.}(2011)\citenamefont
  {Kayis-Topaksu} \emph {et~al.}}]{Kayis-Topaksu:2011ols}%
  \BibitemOpen
  \bibfield  {author} {\bibinfo {author} {\bibfnamefont {A.}~\bibnamefont
  {Kayis-Topaksu}} \emph {et~al.},\ }\href {\doibase
  10.1088/1367-2630/13/9/093002} {\bibfield  {journal} {\bibinfo  {journal}
  {New J. Phys.}\ }\textbf {\bibinfo {volume} {13}},\ \bibinfo {pages} {093002}
  (\bibinfo {year} {2011})}\BibitemShut {NoStop}%
\bibitem [{\citenamefont {Samoylov}\ \emph {et~al.}(2013)\citenamefont
  {Samoylov} \emph {et~al.}}]{NOMAD:2013hbk}%
  \BibitemOpen
  \bibfield  {author} {\bibinfo {author} {\bibfnamefont {O.}~\bibnamefont
  {Samoylov}} \emph {et~al.},\ }\href {\doibase
  10.1016/j.nuclphysb.2013.08.021} {\bibfield  {journal} {\bibinfo  {journal}
  {Nucl. Phys. B}\ }\textbf {\bibinfo {volume} {876}},\ \bibinfo {pages} {339}
  (\bibinfo {year} {2013})}\BibitemShut {NoStop}%
\bibitem [{\citenamefont {Kalantarians}\ \emph {et~al.}(2017)\citenamefont
  {Kalantarians}, \citenamefont {Keppel},\ and\ \citenamefont
  {Christy}}]{Kalantarians:2017mkj}%
  \BibitemOpen
  \bibfield  {author} {\bibinfo {author} {\bibfnamefont {N.}~\bibnamefont
  {Kalantarians}}, \bibinfo {author} {\bibfnamefont {C.~E.}\ \bibnamefont
  {Keppel}}, \ and\ \bibinfo {author} {\bibfnamefont {M.~E.}\ \bibnamefont
  {Christy}},\ }\href {\doibase 10.1103/PhysRevC.96.032201} {\bibfield
  {journal} {\bibinfo  {journal} {Phys. Rev. C}\ }\textbf {\bibinfo {volume}
  {96}},\ \bibinfo {pages} {032201} (\bibinfo {year} {2017})}\BibitemShut
  {NoStop}%
\bibitem [{\citenamefont {Accardi}\ \emph
  {et~al.}(2009{\natexlab{a}})\citenamefont {Accardi}, \citenamefont {Arleo},
  \citenamefont {Brooks}, \citenamefont {D'Enterria},\ and\ \citenamefont
  {Muccifora}}]{Accardi:2009qv}%
  \BibitemOpen
  \bibfield  {author} {\bibinfo {author} {\bibfnamefont {A.}~\bibnamefont
  {Accardi}}, \bibinfo {author} {\bibfnamefont {F.}~\bibnamefont {Arleo}},
  \bibinfo {author} {\bibfnamefont {W.~K.}\ \bibnamefont {Brooks}}, \bibinfo
  {author} {\bibfnamefont {D.}~\bibnamefont {D'Enterria}}, \ and\ \bibinfo
  {author} {\bibfnamefont {V.}~\bibnamefont {Muccifora}},\ }\href {\doibase
  10.1393/ncr/i2009-10048-0} {\bibfield  {journal} {\bibinfo  {journal} {Riv.
  Nuovo Cim.}\ }\textbf {\bibinfo {volume} {32}},\ \bibinfo {pages} {439}
  (\bibinfo {year} {2009}{\natexlab{a}})}\BibitemShut {NoStop}%
\bibitem [{\citenamefont {Airapetian}\ \emph {et~al.}(2008)\citenamefont
  {Airapetian} \emph {et~al.}}]{HERMES:2008pug}%
  \BibitemOpen
  \bibfield  {author} {\bibinfo {author} {\bibfnamefont {A.}~\bibnamefont
  {Airapetian}} \emph {et~al.},\ }\href {\doibase
  10.1016/j.physletb.2008.07.090} {\bibfield  {journal} {\bibinfo  {journal}
  {Phys. Lett. B}\ }\textbf {\bibinfo {volume} {666}},\ \bibinfo {pages} {446}
  (\bibinfo {year} {2008})}\BibitemShut {NoStop}%
\bibitem [{\citenamefont {Airapetian}\ \emph {et~al.}(2014)\citenamefont
  {Airapetian} \emph {et~al.}}]{HERMES:2013ztj}%
  \BibitemOpen
  \bibfield  {author} {\bibinfo {author} {\bibfnamefont {A.}~\bibnamefont
  {Airapetian}} \emph {et~al.},\ }\href {\doibase 10.1103/PhysRevD.89.097101}
  {\bibfield  {journal} {\bibinfo  {journal} {Phys. Rev. D}\ }\textbf {\bibinfo
  {volume} {89}},\ \bibinfo {pages} {097101} (\bibinfo {year}
  {2014})}\BibitemShut {NoStop}%
\bibitem [{\citenamefont {Stolarski}(2015)}]{Stolarski:2014jka}%
  \BibitemOpen
  \bibfield  {author} {\bibinfo {author} {\bibfnamefont {M.}~\bibnamefont
  {Stolarski}},\ }\href {\doibase 10.1103/PhysRevD.92.098101} {\bibfield
  {journal} {\bibinfo  {journal} {Phys. Rev. D}\ }\textbf {\bibinfo {volume}
  {92}},\ \bibinfo {pages} {098101} (\bibinfo {year} {2015})}\BibitemShut
  {NoStop}%
\bibitem [{\citenamefont {Leader}\ \emph {et~al.}(2014)\citenamefont {Leader},
  \citenamefont {Sidorov},\ and\ \citenamefont {Stamenov}}]{Leader:2014oxa}%
  \BibitemOpen
  \bibfield  {author} {\bibinfo {author} {\bibfnamefont {E.}~\bibnamefont
  {Leader}}, \bibinfo {author} {\bibfnamefont {A.~V.}\ \bibnamefont {Sidorov}},
  \ and\ \bibinfo {author} {\bibfnamefont {D.~B.}\ \bibnamefont {Stamenov}},\
  }\href {\doibase 10.1103/PhysRevD.90.054026} {\bibfield  {journal} {\bibinfo
  {journal} {Phys. Rev. D}\ }\textbf {\bibinfo {volume} {90}},\ \bibinfo
  {pages} {054026} (\bibinfo {year} {2014})}\BibitemShut {NoStop}%
\bibitem [{\citenamefont {Leader}\ \emph {et~al.}(2016)\citenamefont {Leader},
  \citenamefont {Sidorov},\ and\ \citenamefont {Stamenov}}]{Leader:2015hna}%
  \BibitemOpen
  \bibfield  {author} {\bibinfo {author} {\bibfnamefont {E.}~\bibnamefont
  {Leader}}, \bibinfo {author} {\bibfnamefont {A.~V.}\ \bibnamefont {Sidorov}},
  \ and\ \bibinfo {author} {\bibfnamefont {D.~B.}\ \bibnamefont {Stamenov}},\
  }\href {\doibase 10.1103/PhysRevD.93.074026} {\bibfield  {journal} {\bibinfo
  {journal} {Phys. Rev. D}\ }\textbf {\bibinfo {volume} {93}},\ \bibinfo
  {pages} {074026} (\bibinfo {year} {2016})}\BibitemShut {NoStop}%
\bibitem [{\citenamefont {Chatrchyan}\ \emph {et~al.}(2011)\citenamefont
  {Chatrchyan} \emph {et~al.}}]{CMS:2011bet}%
  \BibitemOpen
  \bibfield  {author} {\bibinfo {author} {\bibfnamefont {S.}~\bibnamefont
  {Chatrchyan}} \emph {et~al.},\ }\href {\doibase 10.1007/JHEP04(2011)050}
  {\bibfield  {journal} {\bibinfo  {journal} {JHEP}\ }\textbf {\bibinfo
  {volume} {04}},\ \bibinfo {pages} {050} (\bibinfo {year} {2011})}\BibitemShut
  {NoStop}%
\bibitem [{\citenamefont {Chatrchyan}\ \emph {et~al.}(2012)\citenamefont
  {Chatrchyan} \emph {et~al.}}]{CMS:2012ivw}%
  \BibitemOpen
  \bibfield  {author} {\bibinfo {author} {\bibfnamefont {S.}~\bibnamefont
  {Chatrchyan}} \emph {et~al.},\ }\href {\doibase
  10.1103/PhysRevLett.109.111806} {\bibfield  {journal} {\bibinfo  {journal}
  {Phys. Rev. Lett.}\ }\textbf {\bibinfo {volume} {109}},\ \bibinfo {pages}
  {111806} (\bibinfo {year} {2012})}\BibitemShut {NoStop}%
\bibitem [{\citenamefont {Chatrchyan}\ \emph
  {et~al.}(2014{\natexlab{a}})\citenamefont {Chatrchyan} \emph
  {et~al.}}]{CMS:2013pzl}%
  \BibitemOpen
  \bibfield  {author} {\bibinfo {author} {\bibfnamefont {S.}~\bibnamefont
  {Chatrchyan}} \emph {et~al.},\ }\href {\doibase 10.1103/PhysRevD.90.032004}
  {\bibfield  {journal} {\bibinfo  {journal} {Phys. Rev. D}\ }\textbf {\bibinfo
  {volume} {90}},\ \bibinfo {pages} {032004} (\bibinfo {year}
  {2014}{\natexlab{a}})}\BibitemShut {NoStop}%
\bibitem [{\citenamefont {Aaboud}\ \emph {et~al.}(2017)\citenamefont {Aaboud}
  \emph {et~al.}}]{ATLAS:2016nqi}%
  \BibitemOpen
  \bibfield  {author} {\bibinfo {author} {\bibfnamefont {M.}~\bibnamefont
  {Aaboud}} \emph {et~al.},\ }\href {\doibase 10.1140/epjc/s10052-017-4911-9}
  {\bibfield  {journal} {\bibinfo  {journal} {Eur. Phys. J. C}\ }\textbf
  {\bibinfo {volume} {77}},\ \bibinfo {pages} {367} (\bibinfo {year}
  {2017})}\BibitemShut {NoStop}%
\bibitem [{\citenamefont {Khachatryan}\ \emph {et~al.}(2016)\citenamefont
  {Khachatryan} \emph {et~al.}}]{CMS:2016qqr}%
  \BibitemOpen
  \bibfield  {author} {\bibinfo {author} {\bibfnamefont {V.}~\bibnamefont
  {Khachatryan}} \emph {et~al.},\ }\href {\doibase
  10.1140/epjc/s10052-016-4293-4} {\bibfield  {journal} {\bibinfo  {journal}
  {Eur. Phys. J. C}\ }\textbf {\bibinfo {volume} {76}},\ \bibinfo {pages} {469}
  (\bibinfo {year} {2016})}\BibitemShut {NoStop}%
\bibitem [{\citenamefont {Aad}\ \emph {et~al.}(2012)\citenamefont {Aad} \emph
  {et~al.}}]{ATLAS:2012sjl}%
  \BibitemOpen
  \bibfield  {author} {\bibinfo {author} {\bibfnamefont {G.}~\bibnamefont
  {Aad}} \emph {et~al.},\ }\href {\doibase 10.1103/PhysRevLett.109.012001}
  {\bibfield  {journal} {\bibinfo  {journal} {Phys. Rev. Lett.}\ }\textbf
  {\bibinfo {volume} {109}},\ \bibinfo {pages} {012001} (\bibinfo {year}
  {2012})}\BibitemShut {NoStop}%
\bibitem [{\citenamefont {Chatrchyan}\ \emph
  {et~al.}(2014{\natexlab{b}})\citenamefont {Chatrchyan} \emph
  {et~al.}}]{CMS:2013wql}%
  \BibitemOpen
  \bibfield  {author} {\bibinfo {author} {\bibfnamefont {S.}~\bibnamefont
  {Chatrchyan}} \emph {et~al.},\ }\href {\doibase 10.1007/JHEP02(2014)013}
  {\bibfield  {journal} {\bibinfo  {journal} {JHEP}\ }\textbf {\bibinfo
  {volume} {02}},\ \bibinfo {pages} {013} (\bibinfo {year}
  {2014}{\natexlab{b}})}\BibitemShut {NoStop}%
\bibitem [{\citenamefont {Aad}\ \emph {et~al.}(2014)\citenamefont {Aad} \emph
  {et~al.}}]{ATLAS:2014jkm}%
  \BibitemOpen
  \bibfield  {author} {\bibinfo {author} {\bibfnamefont {G.}~\bibnamefont
  {Aad}} \emph {et~al.},\ }\href {\doibase 10.1007/JHEP05(2014)068} {\bibfield
  {journal} {\bibinfo  {journal} {JHEP}\ }\textbf {\bibinfo {volume} {05}},\
  \bibinfo {pages} {068} (\bibinfo {year} {2014})}\BibitemShut {NoStop}%
\bibitem [{\citenamefont {Sirunyan}\ \emph {et~al.}(2019)\citenamefont
  {Sirunyan} \emph {et~al.}}]{CMS:2018dxg}%
  \BibitemOpen
  \bibfield  {author} {\bibinfo {author} {\bibfnamefont {A.~M.}\ \bibnamefont
  {Sirunyan}} \emph {et~al.},\ }\href {\doibase 10.1140/epjc/s10052-019-6752-1}
  {\bibfield  {journal} {\bibinfo  {journal} {Eur. Phys. J. C}\ }\textbf
  {\bibinfo {volume} {79}},\ \bibinfo {pages} {269} (\bibinfo {year}
  {2019})}\BibitemShut {NoStop}%
\bibitem [{\citenamefont {Aaboud}\ \emph {et~al.}(2018)\citenamefont {Aaboud}
  \emph {et~al.}}]{ATLAS:2017rzl}%
  \BibitemOpen
  \bibfield  {author} {\bibinfo {author} {\bibfnamefont {M.}~\bibnamefont
  {Aaboud}} \emph {et~al.},\ }\href {\doibase 10.1140/epjc/s10052-017-5475-4}
  {\bibfield  {journal} {\bibinfo  {journal} {Eur. Phys. J. C}\ }\textbf
  {\bibinfo {volume} {78}},\ \bibinfo {pages} {110} (\bibinfo {year} {2018})},\
  \bibinfo {note} {[Erratum: Eur. Phys. J. C {\bf 78}, 898 (2018)]}\BibitemShut
  {NoStop}%
\bibitem [{\citenamefont {Alekhin}\ \emph {et~al.}(2018)\citenamefont
  {Alekhin}, \citenamefont {Bl\"umlein},\ and\ \citenamefont
  {Moch}}]{Alekhin:2017olj}%
  \BibitemOpen
  \bibfield  {author} {\bibinfo {author} {\bibfnamefont {S.}~\bibnamefont
  {Alekhin}}, \bibinfo {author} {\bibfnamefont {J.}~\bibnamefont {Bl\"umlein}},
  \ and\ \bibinfo {author} {\bibfnamefont {S.}~\bibnamefont {Moch}},\ }\href
  {\doibase 10.1016/j.physletb.2017.12.024} {\bibfield  {journal} {\bibinfo
  {journal} {Phys. Lett. B}\ }\textbf {\bibinfo {volume} {777}},\ \bibinfo
  {pages} {134} (\bibinfo {year} {2018})}\BibitemShut {NoStop}%
\bibitem [{\citenamefont {Borsa}\ \emph {et~al.}(2017)\citenamefont {Borsa},
  \citenamefont {Sassot},\ and\ \citenamefont {Stratmann}}]{Borsa:2017vwy}%
  \BibitemOpen
  \bibfield  {author} {\bibinfo {author} {\bibfnamefont {I.}~\bibnamefont
  {Borsa}}, \bibinfo {author} {\bibfnamefont {R.}~\bibnamefont {Sassot}}, \
  and\ \bibinfo {author} {\bibfnamefont {M.}~\bibnamefont {Stratmann}},\ }\href
  {\doibase 10.1103/PhysRevD.96.094020} {\bibfield  {journal} {\bibinfo
  {journal} {Phys. Rev. D}\ }\textbf {\bibinfo {volume} {96}},\ \bibinfo
  {pages} {094020} (\bibinfo {year} {2017})}\BibitemShut {NoStop}%
\bibitem [{\citenamefont {De~Florian}\ \emph {et~al.}(2019)\citenamefont
  {De~Florian}, \citenamefont {Lucero}, \citenamefont {Sassot}, \citenamefont
  {Stratmann},\ and\ \citenamefont {Vogelsang}}]{DeFlorian:2019xxt}%
  \BibitemOpen
  \bibfield  {author} {\bibinfo {author} {\bibfnamefont {D.}~\bibnamefont
  {De~Florian}}, \bibinfo {author} {\bibfnamefont {G.~A.}\ \bibnamefont
  {Lucero}}, \bibinfo {author} {\bibfnamefont {R.}~\bibnamefont {Sassot}},
  \bibinfo {author} {\bibfnamefont {M.}~\bibnamefont {Stratmann}}, \ and\
  \bibinfo {author} {\bibfnamefont {W.}~\bibnamefont {Vogelsang}},\ }\href
  {\doibase 10.1103/PhysRevD.100.114027} {\bibfield  {journal} {\bibinfo
  {journal} {Phys. Rev. D}\ }\textbf {\bibinfo {volume} {100}},\ \bibinfo
  {pages} {114027} (\bibinfo {year} {2019})}\BibitemShut {NoStop}%
\bibitem [{\citenamefont {Campbell}\ and\ \citenamefont {Ellis}(1999)}]{MCFM}%
  \BibitemOpen
  \bibfield  {author} {\bibinfo {author} {\bibfnamefont {J.~M.}\ \bibnamefont
  {Campbell}}\ and\ \bibinfo {author} {\bibfnamefont {R.~K.}\ \bibnamefont
  {Ellis}},\ }\href {\doibase 10.1103/PhysRevD.60.113006} {\bibfield  {journal}
  {\bibinfo  {journal} {Phys. Rev. D}\ }\textbf {\bibinfo {volume} {60}},\
  \bibinfo {pages} {113006} (\bibinfo {year} {1999})}\BibitemShut {NoStop}%
\bibitem [{\citenamefont {Dokshitzer}(1977)}]{Dokshitzer:1977sg}%
  \BibitemOpen
  \bibfield  {author} {\bibinfo {author} {\bibfnamefont {Y.~L.}\ \bibnamefont
  {Dokshitzer}},\ }\href@noop {} {\bibfield  {journal} {\bibinfo  {journal}
  {Sov. Phys. JETP}\ }\textbf {\bibinfo {volume} {46}},\ \bibinfo {pages} {641}
  (\bibinfo {year} {1977})}\BibitemShut {NoStop}%
\bibitem [{\citenamefont {Gribov}\ and\ \citenamefont
  {Lipatov}(1972)}]{Gribov:1972ri}%
  \BibitemOpen
  \bibfield  {author} {\bibinfo {author} {\bibfnamefont {V.~N.}\ \bibnamefont
  {Gribov}}\ and\ \bibinfo {author} {\bibfnamefont {L.~N.}\ \bibnamefont
  {Lipatov}},\ }\href@noop {} {\bibfield  {journal} {\bibinfo  {journal} {Sov.
  J. Nucl. Phys.}\ }\textbf {\bibinfo {volume} {15}},\ \bibinfo {pages} {438}
  (\bibinfo {year} {1972})}\BibitemShut {NoStop}%
\bibitem [{\citenamefont {Altarelli}\ and\ \citenamefont
  {Parisi}(1977)}]{Altarelli:1977zs}%
  \BibitemOpen
  \bibfield  {author} {\bibinfo {author} {\bibfnamefont {G.}~\bibnamefont
  {Altarelli}}\ and\ \bibinfo {author} {\bibfnamefont {G.}~\bibnamefont
  {Parisi}},\ }\href {\doibase 10.1016/0550-3213(77)90384-4} {\bibfield
  {journal} {\bibinfo  {journal} {Nucl. Phys.}\ }\textbf {\bibinfo {volume}
  {B126}},\ \bibinfo {pages} {298} (\bibinfo {year} {1977})}\BibitemShut
  {NoStop}%
\bibitem [{\citenamefont {Navas}\ \emph {et~al.}(2024)\citenamefont {Navas}
  \emph {et~al.}}]{ParticleDataGroup:2024cfk}%
  \BibitemOpen
  \bibfield  {author} {\bibinfo {author} {\bibfnamefont {S.}~\bibnamefont
  {Navas}} \emph {et~al.} (\bibinfo {collaboration} {Particle Data Group}),\
  }\href {\doibase 10.1103/PhysRevD.110.030001} {\bibfield  {journal} {\bibinfo
   {journal} {Phys. Rev. D}\ }\textbf {\bibinfo {volume} {110}},\ \bibinfo
  {pages} {030001} (\bibinfo {year} {2024})}\BibitemShut {NoStop}%
\bibitem [{\citenamefont {Adamiak}\ \emph {et~al.}(2025)\citenamefont
  {Adamiak}, \citenamefont {Anderson}, \citenamefont {Cocuzza}, \citenamefont
  {Melnitchouk}, \citenamefont {Sato},\ and\ \citenamefont {Zhou}}]{JAMFF}%
  \BibitemOpen
  \bibfield  {author} {\bibinfo {author} {\bibfnamefont {D.}~\bibnamefont
  {Adamiak}}, \bibinfo {author} {\bibfnamefont {T.}~\bibnamefont {Anderson}},
  \bibinfo {author} {\bibfnamefont {C.}~\bibnamefont {Cocuzza}}, \bibinfo
  {author} {\bibfnamefont {W.}~\bibnamefont {Melnitchouk}}, \bibinfo {author}
  {\bibfnamefont {N.}~\bibnamefont {Sato}}, \ and\ \bibinfo {author}
  {\bibfnamefont {Y.}~\bibnamefont {Zhou}},\ }\href@noop {} {}\bibinfo
  {howpublished} {in preparation} (\bibinfo {year} {2025})\BibitemShut
  {NoStop}%
\bibitem [{\citenamefont {Moffat}\ \emph {et~al.}(2021)\citenamefont {Moffat},
  \citenamefont {Melnitchouk}, \citenamefont {Rogers},\ and\ \citenamefont
  {Sato}}]{Moffat:2021dji}%
  \BibitemOpen
  \bibfield  {author} {\bibinfo {author} {\bibfnamefont {E.}~\bibnamefont
  {Moffat}}, \bibinfo {author} {\bibfnamefont {W.}~\bibnamefont {Melnitchouk}},
  \bibinfo {author} {\bibfnamefont {T.~C.}\ \bibnamefont {Rogers}}, \ and\
  \bibinfo {author} {\bibfnamefont {N.}~\bibnamefont {Sato}},\ }\href {\doibase
  10.1103/PhysRevD.104.016015} {\bibfield  {journal} {\bibinfo  {journal}
  {Phys. Rev. D}\ }\textbf {\bibinfo {volume} {104}},\ \bibinfo {pages}
  {016015} (\bibinfo {year} {2021})}\BibitemShut {NoStop}%
\bibitem [{\citenamefont {Whitlow}\ \emph {et~al.}(1992)\citenamefont
  {Whitlow}, \citenamefont {Riordan}, \citenamefont {Dasu}, \citenamefont
  {Rock},\ and\ \citenamefont {Bodek}}]{Whitlow:1991uw}%
  \BibitemOpen
  \bibfield  {author} {\bibinfo {author} {\bibfnamefont {L.~W.}\ \bibnamefont
  {Whitlow}}, \bibinfo {author} {\bibfnamefont {E.~M.}\ \bibnamefont
  {Riordan}}, \bibinfo {author} {\bibfnamefont {S.}~\bibnamefont {Dasu}},
  \bibinfo {author} {\bibfnamefont {S.}~\bibnamefont {Rock}}, \ and\ \bibinfo
  {author} {\bibfnamefont {A.}~\bibnamefont {Bodek}},\ }\href {\doibase
  10.1016/0370-2693(92)90672-Q} {\bibfield  {journal} {\bibinfo  {journal}
  {Phys. Lett. B}\ }\textbf {\bibinfo {volume} {282}},\ \bibinfo {pages} {475}
  (\bibinfo {year} {1992})}\BibitemShut {NoStop}%
\bibitem [{\citenamefont {Benvenuti}\ \emph {et~al.}(1990)\citenamefont
  {Benvenuti} \emph {et~al.}}]{BCDMS:1989ggw}%
  \BibitemOpen
  \bibfield  {author} {\bibinfo {author} {\bibfnamefont {A.~C.}\ \bibnamefont
  {Benvenuti}} \emph {et~al.},\ }\href {\doibase 10.1016/0370-2693(90)91231-Y}
  {\bibfield  {journal} {\bibinfo  {journal} {Phys. Lett. B}\ }\textbf
  {\bibinfo {volume} {237}},\ \bibinfo {pages} {592} (\bibinfo {year}
  {1990})}\BibitemShut {NoStop}%
\bibitem [{\citenamefont {Arneodo}\ \emph
  {et~al.}(1997{\natexlab{a}})\citenamefont {Arneodo} \emph
  {et~al.}}]{NewMuon:1996fwh}%
  \BibitemOpen
  \bibfield  {author} {\bibinfo {author} {\bibfnamefont {M.}~\bibnamefont
  {Arneodo}} \emph {et~al.},\ }\href {\doibase 10.1016/S0550-3213(96)00538-X}
  {\bibfield  {journal} {\bibinfo  {journal} {Nucl. Phys.}\ }\textbf {\bibinfo
  {volume} {B483}},\ \bibinfo {pages} {3} (\bibinfo {year}
  {1997}{\natexlab{a}})}\BibitemShut {NoStop}%
\bibitem [{\citenamefont {Arneodo}\ \emph
  {et~al.}(1997{\natexlab{b}})\citenamefont {Arneodo} \emph
  {et~al.}}]{NewMuon:1996uwk}%
  \BibitemOpen
  \bibfield  {author} {\bibinfo {author} {\bibfnamefont {M.}~\bibnamefont
  {Arneodo}} \emph {et~al.},\ }\href {\doibase 10.1016/S0550-3213(96)00673-6}
  {\bibfield  {journal} {\bibinfo  {journal} {Nucl. Phys.}\ }\textbf {\bibinfo
  {volume} {B487}},\ \bibinfo {pages} {3} (\bibinfo {year}
  {1997}{\natexlab{b}})}\BibitemShut {NoStop}%
\bibitem [{\citenamefont {Abramowicz}\ \emph {et~al.}(2015)\citenamefont
  {Abramowicz} \emph {et~al.}}]{H1:2015ubc}%
  \BibitemOpen
  \bibfield  {author} {\bibinfo {author} {\bibfnamefont {H.}~\bibnamefont
  {Abramowicz}} \emph {et~al.},\ }\href {\doibase
  10.1140/epjc/s10052-015-3710-4} {\bibfield  {journal} {\bibinfo  {journal}
  {Eur. Phys. J. C}\ }\textbf {\bibinfo {volume} {75}},\ \bibinfo {pages} {580}
  (\bibinfo {year} {2015})}\BibitemShut {NoStop}%
\bibitem [{\citenamefont {Adolph}\ \emph
  {et~al.}(2017{\natexlab{a}})\citenamefont {Adolph} \emph
  {et~al.}}]{COMPASS:2016xvm}%
  \BibitemOpen
  \bibfield  {author} {\bibinfo {author} {\bibfnamefont {C.}~\bibnamefont
  {Adolph}} \emph {et~al.},\ }\href {\doibase 10.1016/j.physletb.2016.09.042}
  {\bibfield  {journal} {\bibinfo  {journal} {Phys. Lett. B}\ }\textbf
  {\bibinfo {volume} {764}},\ \bibinfo {pages} {1} (\bibinfo {year}
  {2017}{\natexlab{a}})}\BibitemShut {NoStop}%
\bibitem [{\citenamefont {Adolph}\ \emph
  {et~al.}(2017{\natexlab{b}})\citenamefont {Adolph} \emph
  {et~al.}}]{COMPASS:2016crr}%
  \BibitemOpen
  \bibfield  {author} {\bibinfo {author} {\bibfnamefont {C.}~\bibnamefont
  {Adolph}} \emph {et~al.},\ }\href {\doibase 10.1016/j.physletb.2017.01.053}
  {\bibfield  {journal} {\bibinfo  {journal} {Phys. Lett. B}\ }\textbf
  {\bibinfo {volume} {767}},\ \bibinfo {pages} {133} (\bibinfo {year}
  {2017}{\natexlab{b}})}\BibitemShut {NoStop}%
\bibitem [{\citenamefont {Aaij}\ \emph {et~al.}(2014)\citenamefont {Aaij} \emph
  {et~al.}}]{LHCb:2014liz}%
  \BibitemOpen
  \bibfield  {author} {\bibinfo {author} {\bibfnamefont {R.}~\bibnamefont
  {Aaij}} \emph {et~al.},\ }\href {\doibase 10.1007/JHEP12(2014)079} {\bibfield
   {journal} {\bibinfo  {journal} {JHEP}\ }\textbf {\bibinfo {volume} {12}},\
  \bibinfo {pages} {079} (\bibinfo {year} {2014})}\BibitemShut {NoStop}%
\bibitem [{\citenamefont {Aaij}\ \emph {et~al.}(2016)\citenamefont {Aaij} \emph
  {et~al.}}]{LHCb:2015mad}%
  \BibitemOpen
  \bibfield  {author} {\bibinfo {author} {\bibfnamefont {R.}~\bibnamefont
  {Aaij}} \emph {et~al.},\ }\href {\doibase 10.1007/JHEP01(2016)155} {\bibfield
   {journal} {\bibinfo  {journal} {JHEP}\ }\textbf {\bibinfo {volume} {01}},\
  \bibinfo {pages} {155} (\bibinfo {year} {2016})}\BibitemShut {NoStop}%
\bibitem [{\citenamefont {Abazov}\ \emph {et~al.}(2014)\citenamefont {Abazov}
  \emph {et~al.}}]{D0:2013lql}%
  \BibitemOpen
  \bibfield  {author} {\bibinfo {author} {\bibfnamefont {V.~M.}\ \bibnamefont
  {Abazov}} \emph {et~al.},\ }\href {\doibase 10.1103/PhysRevLett.112.151803}
  {\bibfield  {journal} {\bibinfo  {journal} {Phys. Rev. Lett.}\ }\textbf
  {\bibinfo {volume} {112}},\ \bibinfo {pages} {151803} (\bibinfo {year}
  {2014})},\ \bibinfo {note} {[Erratum: Phys. Rev. Lett. {\bf 114}, 049901
  (2015)]}\BibitemShut {NoStop}%
\bibitem [{\citenamefont {Aaltonen}\ \emph {et~al.}(2009)\citenamefont
  {Aaltonen} \emph {et~al.}}]{CDF:2009cjw}%
  \BibitemOpen
  \bibfield  {author} {\bibinfo {author} {\bibfnamefont {T.}~\bibnamefont
  {Aaltonen}} \emph {et~al.},\ }\href {\doibase 10.1103/PhysRevLett.102.181801}
  {\bibfield  {journal} {\bibinfo  {journal} {Phys. Rev. Lett.}\ }\textbf
  {\bibinfo {volume} {102}},\ \bibinfo {pages} {181801} (\bibinfo {year}
  {2009})}\BibitemShut {NoStop}%
\bibitem [{\citenamefont {Adam}\ \emph {et~al.}(2021)\citenamefont {Adam} \emph
  {et~al.}}]{STAR:2020vuq}%
  \BibitemOpen
  \bibfield  {author} {\bibinfo {author} {\bibfnamefont {J.}~\bibnamefont
  {Adam}} \emph {et~al.},\ }\href {\doibase 10.1103/PhysRevD.103.012001}
  {\bibfield  {journal} {\bibinfo  {journal} {Phys. Rev. D}\ }\textbf {\bibinfo
  {volume} {103}},\ \bibinfo {pages} {012001} (\bibinfo {year}
  {2021})}\BibitemShut {NoStop}%
\bibitem [{\citenamefont {Abazov}\ \emph {et~al.}(2007)\citenamefont {Abazov}
  \emph {et~al.}}]{D0:2007djv}%
  \BibitemOpen
  \bibfield  {author} {\bibinfo {author} {\bibfnamefont {V.~M.}\ \bibnamefont
  {Abazov}} \emph {et~al.},\ }\href {\doibase 10.1103/PhysRevD.76.012003}
  {\bibfield  {journal} {\bibinfo  {journal} {Phys. Rev. D}\ }\textbf {\bibinfo
  {volume} {76}},\ \bibinfo {pages} {012003} (\bibinfo {year}
  {2007})}\BibitemShut {NoStop}%
\bibitem [{\citenamefont {Aaltonen}\ \emph {et~al.}(2010)\citenamefont
  {Aaltonen} \emph {et~al.}}]{CDF:2010vek}%
  \BibitemOpen
  \bibfield  {author} {\bibinfo {author} {\bibfnamefont {T.~A.}\ \bibnamefont
  {Aaltonen}} \emph {et~al.},\ }\href {\doibase 10.1016/j.physletb.2010.06.043}
  {\bibfield  {journal} {\bibinfo  {journal} {Phys. Lett. B}\ }\textbf
  {\bibinfo {volume} {692}},\ \bibinfo {pages} {232} (\bibinfo {year}
  {2010})}\BibitemShut {NoStop}%
\bibitem [{\citenamefont {Abazov}\ \emph {et~al.}(2008)\citenamefont {Abazov}
  \emph {et~al.}}]{PhysRevLett.101.062001}%
  \BibitemOpen
  \bibfield  {author} {\bibinfo {author} {\bibfnamefont {V.~M.}\ \bibnamefont
  {Abazov}} \emph {et~al.},\ }\href {\doibase 10.1103/PhysRevLett.101.062001}
  {\bibfield  {journal} {\bibinfo  {journal} {Phys. Rev. Lett.}\ }\textbf
  {\bibinfo {volume} {101}},\ \bibinfo {pages} {062001} (\bibinfo {year}
  {2008})}\BibitemShut {NoStop}%
\bibitem [{\citenamefont {Abulencia}\ \emph {et~al.}(2007)\citenamefont
  {Abulencia} \emph {et~al.}}]{PhysRevD.75.092006}%
  \BibitemOpen
  \bibfield  {author} {\bibinfo {author} {\bibfnamefont {A.}~\bibnamefont
  {Abulencia}} \emph {et~al.},\ }\href {\doibase 10.1103/PhysRevD.75.092006}
  {\bibfield  {journal} {\bibinfo  {journal} {Phys. Rev. D}\ }\textbf {\bibinfo
  {volume} {75}},\ \bibinfo {pages} {092006} (\bibinfo {year}
  {2007})}\BibitemShut {NoStop}%
\bibitem [{\citenamefont {Abelev}\ \emph {et~al.}(2006)\citenamefont {Abelev}
  \emph {et~al.}}]{PhysRevLett.97.252001}%
  \BibitemOpen
  \bibfield  {author} {\bibinfo {author} {\bibfnamefont {B.~I.}\ \bibnamefont
  {Abelev}} \emph {et~al.},\ }\href {\doibase 10.1103/PhysRevLett.97.252001}
  {\bibfield  {journal} {\bibinfo  {journal} {Phys. Rev. Lett.}\ }\textbf
  {\bibinfo {volume} {97}},\ \bibinfo {pages} {252001} (\bibinfo {year}
  {2006})}\BibitemShut {NoStop}%
\bibitem [{\citenamefont {Albrecht}\ \emph {et~al.}(1989)\citenamefont
  {Albrecht} \emph {et~al.}}]{ARGUS:1989zdf}%
  \BibitemOpen
  \bibfield  {author} {\bibinfo {author} {\bibfnamefont {H.}~\bibnamefont
  {Albrecht}} \emph {et~al.},\ }\href {\doibase 10.1007/BF01549077} {\bibfield
  {journal} {\bibinfo  {journal} {Z. Phys. C}\ }\textbf {\bibinfo {volume}
  {44}},\ \bibinfo {pages} {547} (\bibinfo {year} {1989})}\BibitemShut
  {NoStop}%
\bibitem [{\citenamefont {Lees}\ \emph {et~al.}(2013)\citenamefont {Lees} \emph
  {et~al.}}]{BaBar:2013yrg}%
  \BibitemOpen
  \bibfield  {author} {\bibinfo {author} {\bibfnamefont {J.~P.}\ \bibnamefont
  {Lees}} \emph {et~al.},\ }\href {\doibase 10.1103/PhysRevD.88.032011}
  {\bibfield  {journal} {\bibinfo  {journal} {Phys. Rev. D}\ }\textbf {\bibinfo
  {volume} {88}},\ \bibinfo {pages} {032011} (\bibinfo {year}
  {2013})}\BibitemShut {NoStop}%
\bibitem [{\citenamefont {Leitgab}\ \emph {et~al.}(2013)\citenamefont {Leitgab}
  \emph {et~al.}}]{Belle:2013lfg}%
  \BibitemOpen
  \bibfield  {author} {\bibinfo {author} {\bibfnamefont {M.}~\bibnamefont
  {Leitgab}} \emph {et~al.},\ }\href {\doibase 10.1103/PhysRevLett.111.062002}
  {\bibfield  {journal} {\bibinfo  {journal} {Phys. Rev. Lett.}\ }\textbf
  {\bibinfo {volume} {111}},\ \bibinfo {pages} {062002} (\bibinfo {year}
  {2013})}\BibitemShut {NoStop}%
\bibitem [{\citenamefont {Brandelik}\ \emph
  {et~al.}(1981{\natexlab{a}})\citenamefont {Brandelik} \emph
  {et~al.}}]{BRANDELIK1981357}%
  \BibitemOpen
  \bibfield  {author} {\bibinfo {author} {\bibfnamefont {R.}~\bibnamefont
  {Brandelik}} \emph {et~al.},\ }\href {\doibase
  https://doi.org/10.1016/0370-2693(81)90104-0} {\bibfield  {journal} {\bibinfo
   {journal} {Phys. Lett. B}\ }\textbf {\bibinfo {volume} {100}},\ \bibinfo
  {pages} {357363} (\bibinfo {year} {1981}{\natexlab{a}})}\BibitemShut
  {NoStop}%
\bibitem [{\citenamefont {Althoff}\ \emph {et~al.}(1984)\citenamefont {Althoff}
  \emph {et~al.}}]{TASSO:1983cre}%
  \BibitemOpen
  \bibfield  {author} {\bibinfo {author} {\bibfnamefont {M.}~\bibnamefont
  {Althoff}} \emph {et~al.},\ }\href {\doibase 10.1007/BF01547419} {\bibfield
  {journal} {\bibinfo  {journal} {Z. Phys. C}\ }\textbf {\bibinfo {volume}
  {22}},\ \bibinfo {pages} {307} (\bibinfo {year} {1984})}\BibitemShut
  {NoStop}%
\bibitem [{\citenamefont {Braunschweig}\ \emph {et~al.}(1989)\citenamefont
  {Braunschweig} \emph {et~al.}}]{TASSO:1988jma}%
  \BibitemOpen
  \bibfield  {author} {\bibinfo {author} {\bibfnamefont {W.}~\bibnamefont
  {Braunschweig}} \emph {et~al.},\ }\href {\doibase 10.1007/BF01555856}
  {\bibfield  {journal} {\bibinfo  {journal} {Z. Phys. C}\ }\textbf {\bibinfo
  {volume} {42}},\ \bibinfo {pages} {189} (\bibinfo {year} {1989})}\BibitemShut
  {NoStop}%
\bibitem [{\citenamefont {Lu}(1986)}]{Lu:1986mc}%
  \BibitemOpen
  \bibfield  {author} {\bibinfo {author} {\bibfnamefont {X.-Q.}\ \bibnamefont
  {Lu}},\ }\emph {\bibinfo {title} {{Heavy quark jets from $e^+ e^-$
  annihilation at 29 GeV}}},\ \href@noop {} {\bibinfo {type} {thesis}},\
  \bibinfo  {school} {Johns Hopkins University} (\bibinfo {year}
  {1986})\BibitemShut {NoStop}%
\bibitem [{\citenamefont {Aihara}\ \emph {et~al.}(1988)\citenamefont {Aihara}
  \emph {et~al.}}]{TPCTwoGamma:1988yjh}%
  \BibitemOpen
  \bibfield  {author} {\bibinfo {author} {\bibfnamefont {H.}~\bibnamefont
  {Aihara}} \emph {et~al.},\ }\href {\doibase 10.1103/PhysRevLett.61.1263}
  {\bibfield  {journal} {\bibinfo  {journal} {Phys. Rev. Lett.}\ }\textbf
  {\bibinfo {volume} {61}},\ \bibinfo {pages} {1263} (\bibinfo {year}
  {1988})}\BibitemShut {NoStop}%
\bibitem [{\citenamefont {Itoh}\ \emph {et~al.}(1995)\citenamefont {Itoh} \emph
  {et~al.}}]{TOPAZ:1994voc}%
  \BibitemOpen
  \bibfield  {author} {\bibinfo {author} {\bibfnamefont {R.}~\bibnamefont
  {Itoh}} \emph {et~al.},\ }\href {\doibase 10.1016/0370-2693(94)01685-6}
  {\bibfield  {journal} {\bibinfo  {journal} {Phys. Lett. B}\ }\textbf
  {\bibinfo {volume} {345}},\ \bibinfo {pages} {335} (\bibinfo {year}
  {1995})}\BibitemShut {NoStop}%
\bibitem [{\citenamefont {Buskulic}\ \emph
  {et~al.}(1995{\natexlab{a}})\citenamefont {Buskulic} \emph
  {et~al.}}]{ALEPH:1994cbg}%
  \BibitemOpen
  \bibfield  {author} {\bibinfo {author} {\bibfnamefont {D.}~\bibnamefont
  {Buskulic}} \emph {et~al.},\ }\href {\doibase 10.1007/BF01556360} {\bibfield
  {journal} {\bibinfo  {journal} {Z. Phys. C}\ }\textbf {\bibinfo {volume}
  {66}},\ \bibinfo {pages} {355} (\bibinfo {year}
  {1995}{\natexlab{a}})}\BibitemShut {NoStop}%
\bibitem [{\citenamefont {Abreu}\ \emph {et~al.}(1998)\citenamefont {Abreu}
  \emph {et~al.}}]{DELPHI:1998cgx}%
  \BibitemOpen
  \bibfield  {author} {\bibinfo {author} {\bibfnamefont {P.}~\bibnamefont
  {Abreu}} \emph {et~al.},\ }\href {\doibase 10.1007/s100529800989} {\bibfield
  {journal} {\bibinfo  {journal} {Eur. Phys. J. C}\ }\textbf {\bibinfo {volume}
  {5}},\ \bibinfo {pages} {585} (\bibinfo {year} {1998})}\BibitemShut {NoStop}%
\bibitem [{\citenamefont {Akers}\ \emph {et~al.}(1994)\citenamefont {Akers}
  \emph {et~al.}}]{OPAL:1994zan}%
  \BibitemOpen
  \bibfield  {author} {\bibinfo {author} {\bibfnamefont {R.}~\bibnamefont
  {Akers}} \emph {et~al.},\ }\href {\doibase 10.1007/BF01411010} {\bibfield
  {journal} {\bibinfo  {journal} {Z. Phys. C}\ }\textbf {\bibinfo {volume}
  {63}},\ \bibinfo {pages} {181} (\bibinfo {year} {1994})}\BibitemShut
  {NoStop}%
\bibitem [{\citenamefont {Abe}\ \emph {et~al.}(2004)\citenamefont {Abe} \emph
  {et~al.}}]{SLD:2003ogn}%
  \BibitemOpen
  \bibfield  {author} {\bibinfo {author} {\bibfnamefont {K.}~\bibnamefont
  {Abe}} \emph {et~al.},\ }\href {\doibase 10.1103/PhysRevD.69.072003}
  {\bibfield  {journal} {\bibinfo  {journal} {Phys. Rev. D}\ }\textbf {\bibinfo
  {volume} {69}},\ \bibinfo {pages} {072003} (\bibinfo {year}
  {2004})}\BibitemShut {NoStop}%
\bibitem [{\citenamefont {Althoff}\ \emph {et~al.}(1983)\citenamefont {Althoff}
  \emph {et~al.}}]{TASSO:1982bkc}%
  \BibitemOpen
  \bibfield  {author} {\bibinfo {author} {\bibfnamefont {M.}~\bibnamefont
  {Althoff}} \emph {et~al.},\ }\href {\doibase 10.1007/BF01577813} {\bibfield
  {journal} {\bibinfo  {journal} {Z. Phys. C}\ }\textbf {\bibinfo {volume}
  {17}},\ \bibinfo {pages} {5} (\bibinfo {year} {1983})}\BibitemShut {NoStop}%
\bibitem [{\citenamefont {Brandelik}\ \emph
  {et~al.}(1981{\natexlab{b}})\citenamefont {Brandelik} \emph
  {et~al.}}]{TASSO:1981gag}%
  \BibitemOpen
  \bibfield  {author} {\bibinfo {author} {\bibfnamefont {R.}~\bibnamefont
  {Brandelik}} \emph {et~al.},\ }\href {\doibase 10.1016/0370-2693(81)90104-0}
  {\bibfield  {journal} {\bibinfo  {journal} {Phys. Lett. B}\ }\textbf
  {\bibinfo {volume} {100}},\ \bibinfo {pages} {357} (\bibinfo {year}
  {1981}{\natexlab{b}})}\BibitemShut {NoStop}%
\bibitem [{\citenamefont {Buskulic}\ \emph
  {et~al.}(1995{\natexlab{b}})\citenamefont {Buskulic} \emph
  {et~al.}}]{ALEPH:1995njx}%
  \BibitemOpen
  \bibfield  {author} {\bibinfo {author} {\bibfnamefont {D.}~\bibnamefont
  {Buskulic}} \emph {et~al.},\ }\href {\doibase 10.1016/0370-2693(95)00917-A}
  {\bibfield  {journal} {\bibinfo  {journal} {Phys. Lett. B}\ }\textbf
  {\bibinfo {volume} {357}},\ \bibinfo {pages} {487} (\bibinfo {year}
  {1995}{\natexlab{b}})},\ \bibinfo {note} {[Erratum: Phys. Lett. B {\bf 364},
  247 (1995)]}\BibitemShut {NoStop}%
\bibitem [{\citenamefont {Ackerstaff}\ \emph {et~al.}(1999)\citenamefont
  {Ackerstaff} \emph {et~al.}}]{OPAL:1998arz}%
  \BibitemOpen
  \bibfield  {author} {\bibinfo {author} {\bibfnamefont {K.}~\bibnamefont
  {Ackerstaff}} \emph {et~al.},\ }\href {\doibase 10.1007/s100529901067}
  {\bibfield  {journal} {\bibinfo  {journal} {Eur. Phys. J. C}\ }\textbf
  {\bibinfo {volume} {7}},\ \bibinfo {pages} {369} (\bibinfo {year}
  {1999})}\BibitemShut {NoStop}%
\bibitem [{\citenamefont {Sato}\ \emph {et~al.}(2020)\citenamefont {Sato},
  \citenamefont {Andres}, \citenamefont {Ethier},\ and\ \citenamefont
  {Melnitchouk}}]{JAM19}%
  \BibitemOpen
  \bibfield  {author} {\bibinfo {author} {\bibfnamefont {N.}~\bibnamefont
  {Sato}}, \bibinfo {author} {\bibfnamefont {C.}~\bibnamefont {Andres}},
  \bibinfo {author} {\bibfnamefont {J.~J.}\ \bibnamefont {Ethier}}, \ and\
  \bibinfo {author} {\bibfnamefont {W.}~\bibnamefont {Melnitchouk}},\ }\href
  {\doibase 10.1103/PhysRevD.101.074020} {\bibfield  {journal} {\bibinfo
  {journal} {Phys. Rev. D}\ }\textbf {\bibinfo {volume} {101}},\ \bibinfo
  {pages} {074020} (\bibinfo {year} {2020})}\BibitemShut {NoStop}%
\bibitem [{\citenamefont {Cocuzza}\ \emph
  {et~al.}(2021{\natexlab{b}})\citenamefont {Cocuzza}, \citenamefont {Keppel},
  \citenamefont {Liu}, \citenamefont {Melnitchouk}, \citenamefont {Metz},
  \citenamefont {Sato},\ and\ \citenamefont {Thomas}}]{Cocuzza:2021rfn}%
  \BibitemOpen
  \bibfield  {author} {\bibinfo {author} {\bibfnamefont {C.}~\bibnamefont
  {Cocuzza}}, \bibinfo {author} {\bibfnamefont {C.~E.}\ \bibnamefont {Keppel}},
  \bibinfo {author} {\bibfnamefont {H.}~\bibnamefont {Liu}}, \bibinfo {author}
  {\bibfnamefont {W.}~\bibnamefont {Melnitchouk}}, \bibinfo {author}
  {\bibfnamefont {A.}~\bibnamefont {Metz}}, \bibinfo {author} {\bibfnamefont
  {N.}~\bibnamefont {Sato}}, \ and\ \bibinfo {author} {\bibfnamefont {A.~W.}\
  \bibnamefont {Thomas}},\ }\href {\doibase 10.1103/PhysRevLett.127.242001}
  {\bibfield  {journal} {\bibinfo  {journal} {Phys. Rev. Lett.}\ }\textbf
  {\bibinfo {volume} {127}},\ \bibinfo {pages} {242001} (\bibinfo {year}
  {2021}{\natexlab{b}})}\BibitemShut {NoStop}%
\bibitem [{\citenamefont {Hyndman}\ and\ \citenamefont {Fan}(1996)}]{Hyndman}%
  \BibitemOpen
  \bibfield  {author} {\bibinfo {author} {\bibfnamefont {R.}~\bibnamefont
  {Hyndman}}\ and\ \bibinfo {author} {\bibfnamefont {Y.}~\bibnamefont {Fan}},\
  }\href {\doibase 10.1080/00031305.1996.10473566} {\bibfield  {journal}
  {\bibinfo  {journal} {The American Statistician}\ }\textbf {\bibinfo {volume}
  {50}},\ \bibinfo {pages} {361} (\bibinfo {year} {1996})}\BibitemShut
  {NoStop}%
\bibitem [{\citenamefont {Harland-Lang}\ \emph {et~al.}(2015)\citenamefont
  {Harland-Lang}, \citenamefont {Martin}, \citenamefont {Motylinski},\ and\
  \citenamefont {Thorne}}]{MMHT14}%
  \BibitemOpen
  \bibfield  {author} {\bibinfo {author} {\bibfnamefont {L.~A.}\ \bibnamefont
  {Harland-Lang}}, \bibinfo {author} {\bibfnamefont {A.~D.}\ \bibnamefont
  {Martin}}, \bibinfo {author} {\bibfnamefont {P.}~\bibnamefont {Motylinski}},
  \ and\ \bibinfo {author} {\bibfnamefont {R.~S.}\ \bibnamefont {Thorne}},\
  }\href {https://doi.org/10.1140%2Fepjc%2Fs10052-015-3397-6} {\bibfield
  {journal} {\bibinfo  {journal} {Eur. Phys. J. C}\ }\textbf {\bibinfo {volume}
  {75}},\ \bibinfo {pages} {204} (\bibinfo {year} {2015})}\BibitemShut
  {NoStop}%
\bibitem [{\citenamefont {Boglione}\ \emph {et~al.}(2019)\citenamefont
  {Boglione}, \citenamefont {Dotson}, \citenamefont {Gamberg}, \citenamefont
  {Gordon}, \citenamefont {Gonzalez-Hernandez}, \citenamefont {Prokudin},
  \citenamefont {Rogers},\ and\ \citenamefont {Sato}}]{Boglione:2019nwk}%
  \BibitemOpen
  \bibfield  {author} {\bibinfo {author} {\bibfnamefont {M.}~\bibnamefont
  {Boglione}}, \bibinfo {author} {\bibfnamefont {A.}~\bibnamefont {Dotson}},
  \bibinfo {author} {\bibfnamefont {L.}~\bibnamefont {Gamberg}}, \bibinfo
  {author} {\bibfnamefont {S.}~\bibnamefont {Gordon}}, \bibinfo {author}
  {\bibfnamefont {J.~O.}\ \bibnamefont {Gonzalez-Hernandez}}, \bibinfo {author}
  {\bibfnamefont {A.}~\bibnamefont {Prokudin}}, \bibinfo {author}
  {\bibfnamefont {T.~C.}\ \bibnamefont {Rogers}}, \ and\ \bibinfo {author}
  {\bibfnamefont {N.}~\bibnamefont {Sato}},\ }\href {\doibase
  10.1007/JHEP10(2019)122} {\bibfield  {journal} {\bibinfo  {journal} {JHEP}\
  }\textbf {\bibinfo {volume} {10}},\ \bibinfo {pages} {122} (\bibinfo {year}
  {2019})}\BibitemShut {NoStop}%
\bibitem [{\citenamefont {Bhatt}\ \emph {et~al.}(2025)\citenamefont {Bhatt}
  \emph {et~al.}}]{BHATT2025139485}%
  \BibitemOpen
  \bibfield  {author} {\bibinfo {author} {\bibfnamefont {H.}~\bibnamefont
  {Bhatt}} \emph {et~al.},\ }\href {\doibase
  https://doi.org/10.1016/j.physletb.2025.139485} {\bibfield  {journal}
  {\bibinfo  {journal} {Phys. Lett. B}\ }\textbf {\bibinfo {volume} {865}},\
  \bibinfo {pages} {139485} (\bibinfo {year} {2025})}\BibitemShut {NoStop}%
\bibitem [{\citenamefont {Abdul~Khalek}\ \emph {et~al.}(2022)\citenamefont
  {Abdul~Khalek}, \citenamefont {Bertone}, \citenamefont {Khoudli},\ and\
  \citenamefont {Nocera}}]{AbdulKhalek:2022laj}%
  \BibitemOpen
  \bibfield  {author} {\bibinfo {author} {\bibfnamefont {R.}~\bibnamefont
  {Abdul~Khalek}}, \bibinfo {author} {\bibfnamefont {V.}~\bibnamefont
  {Bertone}}, \bibinfo {author} {\bibfnamefont {A.}~\bibnamefont {Khoudli}}, \
  and\ \bibinfo {author} {\bibfnamefont {E.~R.}\ \bibnamefont {Nocera}},\
  }\href {\doibase 10.1016/j.physletb.2022.137456} {\bibfield  {journal}
  {\bibinfo  {journal} {Phys. Lett. B}\ }\textbf {\bibinfo {volume} {834}},\
  \bibinfo {pages} {137456} (\bibinfo {year} {2022})},\ \Eprint
  {http://arxiv.org/abs/2204.10331} {arXiv:2204.10331 [hep-ph]} \BibitemShut
  {NoStop}%
\bibitem [{\citenamefont {Ball}\ \emph {et~al.}(2022)\citenamefont {Ball} \emph
  {et~al.}}]{NNPDF4.0}%
  \BibitemOpen
  \bibfield  {author} {\bibinfo {author} {\bibfnamefont {R.~D.}\ \bibnamefont
  {Ball}} \emph {et~al.},\ }\href {\doibase 10.1140/epjc/s10052-022-10328-7}
  {\bibfield  {journal} {\bibinfo  {journal} {Eur. Phys. J. C}\ }\textbf
  {\bibinfo {volume} {82}},\ \bibinfo {pages} {428} (\bibinfo {year}
  {2022})}\BibitemShut {NoStop}%
\bibitem [{\citenamefont {Hou}\ \emph {et~al.}(2021)\citenamefont {Hou} \emph
  {et~al.}}]{CT18}%
  \BibitemOpen
  \bibfield  {author} {\bibinfo {author} {\bibfnamefont {T.-J.}\ \bibnamefont
  {Hou}} \emph {et~al.},\ }\href {\doibase 10.1103/PhysRevD.103.014013}
  {\bibfield  {journal} {\bibinfo  {journal} {Phys. Rev. D}\ }\textbf {\bibinfo
  {volume} {103}},\ \bibinfo {pages} {014013} (\bibinfo {year}
  {2021})}\BibitemShut {NoStop}%
\bibitem [{\citenamefont {Accardi}\ \emph
  {et~al.}(2009{\natexlab{b}})\citenamefont {Accardi}, \citenamefont {Hobbs},\
  and\ \citenamefont {Melnitchouk}}]{Accardi:2009md}%
  \BibitemOpen
  \bibfield  {author} {\bibinfo {author} {\bibfnamefont {A.}~\bibnamefont
  {Accardi}}, \bibinfo {author} {\bibfnamefont {T.}~\bibnamefont {Hobbs}}, \
  and\ \bibinfo {author} {\bibfnamefont {W.}~\bibnamefont {Melnitchouk}},\
  }\href {\doibase 10.1088/1126-6708/2009/11/084} {\bibfield  {journal}
  {\bibinfo  {journal} {JHEP}\ }\textbf {\bibinfo {volume} {11}},\ \bibinfo
  {pages} {084} (\bibinfo {year} {2009}{\natexlab{b}})}\BibitemShut {NoStop}%
\bibitem [{\citenamefont {Guerrero}\ \emph {et~al.}(2015)\citenamefont
  {Guerrero}, \citenamefont {Ethier}, \citenamefont {Accardi}, \citenamefont
  {Casper},\ and\ \citenamefont {Melnitchouk}}]{Guerrero:2015wha}%
  \BibitemOpen
  \bibfield  {author} {\bibinfo {author} {\bibfnamefont {J.~V.}\ \bibnamefont
  {Guerrero}}, \bibinfo {author} {\bibfnamefont {J.~J.}\ \bibnamefont
  {Ethier}}, \bibinfo {author} {\bibfnamefont {A.}~\bibnamefont {Accardi}},
  \bibinfo {author} {\bibfnamefont {S.~W.}\ \bibnamefont {Casper}}, \ and\
  \bibinfo {author} {\bibfnamefont {W.}~\bibnamefont {Melnitchouk}},\ }\href
  {\doibase 10.1007/JHEP09(2015)169} {\bibfield  {journal} {\bibinfo  {journal}
  {JHEP}\ }\textbf {\bibinfo {volume} {09}},\ \bibinfo {pages} {169} (\bibinfo
  {year} {2015})}\BibitemShut {NoStop}%
\bibitem [{\citenamefont {Guerrero}\ and\ \citenamefont
  {Accardi}(2018)}]{Guerrero:2017yvf}%
  \BibitemOpen
  \bibfield  {author} {\bibinfo {author} {\bibfnamefont {J.~V.}\ \bibnamefont
  {Guerrero}}\ and\ \bibinfo {author} {\bibfnamefont {A.}~\bibnamefont
  {Accardi}},\ }\href {\doibase 10.1103/PhysRevD.97.114012} {\bibfield
  {journal} {\bibinfo  {journal} {Phys. Rev. D}\ }\textbf {\bibinfo {volume}
  {97}},\ \bibinfo {pages} {114012} (\bibinfo {year} {2018})}\BibitemShut
  {NoStop}%
\bibitem [{\citenamefont {Faura}\ \emph {et~al.}(2020)\citenamefont {Faura},
  \citenamefont {Iranipour}, \citenamefont {Nocera}, \citenamefont {Rojo},\
  and\ \citenamefont {Ubiali}}]{Faura:2020oom}%
  \BibitemOpen
  \bibfield  {author} {\bibinfo {author} {\bibfnamefont {F.}~\bibnamefont
  {Faura}}, \bibinfo {author} {\bibfnamefont {S.}~\bibnamefont {Iranipour}},
  \bibinfo {author} {\bibfnamefont {E.~R.}\ \bibnamefont {Nocera}}, \bibinfo
  {author} {\bibfnamefont {J.}~\bibnamefont {Rojo}}, \ and\ \bibinfo {author}
  {\bibfnamefont {M.}~\bibnamefont {Ubiali}},\ }\href {\doibase
  10.1140/epjc/s10052-020-08749-3} {\bibfield  {journal} {\bibinfo  {journal}
  {Eur. Phys. J. C}\ }\textbf {\bibinfo {volume} {80}},\ \bibinfo {pages}
  {1168} (\bibinfo {year} {2020})}\BibitemShut {NoStop}%
\bibitem [{\citenamefont {Bailey}\ \emph {et~al.}(2021)\citenamefont {Bailey},
  \citenamefont {Cridge}, \citenamefont {Harland-Lang}, \citenamefont
  {Martin},\ and\ \citenamefont {Thorne}}]{Bailey:2020ooq}%
  \BibitemOpen
  \bibfield  {author} {\bibinfo {author} {\bibfnamefont {S.}~\bibnamefont
  {Bailey}}, \bibinfo {author} {\bibfnamefont {T.}~\bibnamefont {Cridge}},
  \bibinfo {author} {\bibfnamefont {L.~A.}\ \bibnamefont {Harland-Lang}},
  \bibinfo {author} {\bibfnamefont {A.~D.}\ \bibnamefont {Martin}}, \ and\
  \bibinfo {author} {\bibfnamefont {R.~S.}\ \bibnamefont {Thorne}},\ }\href
  {\doibase 10.1140/epjc/s10052-021-09057-0} {\bibfield  {journal} {\bibinfo
  {journal} {Eur. Phys. J. C}\ }\textbf {\bibinfo {volume} {81}},\ \bibinfo
  {pages} {341} (\bibinfo {year} {2021})}\BibitemShut {NoStop}%
\bibitem [{\citenamefont {Aad}\ \emph {et~al.}(2022)\citenamefont {Aad} \emph
  {et~al.}}]{ATLAS:2021vod}%
  \BibitemOpen
  \bibfield  {author} {\bibinfo {author} {\bibfnamefont {G.}~\bibnamefont
  {Aad}} \emph {et~al.},\ }\href {\doibase 10.1140/epjc/s10052-022-10217-z}
  {\bibfield  {journal} {\bibinfo  {journal} {Eur. Phys. J. C}\ }\textbf
  {\bibinfo {volume} {82}},\ \bibinfo {pages} {438} (\bibinfo {year}
  {2022})}\BibitemShut {NoStop}%
\bibitem [{\citenamefont {Hirai}\ \emph {et~al.}(2007)\citenamefont {Hirai},
  \citenamefont {Kumano}, \citenamefont {Nagai},\ and\ \citenamefont
  {Sudoh}}]{HKNS}%
  \BibitemOpen
  \bibfield  {author} {\bibinfo {author} {\bibfnamefont {M.}~\bibnamefont
  {Hirai}}, \bibinfo {author} {\bibfnamefont {S.}~\bibnamefont {Kumano}},
  \bibinfo {author} {\bibfnamefont {T.-H.}\ \bibnamefont {Nagai}}, \ and\
  \bibinfo {author} {\bibfnamefont {K.}~\bibnamefont {Sudoh}},\ }\href
  {\doibase 10.1103/PhysRevD.75.094009} {\bibfield  {journal} {\bibinfo
  {journal} {Phys. Rev. D}\ }\textbf {\bibinfo {volume} {75}},\ \bibinfo
  {pages} {094009} (\bibinfo {year} {2007})}\BibitemShut {NoStop}%
\end{thebibliography}%

\end{document}